\documentclass[11pt,a4paper,reqno]{amsart}
\usepackage[T1]{fontenc}
\usepackage{lmodern,microtype}
\usepackage{amsmath,amssymb,amsthm,mathtools}
\usepackage[margin=27mm]{geometry}
\usepackage{enumitem,needspace}
\usepackage{xcolor}
\usepackage[colorlinks=true,linkcolor=red,
 citecolor=black,urlcolor=black,linktoc=page]{hyperref}
\let\OriginalEqref\eqref
\renewcommand{\eqref}[1]{\textcolor{red}{\OriginalEqref{#1}}}
\hypersetup{
 pdftitle={Widom asymptotics for spatial compressions of Schrodinger spectral projections},
 pdfauthor={Pan Ma and Daoyuan Zheng},
 pdfsubject={Semiclassical trace asymptotics and free-fermion entanglement},
 pdfkeywords={Widom conjecture, Schrodinger operators, Schatten classes, entanglement entropy}
}
\numberwithin{equation}{section}
\newtheorem{theorem}{Theorem}[section]
\newtheorem{lemma}[theorem]{Lemma}
\newtheorem{proposition}[theorem]{Proposition}
\newtheorem{corollary}[theorem]{Corollary}
\theoremstyle{definition}

\newtheorem{example}[theorem]{Example}

\theoremstyle{remark}
\newtheorem{remark}[theorem]{Remark}
\DeclareMathOperator{\Tr}{Tr}
\DeclareMathOperator{\Op}{Op}

\newcommand{\R}{\mathbb R}

\setlist{itemsep=3pt,topsep=6pt}
\title[Widom asymptotics for Schr\"odinger projections]{Widom asymptotics for spatial compressions of Schr\"odinger spectral projections}
\author[Pan Ma]{Pan Ma}
\address{School of Mathematics and Statistics, HNP-LAMA, Central South
University, Changsha 410083, China}
\email{pan.ma@csu.edu.cn}
\author[Daoyuan Zheng]{Daoyuan Zheng}
\address{School of Mathematics and Statistics, HNP-LAMA, Central South
University, Changsha 410083, China}
\email{242111058@csu.edu.cn}
\date{}

\begin{document}
\begin{abstract}
We establish a Widom trace formula for spatial compressions of exact
spectral projections of semiclassical Schr\"odinger operators. The potential
is smooth near a compact classically allowed region, and the observation
domain lies strictly inside that region. For every continuous spectral
function with positive power decay at both endpoints, we obtain a trace
formula at scale $h^{1-n}\log h^{-1}$ whose coefficient separates into
a boundary integral and a universal spectral functional. This yields the exact
von Neumann and R\'enyi entanglement asymptotics for every fixed positive
order, and proves their universal ratios to the particle-number variance.
The proof combines commutator estimates in every fixed positive Schatten
class with a local trace formula obtained from a boundary-adapted normal
form and a Hardy model. In dimensions $n\ge2$, a positive partition and
the known variance asymptotic control the grazing contribution; in one
dimension all boundary Fermi points are transverse. A harmonic-oscillator
counterexample shows that continuity and weighted integrability alone
are insufficient for the trace formula.
\end{abstract}
\maketitle
\enlargethispage{2pt}
\section{Introduction}
\label{sec:introduction}

The interaction of sharp localization in energy and in position produces
a transition spectrum that is absent from either projection separately.
Indeed, the Fermi projection $P_h$ has spectrum contained in $\{0,1\}$,
so $g(P_h)=0$ whenever $g(0)=g(1)=0$. By contrast, its spatial
compression $T_h:=\mathbf1_\Omega P_h\mathbf1_\Omega$ generally has
nontrivial spectrum in $(0,1)$. This boundary-generated spectrum governs
both particle-number fluctuations and free-fermion entanglement. We
determine its leading semiclassical nonlinear statistics and identify
the universal local mechanism from which they arise.

For variable potentials, Deleporte and Lambert proved local limits for
the spectral-projector kernel \cite{DeleporteLambertLocalWeyl}, as well
as the particle-number variance asymptotic and entropy bounds
\cite{DeleporteLambert}. The variance determines only the quadratic
statistic $\operatorname{Tr}T_h(I-T_h)$; it does not control the
accumulation of transition eigenvalues at $0$ and $1$, which is essential
for general nonlinear functions and, in particular, for entropy. We
prove the trace formula for every continuous $g$ having fixed positive
power decay at both endpoints, in every dimension $n\ge1$. On compact
subintervals of $(0,1)$, this is a locally finite limiting law whose
density is universal up to a geometric factor. In the
entanglement-energy variable the limiting density is constant.

The two principal ingredients proved here are a one-logarithm commutator
estimate in every fixed positive Schatten class and a local nonlinear
trace formula obtained from a boundary-preserving normal form and the
Hardy compression model. The first controls the spectral endpoints; the
second identifies the universal density. In dimensions $n\ge2$, the
variance asymptotic of Deleporte--Lambert is used only in the final
positive gluing argument to control the grazing remainder. In dimension
one every boundary Fermi point is transverse. The resulting trace law
yields the exact logarithmic term for the von Neumann entropy and for
each R\'enyi entropy of fixed positive order.

\subsection{Assumptions and the main result}

Let $n\ge1$, and let $V\in L^1_{\mathrm{loc}}(\mathbb R^n;\mathbb R)$
be bounded below. Fix a representative of $V$ and an energy
$\mu>\operatorname*{ess\,inf}_{\mathbb R^n}V$, and set
\[
 D=\{x\in\mathbb R^n:V(x)<\mu\}.
\]
We assume that $D$ is relatively compact, that $V$ is smooth on a
neighborhood of $\overline D$, and that
\begin{equation*}
 \partial D=\{x:V(x)=\mu\},\qquad
 \nabla V(x)\ne0\quad (x\in\partial D).
\end{equation*}
The chosen representative has the stated smoothness near $\overline D$.
Outside this neighborhood, only the local integrability and lower-bound
hypotheses are imposed.

For $h>0$, let $H_h=-h^2\Delta+V$ denote the self-adjoint operator
associated with the closed, semibounded quadratic form
\begin{equation}
 \mathfrak h_h[u]=h^2\|\nabla u\|_2^2+\int_{\mathbb R^n}V(x)|u(x)|^2\,dx,
 \qquad
 \mathcal D(\mathfrak h_h)
 =H^1(\mathbb R^n)\cap L^2(\mathbb R^n,V_+(x)\,dx).
 \label{eq:form-realization}
\end{equation}
Here $V_+(x)=\max\{V(x),0\}$. In every dimension, assume that $\Omega\Subset D$. For $n\ge2$, let
$\Omega$ be a bounded domain with smooth boundary; for $n=1$, let
$\Omega$ be a finite union of bounded open intervals. We use the boundary convention
\begin{equation*}
 \partial_*\Omega=
 \begin{cases}
  \partial\Omega,&n\ge2,\\
  \partial(\operatorname{int}\overline\Omega),&n=1.
 \end{cases}
\end{equation*}
In dimension one, this convention counts the essential endpoints;
inserting or deleting finitely many points does not change the spatial
multiplication operator.

Let $\mathsf E_{H_h}$ be the projection-valued spectral measure of
$H_h$. The Fermi projection is
\[
 P_h:=\mathbf1_{(-\infty,\mu]}(H_h)
 =\mathsf E_{H_h}((-\infty,\mu])
 =\int_{\mathbb R}\mathbf1_{(-\infty,\mu]}(\lambda)
       \,d\mathsf E_{H_h}(\lambda).
\]
Thus $P_h=P_h^*=P_h^2$. Let $Q$ be the multiplication operator
$(Qu)(x)=\mathbf1_\Omega(x)u(x)$ on $L^2(\mathbb R^n)$, and set
\[
 T_h=QP_hQ,\qquad E_h=T_h(I-T_h),\qquad
 0\le T_h\le I,\qquad 0\le E_h\le\tfrac14 I.
\]
Here $I$ is the identity and $\operatorname{Tr}$ is the operator trace.
Lemma~\ref{loc:window} proves that $P_h$ has finite rank.

Put
\begin{equation}
 M_h=(2\pi h)^{1-n}\log h^{-1},\qquad
 C_{\Omega,1}=\frac{c_{n-1}}{2\pi^2}
    \int_{\partial_*\Omega}(\mu-V(x))^{(n-1)/2}\,d\sigma(x).
 \label{eq:constants-intro}
\end{equation}
Here
\[
 c_k=\frac{\pi^{k/2}}{\Gamma(k/2+1)},\qquad k\in\mathbb N_0,
\]
is the volume of the unit ball in $\mathbb R^k$, with $c_0=1$; $d\sigma$ is Euclidean surface measure for $n\ge2$ and counting measure for $n=1$.
For a spectral function $g$, let
\begin{equation}
 W(g)=\int_0^1\frac{g(t)}{t(1-t)}\,dt
 \label{eq:W-intro}
\end{equation}
whenever the integral converges absolutely.

\begin{samepage}
\begin{theorem}[Widom trace asymptotics]
\label{thm:main-target}
Under the preceding assumptions, let $g\in C([0,1];\mathbb C)$ satisfy
$g(0)=g(1)=0$ and, for some $\alpha_0,\alpha_1>0$,
\begin{equation}
 |g(t)|=O(t^{\alpha_0}),\qquad
 |g(1-t)|=O(t^{\alpha_1})\qquad(t\downarrow0).
 \label{eq:endpoint-powers}
\end{equation}
Then
\begin{equation}
 \Tr g(T_h)=C_{\Omega,1}W(g)(2\pi h)^{1-n}\log h^{-1}
       +o\bigl(h^{1-n}\log h^{-1}\bigr)
       \qquad(h\downarrow0).
 \label{eq:main-target}
\end{equation}
\end{theorem}
\end{samepage}

The cases $n=1$ and $n\ge2$ have different geometric content. If $n=1$
and $m=\tfrac12\#\partial_*\Omega$, then
\[
 \Tr g(T_h)=\frac{m}{\pi^2}W(g)\log h^{-1}+o(\log h^{-1}).
\]
Thus the leading coefficient counts the component intervals of
$\operatorname{int}\overline\Omega$ and is independent of the potential.
For $n\ge2$, the factor $c_{n-1}(\mu-V(x))^{(n-1)/2}$ in
\eqref{eq:constants-intro} is the volume of the tangential Fermi ball at
$x\in\partial\Omega$. In these dimensions the proof also treats the
grazing set
\[
 \{(x,\xi):x\in\partial\Omega,\ |\xi|^2+V(x)=\mu,
                    \ \xi\cdot\nu(x)=0\},
\]
where $\nu(x)$ is the unit normal. In dimension one the two momenta
$\xi=\pm\sqrt{\mu-V(x)}$ at each essential endpoint are transverse,
since $\Omega\Subset D$.

For $g(t)=t(1-t)$ one has $W(g)=1$, and
\eqref{eq:main-target} agrees with the variance asymptotic
\cite[Theorem~1]{DeleporteLambert} used in the global proof. This
quadratic case fixes the normalization of $C_{\Omega,1}$. The coefficient
printed in \cite[Conjecture~1]{DeleporteLambert} is twice this variance
normalization; see Remark~\ref{rem:DL-normalization}.

\subsection{Entanglement entropy}

For $0<t<1$, define
\begin{equation*}
 s_\alpha(t)=
 \begin{cases}
 \displaystyle\frac{\log(t^\alpha+(1-t)^\alpha)}{1-\alpha},
       &\alpha>0,\ \alpha\ne1,\\[6pt]
 -t\log t-(1-t)\log(1-t),&\alpha=1,
 \end{cases}
 \qquad s_\alpha(0)=s_\alpha(1)=0.
\end{equation*}
The R\'enyi entanglement entropy of order $\alpha$ for the free-fermion
state with correlation projection $P_h$ is
$S_\alpha(h)=\operatorname{Tr}s_\alpha(T_h)$; $S_1(h)$ is the
von Neumann entropy. Let $N_\Omega$ denote the particle count in
$\Omega$ for the determinantal process with correlation projection $P_h$.

\begin{corollary}[Entanglement entropies and particle-number variance]
\label{cor:entropy-intro}
Under the hypotheses of Theorem~\ref{thm:main-target}, assume that
$\Omega$ is nonempty. For every fixed $\alpha>0$,
\begin{equation}
 S_\alpha(h)
 =\frac{\pi^2}{6}(1+\alpha^{-1})C_{\Omega,1}M_h+o(M_h),
 \qquad h\downarrow0.
 \label{eq:entropy-asymptotic-intro}
\end{equation}
Moreover,
\begin{equation}
 \frac{S_\alpha(h)}{\operatorname{Var}N_\Omega}
 \longrightarrow\frac{\pi^2}{6}(1+\alpha^{-1}).
 \label{eq:entropy-variance-intro}
\end{equation}
In particular,
\[
 S_1(h)\sim\frac{\pi^2}{3}C_{\Omega,1}M_h
 \sim\frac{\pi^2}{3}\operatorname{Var}N_\Omega.
\]
\end{corollary}

The proof is given in Section~\ref{sec:applications}. For $n\ge2$,
\eqref{eq:entropy-asymptotic-intro} gives the logarithmic enhancement of
the boundary-area scale in a nonuniform potential. For $n=1$, it gives
\[
 S_\alpha(h)=\frac{m}{6}(1+\alpha^{-1})\log h^{-1}
             +o(\log h^{-1}),
\]
so a single interval has von Neumann coefficient $1/3$.
In every dimension the geometric factor cancels in
\eqref{eq:entropy-variance-intro}. For $\alpha=1$, the resulting
coefficient agrees with the prediction in
\cite[(1.7)]{DeleporteLambert}; the bounds in
\cite[Theorem~3]{DeleporteLambert} differ by a factor of
$\log\log h^{-1}$.

The role of the endpoint estimates is explicit in this application. For fixed
$0<\alpha<1$,
\[
 s_\alpha(t)\sim\frac{t^\alpha}{1-\alpha},\qquad
 \frac{s_1(t)}{t(1-t)}\sim\log(1/t)\qquad(t\downarrow0).
\]
The fractional moment estimates control the spectral endpoints for these
functions. All entropy orders below are fixed independently of $h$.

\subsection{The transition spectrum and entanglement energies}

Let $\lambda_{h,j}\in(0,1)$ be the transition eigenvalues of $T_h$,
counted with multiplicity, and define
\[
 \nu_h:=\frac1{M_h}\sum_j\delta_{\lambda_{h,j}}.
\]
Theorem~\ref{thm:main-target} implies the vague convergence of locally
finite Borel measures on $(0,1)$,
\[
 \nu_h\xrightarrow{\mathrm v}
 C_{\Omega,1}\frac{dt}{t(1-t)}.
\]
Equivalently, for every $\psi\in C_c((0,1))$,
\[
 \frac1{M_h}\sum_j\psi(\lambda_{h,j})
 \longrightarrow
 C_{\Omega,1}\int_0^1\frac{\psi(t)}{t(1-t)}\,dt.
\]
Thus the theorem determines a locally finite transition-spectrum law,
not a probability distribution on $[0,1]$. Under the change of variables
$u=\log((1-t)/t)$, its limiting measure becomes
$C_{\Omega,1}\,du$. This constant density is the common spectral
mechanism behind the leading entropy and fluctuation asymptotics. The
full theorem is stronger than the vague limit because it also applies
to endpoint-sensitive functions satisfying \eqref{eq:endpoint-powers}.

The following fixed-window consequence is proved in
Section~\ref{sec:applications}.

\begin{corollary}[Entanglement energies in a fixed window]
\label{cor:entanglement-spectrum}
Define the single-particle entanglement energies by
\[
 \varepsilon_{h,j}=\log\frac{1-\lambda_{h,j}}{\lambda_{h,j}}.
\]
For every $\varphi\in C_c(\mathbb R)$,
\begin{equation}\label{eq:entanglement-spectrum-vague}
 \frac1{M_h}\sum_j\varphi(\varepsilon_{h,j})
 \longrightarrow C_{\Omega,1}\int_{\mathbb R}\varphi(u)\,du.
\end{equation}
Consequently, for every fixed $a<b$,
\begin{equation}\label{eq:entanglement-window-count}
 \#\{j:a\le\varepsilon_{h,j}\le b\}
 =C_{\Omega,1}(b-a)M_h+o(M_h).
\end{equation}
\end{corollary}

The assertion concerns fixed windows and does not determine individual
level spacings. The positive-power estimates supply the additional
uniform integrability at the endpoints needed for entropy functions.

Section~\ref{sec:applications} also gives exact Schatten asymptotics
for the spatial commutator and a quadratic leading term for the centered
logarithmic moment-generating function, together with explicit geometric
examples.

\subsection{Limitations of the test-function range}

The endpoint condition \eqref{eq:endpoint-powers} implies absolute
convergence of $W(g)$, but continuity and that integrability alone do
not ensure the trace formula.

\begin{theorem}[Failure under weighted integrability alone]
\label{thm:endpoint-counterexample}
Let
\begin{equation}
 \begin{gathered}
 H_h=-h^2\frac{d^2}{dx^2}+x^2,\qquad
 P_h=\mathbf1_{(-\infty,1]}(H_h),\\
 \Omega=(-1/2,1/2),\qquad Q=\mathbf1_\Omega,\qquad T_h=QP_hQ.
 \end{gathered}
 \label{eq:counterexample-model}
\end{equation}
There exists a fixed function $g\in C([0,1])$ satisfying
\[
 g\ge0,\qquad g(t)=g(1-t),\qquad g(0)=g(1)=0,\qquad
 0<W(g)<\infty,
\]
and a sequence $h_k\downarrow0$ such that
\begin{equation*}
 \frac{\operatorname{Tr}g(T_{h_k})}{\log(h_k^{-1})}
 \longrightarrow+\infty.
\end{equation*}
Moreover, $g$ can be chosen so that, for every $\alpha>0$,
\begin{equation}
 \lim_{t\downarrow0}\frac{g(t)}{t^\alpha}
 =
 \lim_{t\downarrow0}\frac{g(1-t)}{t^\alpha}
 =+\infty.
 \label{eq:counterexample-slow-endpoints}
\end{equation}
\end{theorem}

The model satisfies the geometric assumptions of this paper:
$D=(-1,1)$, the energy is regular, and $\Omega\Subset D$.
The proof in Section~\ref{sec:endpoint-counterexample} uses finite-rank
harmonic-oscillator projections and analyticity of Hermite functions.
It disproves the continuous weighted-integrable formulation of
\cite[Conjecture~1]{DeleporteLambert}, even for nonnegative symmetric
functions. Positive-power decay remains a sufficient condition in
Theorem~\ref{thm:main-target}; its necessity for individual test functions
and the maximal admissible function class are separate questions.

The estimates below are pointwise in the moment exponent $q>0$ and
the R\'enyi order $\alpha>0$. For
$c_\alpha=\pi^2(1+\alpha^{-1})/6$, the entropy remainder satisfies
\[
 S_\alpha(h)-c_\alpha\operatorname{Var}N_\Omega=o(M_h).
\]
Uniform estimates as $q\downarrow0$ or $\alpha\downarrow0$,
logarithmic endpoint decay, and remainders of order $h^{1-n}$ are
further questions.

\subsection{Previous results and proof strategy}

Widom's analysis of truncated integral operators
\cite{Widom1982,Widom1990} identifies the functional $W(g)$ in the
logarithmic boundary term. Sobolev proved the multidimensional
Wiener--Hopf asymptotics \cite{Sobolev2013,Sobolev2015} and developed
small-exponent Schatten estimates for nonsmooth spectral functions
\cite{Sobolev2014Schatten,Sobolev2017Functions}. These results also allow
smooth symbols depending on position and momentum, with separate spatial
and momentum truncations. The connection with fermionic entanglement
was developed by Gioev and Klich \cite{GioevKlich}; the
translation-invariant R\'enyi entropy formula was proved by Leschke,
Sobolev, and Spitzer \cite{LeschkeSobolevSpitzer}.

Recent work establishes Widom asymptotics under weak regularity
assumptions for momentum multipliers \cite{FournaisEtAl2026} and sharp
eigenvalue estimates with extended test-function classes for localization
operators \cite{KulikovLarsen2026}.
Here we treat the exact spectral projection of a variable-potential
Schr\"odinger operator, for which the sharp energy cutoff must be
analyzed jointly with the spatial boundary. Related connections between stationary estimates or functional calculus
for Schr\"odinger-type operators and dispersive evolution are developed
in \cite{BRVV09,BVZ13,KPPV2026}.

For $p>0$, let $\mathfrak S_p$ be the ideal of compact operators with
\[
 \|A\|_{\mathfrak S_p}^p=\sum_j s_j(A)^p<\infty,
\]
where $s_j(A)$ are the singular values. For $p<1$ this defines a
quasi-norm; $\|\cdot\|$ denotes the operator norm.

\emph{Fractional transition moments.}
Section~\ref{sec:S6} proves, independently of the trace asymptotic,
\begin{equation}
 \Tr E_h^q=\tfrac12\|[P_h,Q]\|_{\mathfrak S_{2q}}^{2q}
 \le C_qM_h,\qquad q>0,\quad 0<h\le h_0<1.
 \label{eq:moments-intro}
\end{equation}
At each spatial scale $r$, the boundary count $O(r^{1-n})$ and
the block estimate $O((r/h)^{n-1})$ give $O(h^{1-n})$; summation over
dyadic scales produces one logarithm. For a test function satisfying
\eqref{eq:endpoint-powers}, a fixed
$0<q<\min\{1,\alpha_0,\alpha_1\}$ controls the endpoint
approximation in Section~\ref{sec:gluing}.

\emph{Local nonlinear traces.}
The Hardy compression gives $W(g)$ for smooth endpoint-vanishing
functions (Section~\ref{sec:hardy}). To transfer this formula to $T_h$,
we construct a canonical normal form preserving the two sides of the
spatial boundary, and estimate the errors for both the sharp spatial
projection and the sharp spectral projection. These operator estimates
give the local trace formula in
Theorem~\ref{thm:transverse-local-trace}; they retain the information
needed by the nonlinear functional calculus.

\emph{Global localization.}
In dimensions $n\ge2$, a positive partition over transverse charts
leaves a positive remainder. The variance theorem
\cite[Theorem~1]{DeleporteLambert}, together with the local formula for
$g(t)=t(1-t)$, bounds this remainder by the missing transverse variance.
Its normalized coefficient tends to zero as the transverse region
exhausts the boundary Fermi surface. The inequality
$|g(t)|\le C_g t(1-t)$ gives the same control for smooth
endpoint-vanishing $g$. In dimension one a finite transverse cover
contains all boundary Fermi points. Section~\ref{sec:gluing} completes
the argument in both cases and then uses \eqref{eq:moments-intro} to
reach \eqref{eq:endpoint-powers}.

Sections~\ref{sec:localization}--\ref{sec:toolbox} provide the smooth
confining replacement and the trace-ideal comparison estimates.
The semiclassical inputs are the regular-energy Weyl remainder
\cite[Theorem~2.19(i)]{Ivrii}, smooth functional calculus
\cite[Chapter~8]{DimassiSjostrand}, and normal forms and short-time
propagation \cite[Chapters~10--12]{Zworski}.
Appendix~\ref{sec:global-schwartz} supplies the global weighted
remainders for the auxiliary confining operator.
Sections~\ref{sec:applications} and~\ref{sec:endpoint-counterexample}
give the applications and the counterexample, respectively.

\section{Localization and confining replacement}
\label{sec:localization}

\subsection{Operator realization and the comparison theorem}
We retain the hypotheses of Section~\ref{sec:introduction} and the form
realization \eqref{eq:form-realization}. In particular,
$V\ge\mu$ almost everywhere on $\mathbb R^n\setminus D$.

Choose a smooth confining potential $V_{\mathrm c}$ equal to $V$ on a
neighborhood $U$ of $\overline D$, with $V_{\mathrm c}>\mu$ outside $\overline D$,
and equal to a positive quadratic potential plus a sufficiently large
constant outside a compact set. This is obtained by a smooth cutoff
inside the region where $V$ is smooth. Write
\[
 H_h^{\mathrm c}=-h^2\Delta+V_{\mathrm c},\qquad
 P_h^{\mathrm c}=\mathbf1_{(-\infty,\mu]}(H_h^{\mathrm c}).
\]

\begin{theorem}\label{loc:comparison}\label{thm:L7}
For every real $\chi\in C_c^\infty(U)$ and every $p>0$,
\[
 \|P_h\chi-\chi P_h^{\mathrm c}\|_{\mathfrak S_p}^{p}\le C_{p,\chi}h^{1-n},
 \qquad 0<h\le h_0.
\]
The same conclusion holds for complex $\chi$ by linearity. For cutoffs
supported in a common compact subset of $U$, the constant depends on
only finitely many derivatives of $\chi$. In
particular, a family satisfying
$0<\ell_h\le1$ and
$\|\partial^\alpha\chi_h\|_\infty\le C_\alpha\ell_h^{-|\alpha|}$
obeys the estimate with constant $C_p\ell_h^{-K_p}$ for some finite
$K_p$ depending on $p$.
\end{theorem}

\subsection{Spectral counting near the Fermi energy}
For a self-adjoint operator $A$, define
\[
 N_A^{\le}(E)=\operatorname{rank}\mathbf1_{(-\infty,E]}(A),
 \qquad
 N_A^{<}(E)=\operatorname{rank}\mathbf1_{(-\infty,E)}(A).
\]
Let $H_{h,\mathcal O}^{B}(W)$ denote the realization of
$-h^2\Delta+W$ on a bounded smooth domain $\mathcal O$, with
Dirichlet boundary conditions for $B=\mathrm D$ and Neumann boundary
conditions for $B=\mathrm N$. Set
\[
 N_{\mathcal O,B}^{\le}(E;W)
 =N_{H_{h,\mathcal O}^{B}(W)}^{\le}(E),
 \qquad
 N_{\mathcal O,B}^{<}(E;W)
 =N_{H_{h,\mathcal O}^{B}(W)}^{<}(E).
\]
The dependence on $h$ is suppressed; the potential argument is
omitted when $W=V$.

Let
\[
 p_W(x,\xi)=|\xi|^2+W(x).
\]
We use the following consequence of \cite[Theorem~2.19(i),
formula~(2.80)]{Ivrii}. Let $I,J\subset\mathbb R$ be compact intervals
with $I\subset\operatorname{int}J$, and let $U_\partial$ be a fixed
neighborhood of $\partial\mathcal O$. Assume that
\begin{equation}
 \inf_{\substack{x\in\overline{\mathcal O},\,\xi\in\mathbb R^n\\
                   p_W(x,\xi)\in J}}
 |\nabla_{x,\xi}p_W(x,\xi)|>0,
 \qquad
 \inf_{\substack{x\in U_\partial\cap\mathcal O\\ E\in J}}
 (W(x)-E)>0,
 \label{loc:ivrii-hypotheses}
\end{equation}
Then, uniformly for $E\in I$ and
$B\in\{\mathrm D,\mathrm N\}$,
\begin{equation}
 N_{\mathcal O,B}^{\le}(E;W)
 =(2\pi h)^{-n}\operatorname{vol}
 \{(x,\xi)\in T^*\mathcal O:p_W(x,\xi)\le E\}
 +O_I(h^{1-n}).
 \label{loc:weyl}
\end{equation}
Indeed, choose $E_-<\min_{\overline{\mathcal O}}W-1$ and apply
\cite[Theorem~2.19(i)]{Ivrii} with $X=\mathcal O$, $\psi\equiv1$,
$\lambda_1=E_-$, and $\lambda_2=E$. At $E_-$ the operator is
elliptic. At $E\in J$, the first condition in
\eqref{loc:ivrii-hypotheses} gives scalar microhyperbolicity after a
finite microlocal partition of the compact characteristic band. The
choice of $E_-$ gives ellipticity at the lower endpoint, while the
second condition gives boundary ellipticity for $E\in J$; together these
verify the boundary microhyperbolicity hypothesis for both Dirichlet and
Neumann boundary conditions. Compactness makes
the constants uniform for $E\in I$. The same formula holds for
$N_{\mathcal O,B}^{<}(E;W)$: take $E'\uparrow E$ with $E'\in J$ in
the uniform non-strict estimate and use that a regular energy surface
has phase volume zero.

\begin{lemma}[Finite negative rank and uniform shells]\label{loc:window}\label{lem:shell}
For each fixed $C_0>0$, there are constants $C,h_0>0$ such that
\begin{equation}
 \operatorname{rank}P_h+\operatorname{rank}P_h^{\mathrm c}\le Ch^{-n},\label{loc:finite-rank}
\end{equation}
and, for $h\le\delta\le C_0$,
\begin{equation}
 \operatorname{rank}\mathbf1_{[\mu-\delta,\mu]}(H_h)
 +\operatorname{rank}\mathbf1_{[\mu-\delta,\mu]}(H_h^{\mathrm c})
 \le C h^{-n}\delta.\label{loc:shell-rank}
\end{equation}
In particular, $\operatorname{rank}\mathbf1_{[\mu-h,\mu]}(H_h)
=O(h^{1-n})$.
\end{lemma}
\begin{proof}
Choose a bounded smooth domain $\mathcal O$ with
$\overline D\subset\mathcal O\Subset U$ and
$V-\mu\ge c>0$ on a collar of $\partial\mathcal O$.
Such a domain is obtained by enlarging $D$ by a sufficiently small
regular collar; the resulting domain may be disconnected.
Dirichlet--Neumann form bracketing applies to the form realization above.
On the exterior, $V-\mu\ge0$. Its Neumann realization has no nonzero
zero-energy vector: zero form energy forces $\nabla u=0$ and
$(V-\mu)|u|^2=0$; on an unbounded connected component an $L^2$ constant
vanishes, while on a bounded component the positive potential collar
forces that constant to vanish. Hence the exterior Neumann form satisfies
\[
 h^2\|\nabla u\|_2^2+\int_{\mathbb R^n\setminus\overline{\mathcal O}}
 (V-\mu)|u|^2\,dx>0\qquad(u\ne0).
\]

Consequently any subspace on which the full shifted form is nonpositive
has dimension at most $N_{\mathcal O,\mathrm N}^{\le}(\mu)$. Indeed, if it had larger
dimension, a nonzero vector would have interior component orthogonal
to all nonpositive interior Neumann modes, making the bracketed form
strictly positive. 

Bracketing also gives
\[
 N_{H_h}^{\le}(\mu)\le N_{\mathcal O,\mathrm N}^{\le}(\mu),\qquad
 N_{H_h}^{<}(E)\ge N_{\mathcal O,\mathrm D}^{<}(E),\quad E\le\mu.
\]
The same statements hold for $H_h^{\mathrm c}$, with the same interior
operator. Therefore the rank of the closed interval $[\mu-\delta,\mu]$
is bounded by
\[
 N_{\mathcal O,\mathrm N}^{\le}(\mu)
 -N_{\mathcal O,\mathrm D}^{<}(\mu-\delta).
\]
Here is the phase-volume estimate entering this difference. Set
$e(x,\xi)=|\xi|^2+V(x)$ on $T^*\mathcal O$ and
$\mathcal V(E)=\operatorname{vol}\{(x,\xi)\in T^*\mathcal O:e(x,\xi)\le E\}$.
For a sufficiently small fixed $\delta_0>0$, the energy band
$|e-\mu|\le\delta_0$ is compactly contained in $T^*\mathcal O$,
and $|\nabla_{x,\xi}e|$ is bounded below there. These statements follow
from the positive boundary collar, boundedness of the momentum on the
band, and regularity at $\mu$, after decreasing $\delta_0$.
The coarea formula \cite[Section~3.4]{EvansGariepy} gives,
for $0<\delta\le\delta_0$,
\[
 \mathcal V(\mu)-\mathcal V(\mu-\delta)
 =\int_{\mu-\delta}^{\mu}
   \left(\int_{e^{-1}(E)}
    \frac{d\Sigma_E}{|\nabla_{x,\xi}e|}\right)dE
 \le C\delta,
\]
where $d\Sigma_E$ is the induced hypersurface measure in phase space.
A finite cover by regular energy coordinates on the compact band
gives a uniform bound for the inner integral.
Combining this estimate with \eqref{loc:weyl} bounds the counting
difference by $C(h^{-n}\delta+h^{1-n})$.
Since $\delta\ge h$, this is \eqref{loc:shell-rank}.
For larger $\delta$ use \eqref{loc:finite-rank}, which follows from \eqref{loc:weyl} at $\mu$ (or the
usual bounded-domain upper counting estimate). The proof for the
confining operator is identical.
\end{proof}

\subsection{Mixed commutators and Schatten estimates}

Set $\mathsf D_x=-i\partial_x$; the subscript is omitted when the
variable is clear.

\begin{lemma}[Mixed commutators]\label{lem:mixed}
For differential operators $T$ with coefficients supported in $U$,
define
$\mathcal C_h(T)=H_hT-TH_h^{\mathrm c}$ as a local differential
expression.
For every integer $N\ge0$,
\begin{equation}
 \mathcal C_h^N(\chi)=h^N\sum_{|\alpha|\le N}
    a_{N,\alpha}(x;h)(h\mathsf D)^{\alpha},\label{loc:iterated-commutator}
\end{equation}
where the coefficients are smooth, supported in a fixed compact subset
of $U$, and have uniformly bounded derivatives. In particular
\begin{equation}
 \|\mathcal C_h^N(\chi)P_h^{\mathrm c}\|
 +\|P_h\mathcal C_h^N(\chi)\|\le C_Nh^N.
 \label{loc:negative-commutator-bound}
\end{equation}
The second expression in \eqref{loc:negative-commutator-bound} means its bounded extension, equivalently
the adjoint of $\mathcal C_h^N(\chi)^*P_h$.
\end{lemma}
\begin{proof}
On the support of all coefficients, $V=V_{\mathrm c}$ and both are smooth.
The first commutator is $[-h^2\Delta,\chi]$. Commuting a local
semiclassical differential operator of order at most $N$ with
$-h^2\Delta+V$ introduces a further factor $h$ and increases its order
by at most one. Induction proves \eqref{loc:iterated-commutator}, with
all coefficients supported in $U$.

Localized semiclassical elliptic regularity implies
\[
 \|\psi(h\mathsf D)^\alpha P_h\|
 +\|\psi(h\mathsf D)^\alpha P_h^{\mathrm c}\|\le C_{\psi,\alpha},
 \qquad \psi\in C_c^\infty(U).
\]
For completeness, write
$\|v\|_{H_h^s}=\|(1-h^2\Delta)^{s/2}v\|_2$.
For each integer $m\ge0$, nested compact cutoffs in $U$ give
\[
 \|\psi u\|_{H_h^{2m}}
 \le C_{\psi,m}\sum_{j=0}^{m}\|H_h^j u\|_2,
 \qquad u\in\operatorname{Dom}(H_h^m),
\]
and the same estimate holds for $H_h^{\mathrm c}$.
This is the localized elliptic parametrix estimate: extend the smooth
potential from a neighborhood of the cutoffs to a bounded smooth
potential, add a fixed positive constant to make the extended symbol
elliptic in $S(1+|\xi|^2)$, and apply the elliptic construction of
\cite[Theorem~4.29]{Zworski} and the Sobolev mapping property
\cite[Theorem~8.10(i)]{Zworski}, and iterate with cutoffs equal to one
near the preceding supports. The cutoff commutators have lower
differential order and are absorbed by the corresponding interior
estimates. Expanding the powers of the fixed shifted operator gives
the displayed sum. Only smoothness inside $U$ is used.
Both Hamiltonians have an $h$-independent lower bound $-M$.
The spectral theorem therefore bounds
$\|H_h^jP_h\|\le\max\{M,|\mu|\}^j$ for $j\ge1$, with the
analogous bound for $H_h^{\mathrm c}$.
Taking $2m\ge|\alpha|$ and a cutoff equal to one near
$\operatorname{supp}\psi$ proves the displayed derivative bounds.
Applying the bound to \eqref{loc:iterated-commutator} and its formal adjoint
proves \eqref{loc:negative-commutator-bound}. Local regularity and closure
extend these identities to the spectral vectors used below.
\end{proof}

\begin{proof}[Proof of Theorem \ref{loc:comparison}]
Shift both Hamiltonians by $\mu$. Put $P=P_h$, $P_c=P_h^{\mathrm c}$. The identity
\begin{equation}
 P\chi-\chi P_c=P\chi(1-P_c)-(1-P)\chi P_c\label{loc:opposite-sign-decomposition}
\end{equation}
reduces the argument to spectral blocks with opposite energy signs.
For either Hamiltonian, let $\Pi_h$ denote its projection onto
the shifted energy interval $[-h,0]$. Lemma~\ref{loc:window} gives
\[
 \operatorname{rank}\Pi_h\le Ch^{1-n},\qquad
 \|\Pi_h\chi\|_{\mathfrak S_p}^p
 \le\operatorname{rank}\Pi_h\,\|\chi\|^p
 \le C_{p,\chi}h^{1-n}.
\]
The same estimate applies to $\chi\Pi_h$ and to their projected blocks.

For the remaining first term use negative shells
$I_j=[-2^{j+1}h,-2^jh)$, $j\ge0$, and positive shells
$J_{-1}=(0,h)$ and $J_k=[2^kh,2^{k+1}h)$, $k\ge0$.
Let $E_j=\mathbf1_{I_j}(H_h-\mu)$,
$F_k=\mathbf1_{J_k}(H_h^{\mathrm c}-\mu)$,
and $X_{jk}=E_j\chi F_k$. Their spectral gap is at least
$c h2^{\max(j,k,0)}$. Iterated solution of the Sylvester equation gives
\begin{equation}
 \|X_{jk}\|
 \le C_N(h2^{\max(j,k,0)})^{-N}
       \|E_j\mathcal C_h^N(\chi)F_k\|
 \le C_N2^{-N\max(j,k,0)}.\label{loc:sylvester-block-bound}
\end{equation}
For $T\le-a<0$ and $S\ge b\ge0$, the equation $TX-XS=Y$ has solution
\[
 X=-\int_0^\infty e^{tT}Ye^{-tS}\,dt,\qquad
 \|X\|\le(a+b)^{-1}\|Y\|.
\]
Iteration gives \eqref{loc:sylvester-block-bound}. On each finite shell,
the identities hold on operator domains, and Lemma~\ref{lem:mixed}
provides bounds uniform in the positive shell.

Lemma \ref{loc:window} implies
\begin{equation*}
 \operatorname{rank}E_j\le Ch^{1-n}2^j.
\end{equation*}
For the remaining shells, $h2^j\ge\delta_0>0$ and
\[
 \operatorname{rank}E_j\le Ch^{-n}
 \le C\delta_0^{-1}h^{1-n}2^j.
\]
Thus, for $0<p\le1$,
\[
 \|X_{jk}\|_{\mathfrak S_p}^p
 \le Ch^{1-n}2^j2^{-Np\max(j,k,0)}.
\]
Choose $Np>1$. The double series converges, since
\[
 \sum_{j\ge0}\sum_{k\ge-1}
 2^j2^{-Np\max(j,k,0)}
 \le C\sum_{j\ge0}(j+2)2^{-(Np-1)j}<\infty.
\]
The Schatten $p$-triangle inequality proves the estimate for the first
term of \eqref{loc:opposite-sign-decomposition}. The convergent Schatten sum agrees with its strong
spectral-block sum. For the second term of \eqref{loc:opposite-sign-decomposition}, exchange the two
Hamiltonians and apply the same argument to $\chi^*$ and then take
adjoints. This proves the theorem for $0<p\le1$.

Finally $\|P\chi-\chi P_c\|\le2\|\chi\|$, and hence for $p>1$,
\[
 \|P\chi-\chi P_c\|_{\mathfrak S_p}^p
 \le(2\|\chi\|)^{p-1}\|P\chi-\chi P_c\|_{\mathfrak S_1}.
\]
All estimates involve only finitely many cutoff derivatives for fixed
$p$, proving the stated mesoscopic variant as well.
\end{proof}

\paragraph{Localized compressions.}
Set $T_h^{\mathrm c}=QP_h^{\mathrm c}Q$.
Choose $\chi\in C_c^\infty(\mathbb R^n)$ supported in the
region where $V=V_{\mathrm c}$ and equal to one on a
neighborhood of $\overline\Omega$.
Since $Q\chi=\chi Q=Q$, we have
\[
T_h-T_h^{\mathrm c}
=Q(P_h-P_h^{\mathrm c})Q
=Q(P_h\chi-\chi P_h^{\mathrm c})Q.
\]
Consequently, the ideal property and
Theorem~\ref{loc:comparison} yield, for every fixed $p>0$,
\[
\|T_h-T_h^{\mathrm c}\|_{\mathfrak S_p}^{p}
\le
\|P_h\chi-\chi P_h^{\mathrm c}\|_{\mathfrak S_p}^{p}
=O_p(h^{1-n}).
\]

\section{Trace ideals, smooth functional calculus, and spectral regularization}
\label{sec:toolbox}

Write $\|A\|_p=\|A\|_{\mathfrak S_p}$, and let $\|A\|$ denote
the operator norm. The Schatten ideal property and the triangle inequality
in power $r=\min\{p,1\}$ give
\[
 \|A+B\|_p^r\le\|A\|_p^r+\|B\|_p^r,
 \qquad
 \|BAC\|_p\le\|B\|\,\|A\|_p\,\|C\|.
\]
Here $A,B\in\mathfrak S_p$ in the first inequality, whereas $A\in\mathfrak S_p$
and $B,C$ are bounded in the second. These properties include the
quasi-normed range $0<p<1$; see \cite[Section~2.1]{Sobolev2017Functions}.
All Schatten exponents are fixed independently of $h$.

\begin{lemma}[Smooth quasi-intertwining in every Schatten ideal]
\label{toolbox:intertwining}
Let $A,B$ be bounded self-adjoint operators, possibly on different Hilbert
spaces, whose spectra lie in a fixed compact interval, and let $J$ be bounded.
For $g$ smooth on a neighborhood of that interval and $0<p<\infty$,
\[
 AJ-JB\in\mathfrak S_p\quad\Longrightarrow\quad
 \|g(A)J-Jg(B)\|_p\le C_{g,p}\|AJ-JB\|_p.
\]
The constant is controlled by finitely many derivatives of $g$.
\end{lemma}
\begin{proof}
Put $r=\min\{p,1\}$ and $K=AJ-JB$. On a neighborhood of the common
spectral interval, set
\[
 F(x,y)=\int_0^1g'(y+t(x-y))\,dt,
 \qquad (x-y)F(x,y)=g(x)-g(y).
\]
After multiplication by smooth cutoffs, extend $F$ periodically to a
larger square. For a fixed frequency $\omega>0$,
\[
 F(x,y)=\sum_{k,l\in\mathbb Z}c_{kl}e^{i\omega kx}e^{i\omega ly},
 \qquad |c_{kl}|\le C_N(1+|k|+|l|)^{-N}.
\]
Choose $N>\max\{2/r,3\}$, so that both
$\sum_{k,l}|c_{kl}|^r$ and
$\sum_{k,l}(1+|k|+|l|)|c_{kl}|$ are finite. The series
\begin{equation*}
 \mathcal T_{A,B}(K)=\sum_{k,l\in\mathbb Z}
 c_{kl}e^{i\omega kA}Ke^{i\omega lB}
\end{equation*}
converges in $\mathfrak S_p$ and satisfies
\[
 \|\mathcal T_{A,B}(K)\|_p^r
 \le\sum_{k,l}|c_{kl}|^r\|e^{i\omega kA}Ke^{i\omega lB}\|_p^r
 \le\Bigl(\sum_{k,l}|c_{kl}|^r\Bigr)\|K\|_p^r.
\]
Let $A_m,B_m$ be finite-spectrum self-adjoint approximants with spectra
in the same interval, and put $K_m=A_mJ-JB_m$. Then
\[
 \delta_m:=\|A_m-A\|+\|B_m-B\|\longrightarrow0,
 \qquad \|K_m-K\|\le\|J\|\delta_m.
\]
The operators $K_m$ are used only in operator norm; they need not
belong to $\mathfrak S_p$. For bounded inputs the same Fourier series
defines $\mathcal T_{A_m,B_m}$ by absolute convergence in operator norm.
For their spectral projections $E_a,F_b$,
\[
 E_aK_mF_b=(a-b)E_aJF_b,
 \qquad
 \mathcal T_{A_m,B_m}(K_m)
 =\sum_{a,b}(g(a)-g(b))E_aJF_b
 =g(A_m)J-Jg(B_m).
\]
The inequality $\|e^{itA_m}-e^{itA}\|\le|t|\|A_m-A\|$ and the
summability of $(1+|k|+|l|)|c_{kl}|$ give
\[
 \|\mathcal T_{A_m,B_m}(K_m)-\mathcal T_{A,B}(K)\|
 \le C\bigl(\|K_m-K\|+\delta_m\|K\|\bigr)\longrightarrow0.
\]
Continuous functional calculus therefore yields
$\mathcal T_{A,B}(K)=g(A)J-Jg(B)$, proving the estimate.
\end{proof}

\begin{corollary}[Spatial quasi-intertwining]
\label{toolbox:spatial}
Let $P,Q,R$ be orthogonal projections, let $J$ be bounded, and assume
$QJ=JR$. Put $A=QPQ$ and $B=RPR$. Then
\[
 AJ-JB=Q[P,J]R.
\]
Consequently every estimate for $[P,J]$ in $\mathfrak S_p$ gives the
corresponding estimate for $g(A)J-Jg(B)$ for smooth $g$.
For a smooth compactly supported multiplier $\chi$ in the region where
$V$ is smooth, the proof of Theorem~\ref{loc:comparison}, with the same
Hamiltonian on both sides, gives
\[
 \|[P_h,\chi]\|_p^p=O_{p,\chi}(h^{1-n}),\qquad p>0.
\]
\end{corollary}

\begin{proof}
Since $QJ=JR$ and $R^2=R$,
\[
 AJ-JB=QPJR-QJPR=Q(PJ-JP)R.
\]
Lemma~\ref{toolbox:intertwining} gives
$\|g(A)J-Jg(B)\|_p\le C_{g,p}\|[P,J]\|_p$.
The final estimate follows by repeating the proof of
Theorem~\ref{loc:comparison} with the same Hamiltonian on both sides,
using Lemmas~\ref{loc:window} and~\ref{lem:mixed} for the spectral
shells and iterated commutators, respectively.
\end{proof}

We use $\langle\xi\rangle=(1+|\xi|^2)^{1/2}$ and the unitary
Fourier transform on $\mathbb R^m$,
\[
 \widehat u(\xi)=\mathcal Fu(\xi)
 =(2\pi)^{-m/2}\int_{\mathbb R^m}e^{-ix\cdot\xi}u(x)\,dx.
\]
Thus $e(\mathsf D)u=\mathcal F^{-1}(e\widehat u)$ for a bounded multiplier $e$.

For a symbol $b$ on $\mathbb R^{2k}$, its Weyl quantization is
\[
 \operatorname{Op}_h^w(b)u(x)
 =(2\pi h)^{-k}\int_{\mathbb R^{2k}}
 e^{i(x-y)\cdot\xi/h}b\!\left(\frac{x+y}{2},\xi\right)
 u(y)\,dy\,d\xi.
\]
For smooth compactly supported symbols, multiplication of the kernel
by a fixed smooth cutoff equal to one near the diagonal gives a properly
supported operator with error $O_{\mathfrak S_p}(h^\infty)$ for every
fixed $p>0$.

\begin{lemma}[Model trace-ideal estimates]
\label{toolbox:model-ideals}
Let $e\in L^\infty(\mathbb R)$ satisfy
$|e(\xi)|\le C_N\langle\xi\rangle^{-N}$ for every $N$. If $\phi\in C_c^\infty(\mathbb R)$,
then $e(\mathsf D)\phi\in\mathfrak S_p$ for every $p>0$.
Also $[\mathbf1_{\mathsf D<0},\phi]\in\mathfrak S_p$ for every $p>0$.
Finally, a smooth semiclassical pseudodifferential operator in $m$
variables with uniformly bounded symbol seminorms in a fixed compact phase-space
set satisfies $\|A_h\|_p^p\le C_ph^{-m}$.
\end{lemma}
\begin{proof}
Put $r=\min\{p,1\}$.
\emph{(i)} Let $I$ be a fixed interval containing $\operatorname{supp}\phi$
in its interior. With $\ell=|I|$ and $\omega=2\pi/\ell$, put
\[
 u_j(x)=\ell^{-1/2}e^{i\omega jx}\mathbf1_I(x),\qquad j\in\mathbb Z.
\]
These functions form an orthonormal basis of $L^2(I)$, extended
by zero to $\mathbb R$. Set
\[
v_j(\xi)
=\mathcal F(\phi u_j)(\xi)e(\xi)
=\ell^{-1/2}e(\xi)\widehat\phi(\xi-\omega j).
\]
For every $f\in L^2(\mathbb R)$,
\[
e(\mathsf D)(\phi f)
=\sum_{j\in\mathbb Z}
\left(\int_I f(x)\overline{u_j(x)}\,dx\right)
\mathcal F^{-1}v_j,
\]
where the series converges in $L^2(\mathbb R)$.
Each summand defines a rank-one operator whose only
nonzero singular value, when $v_j\ne0$, is $\|v_j\|_2$.
Peetre's inequality gives, for every $L>0$,
\[
 \|v_j\|_2\le C_L\langle j\rangle^{-L}
 \|\langle\xi\rangle^Le(\xi)\|_\infty
 \|\langle\xi\rangle^L\widehat\phi(\xi)\|_2.
\]
Consequently, with $Lr>1$,
\[
 \|e(\mathsf D)\phi\|_p^r\le\sum_j\|v_j\|_2^r
 \le C\sum_j\langle j\rangle^{-Lr}<\infty.
\]
\emph{(ii)} The two off-diagonal Fourier blocks of
$[\mathbf1_{\mathsf D<0},\phi]$ have kernels
$c\widehat\phi(\pm(v+w))$, $v,w>0$. A smooth unit-scale partition
in $v,w$, followed by Fourier expansion on fixed larger squares,
gives rank-one terms with coefficients $c_{ab}^{jk}$ satisfying
\[
 |c_{ab}^{jk}|\le C_{L,M}\langle j+k\rangle^{-L}
 \langle a\rangle^{-M}\langle b\rangle^{-M},
 \quad j,k\ge0,\quad a,b\in\mathbb Z.
\]
The vectors in these rank-one terms have uniformly bounded $L^2$ norms.
Taking $Lr>2$ and $Mr>1$ yields
\[
 \|[\mathbf1_{\mathsf D<0},\phi]\|_p^r
 \le C\sum_{j,k\ge0}\sum_{a,b\in\mathbb Z}|c_{ab}^{jk}|^r<\infty.
\]
\emph{(iii)} On $L^2(\mathbb R^m)$ set
$\Lambda_h=1+|x|^2-h^2\Delta$. Symbol composition with the polynomial
differential operator $\Lambda_h^N$ and the Calder\'on--Vaillancourt
estimate give
\[
 \|A_h\Lambda_h^N\|\le C_N,\qquad
 A_h=(A_h\Lambda_h^N)\Lambda_h^{-N}.
\]
The constants depend on finitely many symbol seminorms and the fixed
phase-space support. The harmonic-oscillator spectrum gives, for $Np>m$,
\[
 \|A_h\|_p^p\le C_N^p\|\Lambda_h^{-N}\|_p^p
 =C_N^p\sum_{\nu\in\mathbb N_0^m}
 [1+h(2|\nu|+m)]^{-Np}\le C_ph^{-m}.
\]
\end{proof}

\begin{lemma}[Global sharp-to-short-time regularization]
\label{toolbox:sharp-smooth}
Suppose $H_h^{\mathrm c}=-h^2\Delta+V_{\mathrm c}$ is smooth and its potential
equals $|x|^2+C$ outside a compact set. Suppose $\mu$ is a regular energy.
Choose $\rho\in\mathcal S(\mathbb R)$ with $\int\rho=1$ and compactly
supported Fourier transform. Set
\[
 F=\rho*\mathbf1_{(-\infty,0]},\qquad
 f_h(E)=F((E-\mu)/h).
\]
For every $p>0$,
\begin{equation*}
 \|P_h^{\mathrm c}-f_h(H_h^{\mathrm c})\|_p^p=O_p(h^{1-n}).
\end{equation*}
\end{lemma}
\begin{proof}
Set $e(u)=\mathbf1_{(-\infty,0]}(u)-F(u)$. Since $\int\rho=1$,
\[
 |e(u)|\le C_N\langle u\rangle^{-N}\quad(N>0),\qquad
 \|P_h^{\mathrm c}-f_h(H_h^{\mathrm c})\|_p^p
 =\sum_j|e((\lambda_j-\mu)/h)|^p,
\]
where $(\lambda_j)$ lists the eigenvalues with multiplicity.
Choose $\delta_0>0$ so that all energies in
$[\mu-\delta_0,\mu+\delta_0]$ are regular. A bounded smooth domain
containing their allowed sets has a classically forbidden boundary
collar. The uniform Weyl estimate in Lemma~\ref{loc:window} gives
\[
 \#\{j:|\lambda_j-\mu|\le\delta\}
 \le C(h^{-n}\delta+h^{1-n}),\qquad 0<\delta\le\delta_0.
\]
For $h<\delta_0/2$, split the sum into the central interval, near shells,
and far shells. On the first two parts,
\begin{align*}
 \sum_{|\lambda_j-\mu|<h}|e((\lambda_j-\mu)/h)|^p
 &\le Ch^{1-n},\\
 \sum_{2^kh\le|\lambda_j-\mu|<2^{k+1}h}
 |e((\lambda_j-\mu)/h)|^p
 &\le C_Nh^{1-n}2^{k(1-Np)},
 \qquad 2^kh<\delta_0/2.
\end{align*}
Quadratic confinement and the min--max principle imply
\[
 N_{H_h^{\mathrm c}}^{\le}(E)\le Ch^{-n}(1+E+C_0)^n
\]
above a fixed lower spectral bound. Thus, for $k\ge0$,
\[
 \sum_{2^{k-1}\delta_0\le|\lambda_j-\mu|<2^k\delta_0}
 |e((\lambda_j-\mu)/h)|^p
 \le C_Nh^{Np-n}2^{k(n-Np)}.
\]
These sets cover the spectrum. Choosing $Np>n+1$ gives
\[
 \|P_h^{\mathrm c}-f_h(H_h^{\mathrm c})\|_p^p
 \le C_Nh^{1-n}\Bigl(1+\sum_{k\ge0}2^{k(1-Np)}\Bigr)
 +C_Nh^{Np-n}\sum_{k\ge0}2^{k(n-Np)}
 =O_p(h^{1-n}).
\]
\end{proof}

\begin{lemma}[Transfer of transition moments and smooth traces]
\label{toolbox:transfer-moments}
Set $E_h^{\mathrm c}=T_h^{\mathrm c}-(T_h^{\mathrm c})^2$.
If $\overline\Omega$ lies in the equality region of the two potentials,
then, for every $0<q\le1$,
\[
 \operatorname{Tr}E_h^q\le\operatorname{Tr}(E_h^{\mathrm c})^q+C_qh^{1-n}.
\]
For every smooth $g$ with $g(0)=0$,
\[
 \|g(T_h)-g(T_h^{\mathrm c})\|_1=O_g(h^{1-n}).
\]
\end{lemma}
\begin{proof}
Choose $\chi=1$ near $\overline\Omega$ in Theorem~\ref{loc:comparison}.
For $\Delta_h=T_h-T_h^{\mathrm c}$, that theorem and the ideal property give
\[
 \|\Delta_h\|_p^p=O_p(h^{1-n}),\qquad p>0.
\]
Since $0\le T_h,T_h^{\mathrm c}\le I$,
\begin{align*}
 E_h-E_h^{\mathrm c}&=\Delta_h-T_h\Delta_h-\Delta_hT_h^{\mathrm c},\\
 \|E_h-E_h^{\mathrm c}\|_q^q
 &\le\|\Delta_h\|_q^q+\|T_h\Delta_h\|_q^q
       +\|\Delta_hT_h^{\mathrm c}\|_q^q
 \le3\|\Delta_h\|_q^q,\\
 \operatorname{Tr}E_h^q
 &=\|E_h\|_q^q
 \le\|E_h^{\mathrm c}\|_q^q+\|E_h-E_h^{\mathrm c}\|_q^q
 \le\operatorname{Tr}(E_h^{\mathrm c})^q+C_qh^{1-n}.
\end{align*}
Finally, Lemma~\ref{toolbox:intertwining}, with $J=I$ and $p=1$, yields
\[
 \|g(T_h)-g(T_h^{\mathrm c})\|_1
 \le C_g\|\Delta_h\|_1=O_g(h^{1-n}).
\]
\end{proof}

\section{Schatten estimates for the spatial commutator}
\label{sec:S6}

We use the confining replacement of Section~\ref{sec:localization},
chosen so that $V_{\mathrm c}(x)=\mu+1+|x|^2$ outside a compact set.
In dimension one, replacing $\Omega$ by
$\operatorname{int}\overline\Omega$ preserves $Q$ and makes the
component intervals have distinct endpoints.
Constants may depend on the fixed Schatten exponent and on finitely
many derivatives of the fixed data.

\begin{theorem}\label{S6:main}
For every fixed $p>0$,
\begin{equation}\label{S6:bound}
 \|[P_h^{\mathrm c},Q]\|_{\mathfrak S_p}^{p}
 \le C_p h^{1-n}(1+\log h^{-1}),
 \qquad 0<h\le h_0.
\end{equation}
Consequently, for every fixed $q>0$,
\begin{equation}\label{S6:moment}
 \operatorname{Tr}(E_h^{\mathrm c})^q
 \le C_q h^{1-n}(1+\log h^{-1}).
\end{equation}
The same estimates hold for $P_h$ and $E_h$, respectively.
\end{theorem}

Analogous single-logarithm bounds for separately truncated
pseudodifferential operators are established in
\cite[Theorem~4.6]{Sobolev2014Schatten}
and \cite[Lemma~4.7]{Sobolev2017Functions}.
Here the estimate is proved for the exact Schr\"odinger projection
through its short-time energy-coordinate kernel.
For $p=1$, \eqref{S6:bound} improves the bound
$O(h^{1-n}\log^2 h^{-1})$ in
\cite[Theorem~4]{DeleporteLambert} to
$O(h^{1-n}\log h^{-1})$.

We use the regular-energy Weyl estimate underlying
Lemma~\ref{loc:window}.  For the present globally confining operator
it holds uniformly for energies on both sides of \(\mu\):
\[
 N_{H_h^{\mathrm c}}^{\le}(E)=(2\pi{h})^{-n}
       \operatorname{vol}\{(x,\xi):|\xi|^2+V_{\mathrm c}(x)\le E\}
       +O({h}^{1-n}),\qquad |E-\mu|\le2a_0.
\]
The phase-volume derivative is uniformly bounded on this regular energy
interval, by the coarea formula on its compact phase-space preimage.
Subtracting the counts at the two endpoints, and allowing the same
remainder for endpoint multiplicities, gives
\begin{equation}\label{S6:window}
 \operatorname{rank}\mathbf1_{[\mu-a,\mu+a]}(H_h^{\mathrm c})
 \le C({h}^{-n}a+{h}^{1-n}),\qquad 0<a<a_0.
\end{equation}
Shrinking \(a_0\) is legitimate: compactness and regularity of the energy
surface imply that every energy in this interval is regular.  The quadratic
behaviour at infinity also gives
\begin{equation}\label{S6:global-count}
 \operatorname{rank}\mathbf1_{(-\infty,E]}(H_h^{\mathrm c})
 \le C{h}^{-n}(1+E+C)^n,\qquad E\ge-C.
\end{equation}
Indeed \(V_{\mathrm c}(x)\ge c|x|^2-C\), and the min--max principle compares
\(H_h^{\mathrm c}\) with the harmonic oscillator, whose eigenvalues are
\({h}\sqrt c(2|\alpha|+n)-C\), \(\alpha\in\mathbb N^n\).

\subsection{Two elementary Schatten estimates}

\begin{lemma}[Kernels on fixed boxes]\label{S6:smooth-kernel}
Let $B_X,B_Y\subset\R^m$ be fixed open boxes. For every $p>0$ there
is an integer $M_p$ such that the integral operator with kernel
$K\in C_c^{M_p}(B_X\times B_Y)$ satisfies
\begin{equation*}
 \|\mathcal K\|_{\mathfrak S_p}^p
 \le C_p\left(\max_{|\alpha|+|\beta|\le M_p}
             \|\partial_X^\alpha\partial_Y^\beta K\|_\infty\right)^p.
\end{equation*}
The bound also holds for $\mathbf1_E\mathcal K\mathbf1_F$ whenever
$E\subset B_X$ and $F\subset B_Y$ are measurable.
\end{lemma}
\begin{proof}
Extend the kernel periodically to larger boxes.  Its double Fourier
coefficients satisfy
\[
 |K_{kl}|\le C_M A_M(1+|k|+|l|)^{-M},
 \qquad
 A_M=\max_{|\alpha|+|\beta|\le M}\|
       \partial_X^\alpha\partial_Y^\beta K\|_\infty.
\]
Each Fourier term is a rank-one operator with its only singular value
bounded by \(C|K_{kl}|\).  For \(0<p\le1\), sum the \(p\)-th powers,
choosing \(Mp>2m\).  For \(p\ge1\), sum the trace norms first.
Smooth cutoffs equal to one on the original boxes justify the periodic
extension.  The last assertion follows from the ideal property, since
multiplication by an indicator has operator norm at most one.
\end{proof}

\begin{lemma}[An energy-surface factorization]\label{S6:extension}
Let $n\ge1$. Let $X,Y\in\R^n$ and $\eta\in\R^{n-1}$ range over fixed compact
sets. Assume that the real phases $\phi_1(X,\eta)$ and
$\phi_2(Y,\eta)$ have uniform $C^\infty$ bounds and that, after a
fixed finite subdivision, $(n-1)$-component coordinate subsets $X',Y'$
satisfy
\[
 |\det\partial_{X'}\partial_\eta\phi_1|\ge c,
 \qquad |\det\partial_{Y'}\partial_\eta\phi_2|\ge c>0.
\]
For amplitudes $a_\lambda$ with fixed compact support and uniform
$C^\infty$ bounds, the kernel
\begin{equation*}
 K_\lambda(X,Y)=
 \lambda^{n-1}\int e^{i\lambda(\phi_1(X,\eta)-\phi_2(Y,\eta))}
                    a_\lambda(X,Y,\eta)\,d\eta,
 \qquad\lambda\ge1,
\end{equation*}
satisfies
\begin{equation}\label{S6:extension-bound}
 \|\mathcal K_\lambda\|_{\mathfrak S_p}^p\le C_p\lambda^{n-1},
 \qquad p>0.
\end{equation}
For fixed $p$, the constant depends on $c^{-1}$ and finitely many
of the stated seminorms. For $n=1$, the integral is a single term.
\end{lemma}
\begin{proof}
First consider a separated amplitude, inserting fixed smooth cutoffs in
\(X,Y,\eta\).  Define
\[
 (E_1v)(X)=\int e^{i\lambda\phi_1(X,\eta)}
                  a_1(X,\eta)v(\eta)\,d\eta,
\]
and define \(E_2\) analogously.  Write \(X=(X',X'')\), where \(X'\)
corresponds to an invertible minor.  At fixed \(X''\), after shrinking the
fixed boxes, the mean-value theorem gives
\[
 |\partial_\eta\{\phi_1(X',X'',\eta)
                  -\phi_1(Z',X'',\eta)\}|
 \ge c|X'-Z'|.
\]
Integration by parts in \(\eta\) in the \(TT^*\) kernel, including the
region \(|X'-Z'|\le\lambda^{-1}\), gives
\[
 |(TT^*)(X',Z')|\le C_N(1+\lambda|X'-Z'|)^{-N}.
\]
Schur's test, followed by integration over the bounded \(X''\)-interval,
proves
\begin{equation}\label{S6:extension-L2}
 \|E_i\|\le C\lambda^{-(n-1)/2}.
\end{equation}
Put the \(\eta\)-support in the interior of a fixed torus and set
\[
 L_\lambda=1-\lambda^{-2}\Delta_\eta.
\]
Integration by parts shows that \(E_1L_\lambda^N\), initially defined
on smooth functions, extends boundedly to \(L^2\).  Its kernel has the
same phase as \(E_1\); its new amplitude is a finite sum of products of
derivatives of \(a_1,\phi_1\) and nonpositive powers of \(\lambda\).
It therefore still satisfies \eqref{S6:extension-L2}.  Consequently
\[
 \lambda^{n-1}E_1E_2^*
 =(\lambda^{(n-1)/2}E_1L_\lambda^N)\,
       L_\lambda^{-N}\,(\lambda^{(n-1)/2}E_2)^* .
\]
The exterior factors are uniformly bounded.  The eigenvalues of the
middle factor yield
\[
 \|L_\lambda^{-N}\|_{\mathfrak S_p}^p
 =\sum_{k\in\mathbb Z^{n-1}}(1+c|k|^2/\lambda^2)^{-Np}
 \le C_{N,p}\lambda^{n-1},\qquad 2Np>n-1.
\]
This proves the estimate for separated amplitudes.

For the general amplitude, take its Fourier series in \(X,Y\) on larger
fixed boxes.  For every multi-index \(\gamma\) and every \(L\),
the coefficients satisfy
\[
 \|\partial_\eta^\gamma a_{kl}(\eta)\|_\infty
 \le C_{\gamma,L}(1+|k|+|l|)^{-L}.
\]
Assign \(a_{kl}\) to the first extension operator and a fixed
\(\eta\)-cutoff to the second.  The Fourier factors in \(X,Y\) act unitarily.
More explicitly, let \(\mathcal K_{\lambda,kl}\) denote the operator
associated with this Fourier term, including its factor
\(\lambda^{n-1}\). The separated factorization above gives
\[
 \|\mathcal K_{\lambda,kl}\|_{\mathfrak S_p}
 \le C_p\lambda^{(n-1)/p}
       \sum_{|\gamma|\le M_p}
          \|\partial_\eta^\gamma a_{kl}\|_\infty
 \le C_{p,L}\lambda^{(n-1)/p}(1+|k|+|l|)^{-L}.
\]
Here \(M_p\) is a fixed finite derivative order, chosen after the
integer \(N\) in \(L_\lambda^{-N}\); all spatial cutoffs and phases
are fixed. In particular, the unitary Fourier multipliers introduce
no factors growing in \(k,l\).
For \(p\le1\) choose \(Lp>2n\) and sum \(p\)-th powers; for \(p>1\)
use the norm triangle inequality with \(L>2n\).
In the latter case the summation first gives
\(\|\mathcal K_\lambda\|_{\mathfrak S_p}\le C_p\lambda^{(n-1)/p}\);
raising this bound to the power \(p\) gives \eqref{S6:extension-bound}.
All factors involving derivatives of \(a_{kl}\) obey the same rapid
decay.  This proves \eqref{S6:extension-bound}.  For \(n=1\) the same
Fourier argument is precisely Lemma~\ref{S6:smooth-kernel}, after
removing the two scalar phase multipliers.
\end{proof}

\subsection{Regularization and the exact energy-coordinate kernel}

Shift the energy origin by $\mu$, and write
$A_h=H_h^{\mathrm c}-\mu$. Although the fixed objects used below are
introduced where they enter the argument, they are chosen in the
following order: first the compact energy band, the finite family of
nested canonical charts, the compact input supports, and the spatial
scales; next a common propagation time $t_0>0$ satisfying all flow-tube
and spatial-displacement conditions below; then the associated output
cutoffs; and finally the regularizing function $\rho$. Having fixed
$t_0$, choose a real Schwartz function $\rho$ such that
\[
 \int_{\mathbb R}\rho(u)\,du=1,
 \qquad \operatorname{supp}\widehat\rho\subset(-t_0,t_0).
\]
It is possible to choose $\rho\ge0$ by starting with the squared
absolute value of the inverse Fourier transform of a compactly
supported smooth function. Define
\[
 F(v)=\int_v^\infty\rho(u)\,du,\qquad F_{h}(E)=F(E/{h}).
\]
For \(E\ne0\), for every \(N\),
\begin{equation}\label{S6:tail}
 |F_{h}(E)-\mathbf1_{\{E\le0\}}|
 \le C_N(1+|E|/{h})^{-N}.
\end{equation}
The value of the indicator at zero changes only a spectral cluster of
rank \(O({h}^{1-n})\).  By \eqref{S6:window}, dyadic shells
\({h}2^{j-1}<|E|\le{h}2^j<a_0\) contribute at most
\[
 C_{N,p}{h}^{1-n}2^j2^{-Npj}.
\]
Their sum is \(O({h}^{1-n})\) if \(Np>1\).
For the remaining energies use \eqref{S6:global-count} and take \(N\)
larger if necessary.  The spectral theorem therefore gives
\begin{equation}\label{S6:sharp-smooth}
 \|P_h^{\mathrm c}-F_{h}(A_{h})\|_{\mathfrak S_p}^p
 \le C_p{h}^{1-n},\qquad p>0.
\end{equation}
The spectral sum is taken over the discrete spectrum of the auxiliary
quadratically confining operator.

We next construct the short-time functional calculus for the
$h$-dependent function $F_h$.
For a fixed \(f\in C_c^\infty(\mathbb R)\), smooth functional calculus
gives \(f(A_h)\) as a semiclassical pseudodifferential operator.
Indeed, the real symbol \(a(x,\xi)=|\xi|^2+V_{\mathrm c}(x)-\mu\) belongs to
the class associated with the tempered order function
\(m(x,\xi)=1+|x|^2+|\xi|^2\), and \(|a\pm i|\ge c m\).
The fixed-function theorem in \cite[Section~12.6, Theorem~83]{GuilleminSternberg}
therefore applies; see also \cite[Theorem~8.7]{DimassiSjostrand}.
Lemma~\ref{app:fixed-calculus} gives the precise form used here:
a classical symbol supported in a fixed compact energy region and a
remainder whose kernel is \(O(h^\infty)\) in every Schwartz seminorm.
The global remainder follows from the oscillator graph-norm estimates
in Lemma~\ref{app:graphnorms}.

Choose a fixed real-valued smooth function \(b(E)\), supported in a small interval
about zero and equal to one in a smaller interval.  On the spectrum of
\(A_{h}\),
\[
 F_{h}(E)=b(E)F_{h}(E)+f_-(E)+r_{h}(E),
\]
where \(f_-\in C_c^\infty(\mathbb R)\) is fixed and agrees with
\((1-b(E))\mathbf1_{\{E<0\}}\) above a fixed lower bound for the spectrum.
Equation~\eqref{S6:tail}, with arbitrarily large \(N\), and
\eqref{S6:global-count} show that
\begin{equation}\label{S6:energy-decomp}
 F_{h}(A_{h})
 =b(A_{h})F_{h}(A_{h})+f_-(A_{h})
       +O_{\mathfrak S_p}({h}^\infty),\qquad p>0.
\end{equation}
Here $O_{\mathfrak S_p}(h^\infty)$ denotes an operator family whose
$\mathfrak S_p$ quasi-norm is $O(h^N)$ for every $N$, with $p$ fixed.
Thus only the fixed compact energy band containing zero requires the
singular analysis below.

Let \(W\) be a fixed relatively compact neighborhood of
\(\overline\Omega\) contained in \(\{V_{\mathrm c}<\mu\}\).  By reducing the energy
band, there is \(c>0\) such that
\[
 |\xi|\ge c
 \quad\text{on }\{x\in\overline W,\ 
          ||\xi|^2+V_{\mathrm c}(x)-\mu|\le2a_0\}.
\]
In every sufficiently small phase-space chart in this compact set,
some component \(\partial_{\xi_j}(|\xi|^2+V_{\mathrm c}-\mu)\) is bounded away
from zero.  After a rotation, the Hamilton--Jacobi construction gives
a smooth real function \(S(x,\eta,E)\), \(\eta\in\mathbb R^{n-1}\), such that
\begin{equation}\label{S6:HJ}
 |\partial_xS(x,\eta,E)|^2+V_{\mathrm c}(x)-\mu=E,\qquad
 \det\partial_x\partial_{(\eta,E)}S\ne0.
\end{equation}
One may prescribe \(S(x',0;\eta,E)=x'\cdot\eta\);
the implicit-function theorem solves for the remaining momentum
component, and the noncharacteristic Hamilton--Jacobi equation
then supplies \(S\) on a fixed small chart.
The finite charts are chosen first on the zero-energy surface and
then extended in \(E\).  Reduce the common energy interval after
this finite choice.  Thus, for every \((x,\eta)\) in an amplitude
support, \(S(x,\eta,0)\) is defined in the same chart, and the
required mixed minors remain uniformly nonzero throughout that
energy interval.

In model canonical coordinates $(z',z_n;\eta,E)$, fix nested
charts $\mathcal V_0\Subset\mathcal V_1$ and let
$\kappa$ be the canonical transformation generated by $S$ in
\eqref{S6:HJ}, defined on a neighborhood of $\overline{\mathcal V_1}$.
The corresponding physical charts use the common energy interval
containing zero chosen above. We write $\Psi_h^0$ for semiclassical pseudodifferential operators
of order zero, using the Weyl convention above; the symbols below
are classical and have fixed compact phase-space support.
Let $B_h,L_h\in\Psi_h^0$ be classical
cutoffs with full symbols supported in fixed compact subsets of
$\kappa(\mathcal V_1)$ and uniformly bounded symbol seminorms.
They have fixed compact spatial cutoffs on both sides, modulo kernels
in $h^\infty\mathcal S(\R^{2n})$, where $\mathcal S$ denotes the
Schwartz space. This notation means that, for every
$N\ge0$ and all multi-indices,
\[
 \sup_{x,y}|x^\alpha y^\beta
 \partial_x^\gamma\partial_y^\delta R_h(x,y)|=O(h^N).
\]
For a fixed compact neighborhood $K_B$ of the full-symbol support of
$B_h$ satisfying $\kappa^{-1}(K_B)\Subset\mathcal V_0$, the common time
$t_0$ is fixed so that
\[
 \bigcup_{|t|\le t_0}
 \exp(tH_E)\bigl(\kappa^{-1}(K_B)\bigr)
 \Subset\mathcal V_0,
 \qquad H_E=\partial_{z_n}.
\]
The normal form is used on $\mathcal V_0$, with its FIO cutoffs
supported in the larger chart $\mathcal V_1$.
All charts and support sets are independent of $h$. By the preceding
choice of $\rho$, $\operatorname{supp}\widehat\rho\subset(-t_0,t_0)$.
Put $D_h=h\mathsf D_{z_n}$.

\begin{proposition}[Localized energy-coordinate kernel]
\label{prop:localized-energy-kernel}
Under these conventions, there is a uniformly bounded order-zero FIO
$U_h$ with phase $S(x,\eta,E)-z'\cdot\eta-z_nE$ such that
\begin{equation}\label{S6:function-intertwine}
 F_h(A_h)B_h=U_hF_h(D_h)U_h^*B_h+R_h^{(1)}.
\end{equation}
The kernel of $L_hF_h(A_h)B_h$ is
\begin{equation}\label{S6:energy-kernel}
 \begin{split}
 K_h(x,y)={}&(2\pi h)^{-n}
 \int e^{i(S(x,\eta,E)-S(y,\eta,E))/h}
       F_h(E)a_h(x,y,\eta,E)\,d\eta\,dE\\
 &+R_h^{(2)}(x,y),
 \end{split}
\end{equation}
where $a_h$ has fixed compact support in this single phase chart and
$\sup_{0<h\le h_0}\|a_h\|_{C^N}<\infty$ for every $N$.
The values $S(x,\eta,0)$ and $S(y,\eta,0)$ are defined throughout
that support. Both residual kernels belong to
$h^\infty\mathcal S(\R^{2n})$; their operators are
$O_{\mathfrak S_p}(h^\infty)$ for every fixed $p>0$.
\end{proposition}

\begin{proof}
The prescribed canonical graph can be quantized by
\cite[Theorem~11.6]{Zworski}; the transport construction in the proof
of \cite[Theorem~12.3]{Zworski} removes the lower-order errors while
preserving this graph. The symbol is real and classical, and its
differential is nonzero by \eqref{S6:HJ}. Shrinking the fixed charts
ensures the local symplectic hypotheses of the quantization theorem.
For the short-time parametrix we use
\cite[Theorem~10.4(i)--(ii)]{Zworski}; the earlier formulations are
\cite[Theorems~10.9 and~10.18]{EvansZworski}. The comparison with the
exact propagator of the unbounded auxiliary Hamiltonian is proved
below using Lemma~\ref{app:propagation}.

Equation~\eqref{S6:HJ} gives
\begin{equation}\label{S6:schrodinger-transport}
 (A_h-E)(e^{iS/h}a)
 =e^{iS/h}\left[-ih\bigl(2\nabla_xS\cdot\nabla_xa
       +(\Delta_xS)a\bigr)-h^2\Delta_xa\right].
\end{equation}
The transport field $2\nabla_xS$ is transverse to the initial
hypersurface on the chosen chart. Put
$\mathcal T=2\nabla_xS\cdot\nabla_x+\Delta_xS$ and seek
$a\sim\sum_{j\ge0}h^ja_j$. Equation~\eqref{S6:schrodinger-transport}
gives
\[
 \mathcal T a_0=0,\qquad
 \mathcal T a_j=i\Delta_xa_{j-1}\quad(j\ge1).
\]
Nonvanishing initial data for $a_0$ give an elliptic leading amplitude.
Solving these transport equations and applying Borel summation removes
the intertwining defect to every order, with the phase $S$ fixed.

More precisely, let $C_h\in\Psi_h^0(\mathbb R_z^n)$ be equal to the
identity microlocally on a neighborhood of
\[
 \bigcup_{|t|\le t_0}
 \exp(tH_E)\bigl(\kappa^{-1}(K_B)\bigr)
\]
and have full-symbol support in $\mathcal V_0$. There is a local Fourier
integral operator $U_h$ with the \emph{left-amplitude} representation
\begin{equation}\label{S6:left-U}
 U_h(x,z)=(2\pi h)^{-n}
 \int e^{i(S(x,\eta,E)-z'\!\cdot\eta-z_nE)/h}
        u_h(x,\eta,E)\,d\eta\,dE,
\end{equation}
whose amplitude is classical and has uniformly bounded derivatives of
every fixed order. Its fixed outer cutoff is equal to one on the
canonical graph over the preceding model flow tube. With
\[
 \mathcal R_h:=A_hU_h-U_hD_h,
\]
the construction gives
\begin{equation}\label{S6:localized-normal-form}
 \mathcal R_hC_h=O(h^\infty),\qquad
 (U_h^*U_h-I)C_h=O(h^\infty),\qquad
 (U_hU_h^*-I)B_h=O(h^\infty)
\end{equation}
as operators $\mathcal S'\to\mathcal S$.

For completeness, start with an elliptic intertwiner $U_{0,h}$ and a
physical cutoff $\Lambda_h$ that is microlocally equal to $I$ on the
canonical image of the full-symbol support of $C_h$. The transport
construction gives
\begin{align*}
 \mathcal R_{0,h}&:=A_hU_{0,h}-U_{0,h}D_h,\\
 \mathcal R_{0,h}C_h&=O(h^\infty),&
 \mathcal R_{0,h}^*\Lambda_h&=O(h^\infty),&
 (I-\Lambda_h)U_{0,h}C_h&=O(h^\infty)
\end{align*}
as operators $\mathcal S'\to\mathcal S$. Set
\[
 \Gamma_h:=U_{0,h}^*U_{0,h}.
\]
On the inner model chart,
\[
 [D_h,\Gamma_h]C_h
 =\bigl(U_{0,h}^*\mathcal R_{0,h}
       -\mathcal R_{0,h}^*U_{0,h}\bigr)C_h
 =O(h^\infty):\mathcal S'\to\mathcal S.
\]
Choose a self-adjoint $N_h\in\Psi_h^0$ to be a microlocal inverse
square root of $\Gamma_h$ on a neighborhood of the full-symbol support
of $C_h$. Then
\[
 (N_h\Gamma_hN_h-I)C_h=O(h^\infty),\qquad
 [D_h,N_h]C_h=O(h^\infty).
\]
Consequently,
\[
 \bigl(A_hU_{0,h}N_h-U_{0,h}N_hD_h\bigr)C_h
 =\mathcal R_{0,h}N_hC_h+U_{0,h}[D_h,N_h]C_h
 =O(h^\infty).
\]
Replacing $U_{0,h}$ by $U_{0,h}N_h$ proves the first two relations in
\eqref{S6:localized-normal-form}; the elliptic FIO parametrix on the
corresponding physical chart gives the third. Composition with $N_h$
preserves the phase and the representation \eqref{S6:left-U}.

To prove \eqref{S6:function-intertwine}, use the cutoff $C_h$ fixed
above and set $V_h(t)=e^{-itD_h/h}U_h^*B_h$. Then
\[
 \mathcal R_hV_h(t)
 =\mathcal R_hC_hV_h(t)+\mathcal R_h(I-C_h)V_h(t).
\]
The first term is residual by the normal-form construction. The
cutoff choice gives $(I-C_h)V_h(t)=O(h^\infty):\mathcal S'\to\mathcal S$,
which yields the same conclusion for the second term.
Both estimates hold in every Schwartz kernel
seminorm, uniformly with any fixed number of $t$-derivatives.
The initial discrepancy is residual by microlocal unitarity.
Duhamel's formula gives
\begin{align*}
 e^{-itA_h/h}B_h-U_he^{-itD_h/h}U_h^*B_h
 ={}&e^{-itA_h/h}(I-U_hU_h^*)B_h\\
 &-\frac{i}{h}\int_0^t e^{-i(t-v)A_h/h}\mathcal R_hV_h(v)\,dv.
\end{align*}
By Lemma~\ref{app:propagation}, the kernel of this difference satisfies
\[
 \sup_{|t|\le t_0}
 \bigl\|x^\alpha y^\beta
 \partial_x^\gamma\partial_y^\delta\partial_t^j
 \bigl(e^{-itA_h/h}B_h-U_he^{-itD_h/h}U_h^*B_h\bigr)(x,y)
 \bigr\|_\infty
 =O(h^N)
\]
for all multi-indices $\alpha,\beta,\gamma,\delta$ and integers $j,N\ge0$.

The Fourier transform of \(F\) is a distribution of finite order
supported in \([-t_0,t_0]\): this follows from \(F'=-\rho\);
the possible additional term is supported at \(t=0\).
Denote the preceding propagator discrepancy by $Z_h(t)$.
For every Schwartz kernel seminorm $\mathfrak p$, finite-order
continuity gives an integer $m$ such that
\[
 \mathfrak p\!\left(
 (2\pi)^{-1/2}\langle\widehat F(t),Z_h(-t)\rangle\right)
 \le C_F\sum_{j=0}^m\sup_{|t|\le t_0}
 \mathfrak p(\partial_t^jZ_h(t))=O(h^N)
\]
for every $N$. Fourier inversion identifies the paired operator as
\[
 F_h(A_h)B_h-U_hF_h(D_h)U_h^*B_h,
\]
which proves \eqref{S6:function-intertwine}.

Stationary phase writes $L_hU_h$ and $B_h^*U_h$ with the common
phase $S$ and compactly supported left amplitudes $\ell_h$ and $b_h$,
respectively, with uniformly bounded derivatives. Since
$U_h^*B_h=(B_h^*U_h)^*$, exact Fourier integration in the model
variables gives \eqref{S6:energy-kernel} with
\[
 a_h(x,y,\eta,E)
 =\ell_h(x,\eta,E)\overline{b_h(y,\eta,E)}.
\]
The chart choice following \eqref{S6:HJ} ensures that $S(x,\eta,0)$
and $S(y,\eta,0)$ are defined throughout this support.
The localized composition remainders remain globally Schwartz and
$O(h^\infty)$ after multiplication by $F_h(D_h)$. Indeed,
\[
 F_h(D_h)=F(\mathsf D_{z_n})
\]
is an $h$-independent ordinary Fourier multiplier, continuous on
$\mathcal S$ and $\mathcal S'$. The other localized FIO factors
have at most polynomial losses in $h^{-1}$ in these spaces.

For a residual kernel $R_h$, its coefficients in a fixed Hermite
basis satisfy
\[
 |r_{\alpha\beta}(h)|
 \le C_{L,N}h^L(1+|\alpha|+|\beta|)^{-N}
 \qquad(L,N\ge0).
\]
For $0<p\le1$ and $Np>2n$, rank-one summation gives
\[
 \|R_h\|_{\mathfrak S_p}^p
 \le\sum_{\alpha,\beta}|r_{\alpha\beta}(h)|^p
 \le C_{L,N,p}h^{Lp}.
\]
For $p>1$, use $\|R_h\|_{\mathfrak S_p}\le\|R_h\|_{\mathfrak S_1}$.

\end{proof}

Choose $\chi\in C_c^\infty(W)$ equal to one on a neighborhood of
$\overline\Omega$. Since $\chi Q=Q$ and
$b(A_h)F_h(A_h)=F_h(A_h)b(A_h)$,
\[
(I-Q)b(A_h)F_h(A_h)Q
=(I-Q)F_h(A_h)b(A_h)\chi Q.
\]
A finite microlocal partition of the compact energy localization gives
\[
b(A_h)\chi=\sum_{\nu=1}^{J}B_{h,\nu}+R_h^{(0)},
\qquad
R_h^{(0)}\in h^\infty\mathcal S(\mathbb R^{2n}),
\]
where $J$ is independent of $h$ and each $B_{h,\nu}$ satisfies
the input localization hypotheses of
Proposition~\ref{prop:localized-energy-kernel}.

Let $\Phi_t$ denote the Hamilton flow of
$a(x,\xi)=|\xi|^2+V_{\mathrm c}(x)-\mu$.
For each $\nu$, choose a compact neighborhood $K_\nu$ of the
full-symbol support of $B_{h,\nu}$ and a nested outer canonical chart
$\mathcal U_\nu$. Since the family is finite, the common time $t_0$,
fixed before $\rho$ is chosen, was chosen so that
\[
\bigcup_{|t|\le t_0}\Phi_t(K_\nu)\Subset\mathcal U_\nu
\qquad(1\le\nu\le J).
\]
After $t_0$ has been fixed, choose $L_{h,\nu}$ supported in
$\mathcal U_\nu$ and satisfying
\[
L_{h,\nu}=I+O(h^\infty)
\quad\text{microlocally near }
\bigcup_{|t|\le t_0}\Phi_t(K_\nu).
\]
The global remainder estimates in the preceding proof give
\[
R_{h,\nu}^{(1)}
:=(I-L_{h,\nu})F_h(A_h)B_{h,\nu}
\in h^\infty\mathcal S(\mathbb R^{2n}).
\]
Consequently,
\[
\begin{aligned}
	(I-Q)b(A_h)F_h(A_h)Q
	&=
	\sum_{\nu=1}^{J}
	(I-Q)L_{h,\nu}F_h(A_h)B_{h,\nu}Q
	+\mathcal R_h,\\
	\mathcal R_h
	&:=(I-Q)
	\left(
	F_h(A_h)R_h^{(0)}
	+\sum_{\nu=1}^{J}R_{h,\nu}^{(1)}
	\right)Q.
\end{aligned}
\]
Since $\sup_h\|F_h(A_h)\|\le\|F\|_\infty$, the ideal property
and the finite-sum inequality in $\mathfrak S_p$ imply
\[
\|\mathcal R_h\|_{\mathfrak S_p}
=O(h^N)
\qquad\text{for every fixed }p>0
\text{ and every }N>0.
\]
By \eqref{S6:function-intertwine} and
Proposition~\ref{prop:localized-energy-kernel}, each
$L_{h,\nu}F_h(A_h)B_{h,\nu}$ has the kernel representation
\eqref{S6:energy-kernel}, with both cutoffs in the same canonical
chart. Since $b$ and $F_h$ are real-valued, taking adjoints gives
the corresponding decomposition of
$Qb(A_h)F_h(A_h)(I-Q)$.
\Needspace{6\baselineskip}
\subsection{The uniform scaled-kernel lemma}

Let $\mathcal K_{h,r}$ be the localization of the oscillatory term in
\eqref{S6:energy-kernel} to two boxes of side $O(r)$, by a smooth
cutoff $\vartheta_{h,r}$ satisfying
\[
 \|\partial_x^\alpha\partial_y^\beta\vartheta_{h,r}\|_\infty
 \le C_{\alpha\beta}r^{-|\alpha|-|\beta|}.
\]
The constants below are uniform in the boxes, $h$, and $r$.
For fixed $p$, they depend on the fixed supports, the nondegeneracy
bounds in \eqref{S6:HJ}, finitely many seminorms of the phase,
amplitude and rescaled cutoffs, and finitely many moments
$\int_\R|\rho(u)|(1+|u|)^M\,du$.

\begin{lemma}\label{S6:scaled}
If $h\le r\le r_0$ and
$cr\le|x-y|\le Cr$ on $\operatorname{supp}\vartheta_{h,r}$, then
\begin{equation}\label{S6:scaled-bound}
 \|\mathcal K_{h,r}\|_{\mathfrak S_p}^p
 \le C_p(r/h)^{n-1},\qquad p>0.
\end{equation}
For $r=h$ and
$\operatorname{supp}\vartheta_{h,h}\subset\{|x-y|\le Ch\}$,
the bound is $C_p$.
These estimates remain valid after restricting either box to
$\Omega$ or $\Omega^c$.
\end{lemma}
\begin{proof}
Use the fixed product phase charts from the construction following
\eqref{S6:HJ}, with an energy interval \(I\) containing zero and
the compact energy support of the amplitude. Choose
\(\chi_E\in C_c^\infty(I)\) equal to one near both this support and
zero. Replace \(S\) outside that neighborhood by \(\chi_E(E)S\),
extended by zero to all \(E\in\mathbb R\), and retain the notation \(S\).
This fixed smooth extension has bounded derivatives of every order.
It leaves the kernel and its \(E=0\) phase unchanged, and preserves
the full mixed nondegeneracy in \eqref{S6:HJ} on the original
amplitude support and near \(E=0\).

Write \(x=x_0+rX,\ y=y_0+rY,\ \lambda=r/{h}\).
The unitary change of scale multiplies the kernel by \(r^n\).
Choose the \emph{same} reference point \(z_0\), at distance \(O(r)\)
from both boxes, and set
\[
 \begin{aligned}
 \phi_1(X,\eta,E)&=
 \frac{S(x_0+rX,\eta,E)-S(z_0,\eta,E)}r,\\
 \phi_2(Y,\eta,E)&=
 \frac{S(y_0+rY,\eta,E)-S(z_0,\eta,E)}r .
 \end{aligned}
\]
Both families have uniform smooth seminorms.  The common subtraction preserves the kernel phase and gives uniform
seminorms also for derivatives in $\eta$.  Put $\Phi=\phi_1-\phi_2$. Choose the physical coordinate boxes
convex and sufficiently small. On the original amplitude support,
uniform invertibility of $\partial_x\partial_{(\eta,E)}S$ and the
mean-value formula imply
\begin{equation}\label{S6:full-gradient}
 |\partial_{(\eta,E)}\Phi|\ge c_1>0.
\end{equation}
More explicitly, the matrix averaged along the segment from \(y\)
to \(x\) stays within a small fixed distance of one invertible matrix,
and is therefore uniformly invertible.  The spatial separation
assumption supplies the lower bound in \eqref{S6:full-gradient}.
Thus the rescaled kernel is a fixed constant times
\begin{equation*}
 \lambda^n\int e^{i\lambda\Phi(X,Y,\eta,E)}
       F_{h}(E)a_{{h},r}(X,Y,\eta,E)\,d\eta\,dE .
\end{equation*}
All derivatives of \(a_{{h},r}\) are uniform.

Use a smooth partition, with uniform seminorms, into
\[
 |\partial_\eta\Phi|\ge c_2
 \quad\text{and}\quad
 |\partial_E\Phi|\ge c_2 .
\]
For \(n=1\) only the second part occurs.  On the first part, integrate
in \(\eta\) with
\[
 \frac{\partial_\eta\Phi\cdot\partial_\eta}
      {i\lambda|\partial_\eta\Phi|^2}.
\]
Since $\partial_\eta F_h(E)=0$ and
$\|F_h\|_\infty\le\|\rho\|_1$, the resulting kernel $K_\eta$ satisfies
\[
 \|K_\eta\|_{C^M(X,Y)}
 \le C_{M,N}\lambda^{n+M-N}
 \le C_{M,L}\lambda^{-L},\qquad N\ge n+M+L.
\]

On the second part set \(D_E=\partial_E\Phi\), and use the exact identity
\begin{equation*}
 F_{h}(E)=\int_{\mathbb R}\rho(u)\,
                   \mathbf1_{\{E\le{h} u\}}\,du.
\end{equation*}
The compact amplitude support and $\rho\in L^1$ justify Fubini's
theorem. Successive integrations in \(E\) up to the moving endpoint
give, for every integer \(N\ge1\),
\begin{align}
 \int_{-\infty}^{{h} u}e^{i\lambda\Phi(E)}a(E)\,dE
 &=
 e^{i\lambda\Phi({h} u)}
 \sum_{k=0}^{N-1}(i\lambda)^{-k-1} b_k({h} u)
 +(i\lambda)^{-N}
 \int_{-\infty}^{{h} u}e^{i\lambda\Phi(E)}a_N(E)\,dE,
 \label{S6:IBP}\\
 a_0&=a,\qquad b_k=a_k/D_E,\qquad
 a_{k+1}=-\partial_Eb_k\quad(k\ge0).
 \nonumber
\end{align}
The amplitudes are extended by zero off their fixed support;
the reciprocal of \(D_E\) is used only on a slightly larger
nonstationary support.  Consequently every \(a_k,b_k\) has uniform
smooth seminorms and fixed compact support.

The phase at the moving energy endpoint satisfies the exact identity
\begin{equation}\label{S6:moving-phase}
 \lambda\{\Phi(X,Y,\eta,{h} u)-\Phi(X,Y,\eta,0)\}
 =r u\int_0^1\partial_E\Phi(X,Y,\eta,\theta{h} u)\,d\theta.
\end{equation}
The fixed energy extension makes this identity meaningful for every
\(u\in\mathbb R\); only endpoints in the fixed support of \(b_k\)
contribute to the boundary terms.
For every \(M\), the \(X,Y,\eta\) derivatives through order \(M\) of
the exponential of \eqref{S6:moving-phase} are bounded by
\(C_M(1+|u|)^M\), uniformly for \(r\le r_0\).
Set $Z=(X,Y,\eta)$ and
\[
 \Psi_{h,r}(Z,u)=ru\int_0^1\partial_E\Phi(Z,\theta hu)\,d\theta.
\]
The fixed smooth extension of $\Phi$ gives, for $|\gamma|\le M$,
\[
 |\partial_Z^\gamma\Psi_{h,r}(Z,u)|\le C_M|u|,
 \qquad
 \left|\partial_Z^\gamma
 \bigl(e^{i\Psi_{h,r}(Z,u)}b_k(Z,hu)\bigr)\right|
 \le C_{M,k}(1+|u|)^M.
\]
The Schwartz moments of \(\rho\) therefore show that
\begin{equation}\label{S6:averaged-amplitude}
 \widetilde b_{{h},r,\lambda}(X,Y,\eta)
 =
 \sum_{k=0}^{N-1}i^{-k-1}\lambda^{-k}
 \int \rho(u)\,
 e^{i\lambda(\Phi({h} u)-\Phi(0))}
 b_k(X,Y,\eta,{h} u)\,du
\end{equation}
has uniform smooth seminorms. More precisely, for every fixed \(M,N\),
\[
 \|\widetilde b_{{h},r,\lambda}\|_{C^M(X,Y,\eta)}
 \le C_{M,N}\sum_{k=0}^{N-1}\lambda^{-k}
       \int_{\mathbb R}|\rho(u)|(1+|u|)^M\,du
 \le C_{M,N,\rho}.
\]
The contribution of $|u|>U$ has $C^M$ norm bounded by
\[
 C_{M,N}\int_{|u|>U}|\rho(u)|(1+|u|)^M\,du
 \longrightarrow0\qquad(U\to\infty),
\]
uniformly in $h,r$. This also justifies differentiation under the
integral.

Multiplying \eqref{S6:IBP} by \(\lambda^n\) thus gives
\begin{equation}\label{S6:final-scaled}
 \lambda^{n-1}\int
 e^{i\lambda(\phi_1(X,\eta,0)-\phi_2(Y,\eta,0))}
 \widetilde b_{{h},r,\lambda}(X,Y,\eta)\,d\eta
 +\mathcal R_{{h},r,\lambda}(X,Y).
\end{equation}
For prescribed \(M,L\), take \(N>n+M+L\) in
\eqref{S6:IBP}.  Differentiating its integral remainder at most \(M\)
times loses at most \(\lambda^M\); the integration region remains
bounded, and \(\int|\rho|<\infty\).
It follows that
\begin{equation}\label{S6:scaled-remainder}
 \|\mathcal R_{{h},r,\lambda}\|_{C^M(X,Y)}
 \le C_{M,L}\lambda^{-L}.
\end{equation}
This includes the earlier \(\eta\)-nonstationary part.

The full invertibility in \eqref{S6:HJ} implies that the \(n\)-by-\(n-1\)
matrix \(\partial_x\partial_\eta S\) has rank \(n-1\).
Choose an invertible \(n-1\)-by-\(n-1\) minor on each of finitely many
fixed subcharts.  The phases in \eqref{S6:final-scaled} satisfy the
hypotheses of Lemma~\ref{S6:extension}, uniformly in \(r\).
That lemma bounds the leading term by \(C_p\lambda^{n-1}\).
Choose \(M\ge M_p\) in \eqref{S6:scaled-remainder} and apply
Lemma~\ref{S6:smooth-kernel} to the remainder.
This proves \eqref{S6:scaled-bound}.

At $r=h$, the rescaled kernel in \eqref{S6:energy-kernel} satisfies
\[
 \bigl\|h^nK_h(x_0+hX,y_0+hY)\bigr\|_{C^M(X,Y)}\le C_M
 \qquad(M\ge0),
\]
since $r^n(2\pi h)^{-n}=(2\pi)^{-n}$ and $F_h$ is bounded and
independent of $X,Y$. Lemma~\ref{S6:smooth-kernel} and the ideal
property give the bound $C_p$ for each restricted pair.
\end{proof}

\subsection{Spatial counting, distant pairs, and completion of the estimate}

\begin{proof}[Proof of Theorem~\ref{S6:main}]
Put
\[
 A_h=H_h^{\mathrm c}-\mu,\qquad
 \ell(h)=1+\log h^{-1},\qquad 0<h\le h_0<1.
\]
We first fix $0<p\le1$.

\smallskip\noindent
\emph{Spectral regularization.}
By \eqref{S6:sharp-smooth} and the ideal property,
\[
 \begin{aligned}
 \|QP_h^{\mathrm c}(I-Q)\|_{\mathfrak S_p}^p
 &\le\|QF_h(A_h)(I-Q)\|_{\mathfrak S_p}^p
   +\|P_h^{\mathrm c}-F_h(A_h)\|_{\mathfrak S_p}^p\\
 &\le\|QF_h(A_h)(I-Q)\|_{\mathfrak S_p}^p+C_ph^{1-n}.
 \end{aligned}
\]
Equation~\eqref{S6:energy-decomp} gives
\[
 F_h(A_h)=G_h+S_h+R_h,\qquad
 G_h=b(A_h)F_h(A_h),\qquad S_h=f_-(A_h),
 \qquad \|R_h\|_{\mathfrak S_p}=O(h^N)
\]
for every $N$. We estimate the two off-diagonal blocks separately.

\smallskip\noindent
\emph{Spatial localization.}
For $\delta>0$, set
$\mathcal U_\delta=\{x:\operatorname{dist}(x,\partial\Omega)<\delta\}$.
Choose $\delta$ and a fixed radius $r_*$ so that
\[
 \overline{\mathcal U_{4\delta}}\subset W,\qquad
 0<r_*<\delta,
\]
with $r_*$ no larger than the radius allowed in
Lemma~\ref{S6:scaled}. Choose $\chi\in C_c^\infty(\mathcal U_{2\delta})$
equal to one on $\mathcal U_\delta$. If $x\in\Omega$, $y\notin\Omega$,
then
\[
 \max\{\operatorname{dist}(x,\partial\Omega),
        \operatorname{dist}(y,\partial\Omega)\}\le |x-y|.
\]
Thus $|x-y|<r_*$ implies $\chi(x)=\chi(y)=1$.

Let $a(x,\xi)=|\xi|^2+V_{\mathrm c}(x)-\mu$, let $H_a$ be its
Hamilton vector field, and write $\Phi_t=\exp(tH_a)$. On the fixed
compact energy region containing $\operatorname{supp}b(a)$, choose
$v_0$ such that $|\partial_\xi a|\le v_0$. In the preceding choice of
the common time $t_0$, impose also
\[
 v_0t_0<r_*/4
\]
and require all relevant short trajectories to remain in their
prescribed outer charts. Hence
\[
 (x,\xi)=\Phi_t(y,\eta),\quad |t|\le t_0
 \quad\Longrightarrow\quad |x-y|\le v_0|t|<r_*/4.
\]
Choose $\zeta\in C_c^\infty(\mathbb R^n)$ equal to one for
$|z|\le r_*/2$ and zero for $|z|\ge r_*$.
On the support of $1-\zeta(x-y)$ the localized propagator phase
is nonstationary. Integration by parts gives, for every Schwartz
kernel seminorm $\mathfrak p$ and all fixed integers $k,N\ge0$,
\[
 \sup_{|t|\le t_0}
 \mathfrak p\!\left(
 \partial_t^k[(1-\zeta(x-y))K_{b(A_h)e^{-itA_h/h}}(x,y)]
 \right)=O(h^N).
\]
The compact energy localization and Lemma~\ref{app:propagation}
control the global remainders. Pairing with the finite-order
distribution $\widehat F$ therefore gives the same estimate for
$(1-\zeta(x-y))K_{G_h}(x,y)$.

A finite microlocal partition and
Proposition~\ref{prop:localized-energy-kernel} yield
\[
 QG_h(I-Q)=\sum_{\ell=1}^{L_0}
 Q\mathcal K_h^{(\ell)}(I-Q)+R_h^{\mathrm{loc}},
 \qquad
 \|R_h^{\mathrm{loc}}\|_{\mathfrak S_p}=O(h^N)
\]
for every $N$. Here $L_0$ is independent of $h$, and each
$\mathcal K_h^{(\ell)}$ has the kernel \eqref{S6:energy-kernel}
multiplied by $\chi(x)\chi(y)\zeta(x-y)$.

\smallskip\noindent
\emph{Dyadic scales and boundary blocks.}
Choose $J=J(h)$ with $2^Jh\le r_*<2^{J+1}h$, and put
\[
 r_j=2^jh\quad(0\le j\le J),\qquad J+1\le C\ell(h).
\]
Choose a smooth partition $\{\vartheta_{j,h}\}_{j=0}^J$ on
$\operatorname{supp}\zeta$ such that
\[
 \begin{gathered}
 \sum_{j=0}^J\vartheta_{j,h}=1,\qquad
 \operatorname{supp}\vartheta_{0,h}\subset\{|z|\le Ch\},\\
 \operatorname{supp}\vartheta_{j,h}
 \subset\{cr_j\le |z|\le Cr_j\}\quad(j\ge1),\qquad
 |\partial^\alpha\vartheta_{j,h}|\le C_\alpha r_j^{-|\alpha|}.
 \end{gathered}
\]
Let $\{\psi_{j,a}\}_a$ be a smooth partition subordinate to cubes
of side comparable to $r_j$, with bounded overlap and
$|\partial^\alpha\psi_{j,a}|\le C_\alpha r_j^{-|\alpha|}$.
Let $\mathcal I_j$ be the pairs $(a,b)$ for which
\[
 (\operatorname{supp}\psi_{j,a}\cap\Omega)
 \times(\operatorname{supp}\psi_{j,b}\cap\Omega^c)
 \cap\operatorname{supp}[\vartheta_{j,h}(x-y)]\ne\varnothing.
\]
Each cube in such a pair lies in $\mathcal U_{Cr_j}$ and has at most
$C$ possible partners. The tubular-neighborhood estimate gives
\[
 |\mathcal U_{Cr_j}|\le Cr_j,\qquad
 \#\mathcal I_j\le Cr_j^{-n}|\mathcal U_{Cr_j}|
 \le Cr_j^{1-n}.
\]
For $n=1$, this follows from the finite number of essential endpoints.

Define $\mathcal K_{j,a,b,h}^{(\ell)}$ by its kernel
\[
 \psi_{j,a}(x)\vartheta_{j,h}(x-y)
 K_h^{(\ell)}(x,y)\psi_{j,b}(y).
\]
The decomposition is
\[
 Q\mathcal K_h^{(\ell)}(I-Q)
 =\sum_{j=0}^J\sum_{(a,b)\in\mathcal I_j}
 Q\mathcal K_{j,a,b,h}^{(\ell)}(I-Q).
\]
After unitary dilation of the cubes to fixed boxes,
Lemma~\ref{S6:scaled} and the ideal property give
\[
 \|Q\mathcal K_{j,a,b,h}^{(\ell)}(I-Q)\|_{\mathfrak S_p}^p
 \le C_p(r_j/h)^{n-1}.
\]
The near-diagonal estimate covers $j=0$. Hence
\begin{equation*}
 \begin{aligned}
 \sum_{(a,b)\in\mathcal I_j}
 \|Q\mathcal K_{j,a,b,h}^{(\ell)}(I-Q)\|_{\mathfrak S_p}^p
 &\le C_p\#\mathcal I_j(r_j/h)^{n-1}\\
 &\le C_pr_j^{1-n}(r_j/h)^{n-1}=C_ph^{1-n}.
 \end{aligned}
\end{equation*}
The $p$th-power triangle inequality now yields
\[
 \begin{aligned}
 \|QG_h(I-Q)\|_{\mathfrak S_p}^p
 &\le\sum_{\ell=1}^{L_0}\sum_{j=0}^J
 \sum_{(a,b)\in\mathcal I_j}
 \|Q\mathcal K_{j,a,b,h}^{(\ell)}(I-Q)\|_{\mathfrak S_p}^p+O(h^N)\\
 &\le C_ph^{1-n}(J+1)\le C_ph^{1-n}\ell(h).
 \end{aligned}
\]
The energy integration has already been incorporated into the
amplitude in \eqref{S6:averaged-amplitude}.

\smallskip\noindent
\emph{The smooth energy term.}
Modulo a kernel in $h^\infty\mathcal S$, the compact symbol of
$S_h=f_-(A_h)$ gives
\begin{equation}\label{S6:smooth-energy-kernel}
 |\partial_x^\alpha\partial_y^\beta K_{S_h}(x,y)|
 \le C_{\alpha,\beta,N}h^{-n-|\alpha|-|\beta|}
 (1+|x-y|/h)^{-N}.
\end{equation}
This follows by integration by parts in the momentum variable.
The part $|x-y|\ge r_*/2$ is $O_{\mathfrak S_p}(h^\infty)$.
Use the same partitions and let $S_{j,a,b,h}$ have kernel
\[
 \psi_{j,a}(x)\vartheta_{j,h}(x-y)
 K_{S_h}(x,y)\psi_{j,b}(y).
\]
For cube centers $x_{j,a},x_{j,b}$, its unitarily rescaled kernel is
\[
 \widetilde K_{j,a,b,h}(X,Y)
 =r_j^n S_{j,a,b,h}(x_{j,a}+r_jX,x_{j,b}+r_jY).
\]
For every $M,L\ge0$, choosing $N$ sufficiently large in
\eqref{S6:smooth-energy-kernel} gives
\[
 \|\widetilde K_{j,a,b,h}\|_{C^M}
 \le C_{M,L}(r_j/h)^{-L}.
\]
Lemma~\ref{S6:smooth-kernel} implies
\[
 \|S_{j,a,b,h}\|_{\mathfrak S_p}^p
 \le C_{p,L}(r_j/h)^{-Lp}.
\]
For any fixed $L>0$,
\[
 \begin{aligned}
 \|QS_h(I-Q)\|_{\mathfrak S_p}^p
 &\le C_{p,L}\sum_{j=0}^Jr_j^{1-n}(r_j/h)^{-Lp}+O(h^N)\\
 &\le C_{p,L}h^{1-n}\sum_{j\ge0}2^{-j(n-1+Lp)}
 \le C_ph^{1-n}.
 \end{aligned}
\]
Combining the two energy terms gives
\begin{equation*}
 \|QF_h(A_h)(I-Q)\|_{\mathfrak S_p}^p
 \le C_ph^{1-n}\ell(h),
\end{equation*}
and the spectral regularization then gives the same bound for
$QP_h^{\mathrm c}(I-Q)$.

\smallskip\noindent
\emph{Commutators and transition moments.}
Set $X_h=(I-Q)P_h^{\mathrm c}Q$. Relative to
$L^2(\mathbb R^n)=\operatorname{Ran}Q\oplus\ker Q$,
\[
 [P_h^{\mathrm c},Q]
 =\begin{pmatrix}0&-X_h^*\\X_h&0\end{pmatrix},\qquad
 \|[P_h^{\mathrm c},Q]\|_{\mathfrak S_p}^p
 =2\|X_h\|_{\mathfrak S_p}^p
 \le C_ph^{1-n}\ell(h).
\]
This proves \eqref{S6:bound} for $0<p\le1$.
For $p>1$, use $\|[P_h^{\mathrm c},Q]\|\le2$ and the case $p=1$:
\begin{equation}\label{S6:large-p}
 \begin{aligned}
 \|[P_h^{\mathrm c},Q]\|_{\mathfrak S_p}^p
 &\le\|[P_h^{\mathrm c},Q]\|^{p-1}
       \|[P_h^{\mathrm c},Q]\|_{\mathfrak S_1}\\
 &\le2^{p-1}\|[P_h^{\mathrm c},Q]\|_{\mathfrak S_1}
 \le C_ph^{1-n}\ell(h).
 \end{aligned}
\end{equation}
Projection algebra gives
\begin{equation*}
 \begin{aligned}
 E_h^{\mathrm c}
 &=QP_h^{\mathrm c}Q-QP_h^{\mathrm c}QP_h^{\mathrm c}Q\\
 &=QP_h^{\mathrm c}(I-Q)P_h^{\mathrm c}Q=X_h^*X_h,\\
 \|[P_h^{\mathrm c},Q]\|_{\mathfrak S_{2q}}^{2q}
 &=2\operatorname{Tr}(E_h^{\mathrm c})^q.
 \end{aligned}
\end{equation*}
Taking $p=2q$ proves \eqref{S6:moment} for every fixed $q>0$.

\smallskip\noindent
\emph{Transfer to the original potential.}
Choose a real cutoff $\chi$ supported in the equality region of
$V$ and $V_{\mathrm c}$, with $\chi=1$ near $\overline\Omega$.
Then
\[
 \Delta_h:=T_h-T_h^{\mathrm c}
 =Q(P_h\chi-\chi P_h^{\mathrm c})Q.
\]
Theorem~\ref{loc:comparison} implies
\begin{equation*}
 \|\Delta_h\|_{\mathfrak S_q}^q\le C_qh^{1-n},\qquad q>0.
\end{equation*}
For $0<q\le1$, the identities and estimates
\[
 \begin{aligned}
 E_h-E_h^{\mathrm c}
 &=\Delta_h-T_h\Delta_h-\Delta_hT_h^{\mathrm c},\\
 \|E_h-E_h^{\mathrm c}\|_{\mathfrak S_q}^q
 &\le3\|\Delta_h\|_{\mathfrak S_q}^q\le C_qh^{1-n},\\
 \operatorname{Tr}E_h^q
 &\le\operatorname{Tr}(E_h^{\mathrm c})^q
       +\|E_h-E_h^{\mathrm c}\|_{\mathfrak S_q}^q
 \le C_qh^{1-n}\ell(h)
 \end{aligned}
\]
follow from $0\le T_h,T_h^{\mathrm c}\le I$ and the ideal property.
For $q>1$, use
\[
 0\le E_h\le I/4,\qquad E_h^q\le4^{1-q}E_h,
 \qquad\operatorname{Tr}E_h^q\le C_qh^{1-n}\ell(h).
\]
Finally,
\[
 \|[P_h,Q]\|_{\mathfrak S_p}^p
 =2\operatorname{Tr}E_h^{p/2}
 \le C_ph^{1-n}\ell(h),\qquad p>0.
\]
\end{proof}

\begin{remark}[Scale dependence]\label{S6:scope}
For each fixed $p>0$, \eqref{S6:scaled-bound} and the boundary block
count give
\[
 r^{1-n}\|\mathcal K_{h,r}\|_{\mathfrak S_p}^p
 \le C_p h^{1-n},\qquad h\le r\le r_0.
\]
For $0<p\le1$, summation over $r=2^jh$ produces the logarithm in \eqref{S6:bound}; the case $p>1$ follows from \eqref{S6:large-p}.
\end{remark}

\section{The weighted Hardy model and the two Fermi branches}
\label{sec:weighted-hardy-model}
\label{sec:hardy}

In the one-dimensional model, write
$\mathsf D=\mathsf D_s=-i\partial_s$. On $L^2(\mathbb R)$ let
$Q=\mathbf1_{(0,\infty)}$ and
\[
 P_- =\mathbf1_{(-\infty,0)}(\mathsf D),\qquad
 B=(QP_-Q)\big|_{\operatorname{Ran}Q}.
\]
We identify $\operatorname{Ran}Q$ with $L^2(\mathbb R_+)$. If $A$ acts
on $L^2(\mathbb R)$, the notation $QAQ$ in a trace over
$L^2(\mathbb R_+)$ denotes its compression to $\operatorname{Ran}Q$.
For $g\in C^\infty([0,1])$ with $g(0)=g(1)=0$, let $W(g)$ be as
in \eqref{eq:W-intro}, and set
\begin{equation*}
 c_g=\frac{W(g)}{4\pi^2}.
\end{equation*}
Complex-valued functions and symbols are allowed. Smoothness on the closed
interval means the restriction of a smooth function on a neighborhood of it.

\subsection{Mellin diagonalization and the exact kernel}

\begin{lemma}[Mellin multiplier]
\label{lem:hardy-mellin-kernel}
The unitary map
\[
 (\mathcal Mu)(x)=e^{x/2}u(e^x),\qquad
 \mathcal M:L^2(\mathbb R_+,ds)\longrightarrow L^2(\mathbb R,dx),
\]
diagonalizes $B$ as a Fourier multiplier:
\begin{equation*}
 \mathcal M B\mathcal M^{-1}=\beta(\mathsf D_x),\qquad
 \beta(k)=\frac1{1+e^{2\pi k}}.
\end{equation*}
If
\begin{equation*}
 k_g(z)=\frac1{2\pi}\int_{\mathbb R}e^{ikz}g(\beta(k))\,dk,
\end{equation*}
then $k_g\in\mathcal S(\mathbb R)$, $k_g(0)=c_g$, and the kernel of
$g(B)$ on $\mathbb R_+\times\mathbb R_+$ is
\begin{equation}
 G_g(s,t)=(st)^{-1/2}k_g(\log(s/t)).
 \label{eq:hardy-functional-kernel}
\end{equation}
\end{lemma}

\begin{proof}
In the sense of distributions, the kernel of $P_-$ is
\[
 \frac12\delta(s-t)-\frac{i}{2\pi}\operatorname{pv}\frac1{s-t}.
\]
After the logarithmic change of variables, its compression has convolution
kernel
\[
 \frac12\delta(x-y)-\frac{i}{4\pi}
       \operatorname{pv}\frac1{\sinh((x-y)/2)}.
\]
The Fourier identity
\[
 \int_{\mathbb R}e^{-ikx}
       \operatorname{pv}\frac1{\sinh(x/2)}\,dx
 =-2\pi i\tanh(\pi k)
\]
gives the multiplier $(1-\tanh(\pi k))/2=\beta(k)$.
By oddness, this identity follows from
\[
 \int_0^\infty\frac{\sin(kx)}{\sinh(x/2)}\,dx
 =2\sum_{m=0}^\infty\frac{k}{(m+1/2)^2+k^2}
 =\pi\tanh(\pi k).
\]
The geometric expansion of $1/\sinh(x/2)$ may be integrated termwise,
since
\[
 2\sum_{m=0}^\infty\int_0^\infty
 e^{-(m+1/2)x}|\sin(kx)|\,dx
 \le 2|k|\sum_{m=0}^\infty(m+1/2)^{-2}<\infty.
\]
The second equality is the partial-fraction identity for $\tanh$.

Because $g$ vanishes at both endpoints, $g(\beta(k))$, together with all of
its derivatives, decays exponentially as $k\to\pm\infty$. It is therefore
a Schwartz function, and so is its inverse Fourier transform $k_g$.
Functional calculus for the Fourier multiplier and the inverse Mellin
transformation give \eqref{eq:hardy-functional-kernel}. Finally,
\[
 k_g(0)=\frac1{2\pi}\int_{\mathbb R}g(\beta(k))\,dk
 =\frac1{4\pi^2}\int_0^1\frac{g(\lambda)}{\lambda(1-\lambda)}\,d\lambda,
\]
since $\beta'(k)=-2\pi\beta(k)(1-\beta(k))$.
\end{proof}

\subsection{An arbitrary smooth weight in normal phase space}

\begin{theorem}[Hardy trace with a general pseudodifferential weight]
\label{thm:hardy-general-weight}
Let $b\in C_c^\infty(\mathbb R_s\times\mathbb R_\tau)$, and let $g$ be as
above. Then
\begin{equation*}
 \operatorname{Tr}_{L^2(\mathbb R_+)}
 \left[Q\operatorname{Op}_h^w(b)Q\,g(B)\right]
 =c_g b(0,0)\log h^{-1}+O_{b,g}(1),\qquad h\downarrow0.
\end{equation*}
The remainder is uniform for symbols supported in a fixed compact set
with uniform bounds on finitely many symbol seminorms.
The same formula holds for
$B_+=Q\mathbf1_{(0,\infty)}(\mathsf D)Q$, with the same
coefficient $c_g$.
\end{theorem}

\begin{proof}
Define the partial Fourier transform
\[
 \mathfrak b(v,z)=\frac1{2\pi}\int_{\mathbb R}e^{-iz\tau}b(v,\tau)\,d\tau.
\]
Choose $L\ge1$ such that $b(v,\tau)=0$ for $v\ge L$. Integration by parts
in $\tau$ gives, uniformly in $v$,
\begin{equation}
 |\mathfrak b(v,z)|\le C_{b,N}(1+|z|)^{-N},
 \qquad \int_{\mathbb R}\mathfrak b(v,z)\,dz=b(v,0).
 \label{eq:hardy-weight-fourier-bounds}
\end{equation}

Put
\begin{equation*}
 L_g(z)=\left(1-\frac{z^2}{4}\right)^{-1/2}
 k_g\!\left(\log\frac{1+z/2}{1-z/2}\right),\qquad -2<z<2.
\end{equation*}
The function $L_g$ is smooth near zero and
\begin{equation}
 L_g(0)=c_g,\qquad
 |L_g(z)-c_g|\le C_g|z|\quad(|z|\le1),\qquad
 \int_{-2}^2|L_g(z)|\,dz<\infty.
 \label{eq:hardy-Lg-bounds}
\end{equation}
The last bound follows already from the boundedness of $k_g$ and the
integrability of $(1-z^2/4)^{-1/2}$.

The exact trace identity is
\begin{equation}
 \operatorname{Tr}\left[Q\operatorname{Op}_h^w(b)Q\,g(B)\right]
 =\int_0^L\frac{dv}{v}
   \int_{|w|<2v}h^{-1}\mathfrak b(v,w/h)L_g(w/v)\,dw.
 \label{eq:hardy-exact-weighted-trace}
\end{equation}
Indeed, the kernel of the Weyl operator at $(s,t)$ is
$h^{-1}\mathfrak b((s+t)/2,(t-s)/h)$. In the trace product it is multiplied
by $G_g(t,s)$. The variables
\[
 v=(s+t)/2,\qquad w=t-s
\]
have Jacobian one and range $v>0$, $|w|<2v$; moreover
$G_g(t,s)=v^{-1}L_g(w/v)$.

A compactly supported smooth Weyl symbol satisfies
$\|\operatorname{Op}_h^w(b)\|_{\mathfrak S_1}\le C_bh^{-1}$.
To obtain this bound, set
\[
 R_h^{\mathrm{wt}}=(1+s^2)^{-1}(1+(h\mathsf D)^2)^{-1},\qquad
 \|R_h^{\mathrm{wt}}\|_{\mathfrak S_2}^2=C h^{-1}.
\]
Writing $W_h=\operatorname{Op}_h^w(b)$, the factorization is
\[
 W_h=R_h^{\mathrm{wt}} C_h(R_h^{\mathrm{wt}})^*,\qquad
 C_h=(1+(h\mathsf D)^2)(1+s^2)W_h(1+s^2)(1+(h\mathsf D)^2).
\]
The middle operator $C_h$ is bounded uniformly, by the smooth-symbol
operator bound after applying these differential and polynomial weights.
Consequently the product with the bounded operator $g(B)$ is trace class.
Let $Q_{\delta,R}=\mathbf1_{(\delta,R)}$, $0<\delta<R<\infty$.
The kernel of $Q_{\delta,R}g(B)Q_{\delta,R}$ is square integrable, so the
Hilbert--Schmidt kernel formula proves the truncated trace identity.
As $\delta\downarrow0$ and $R\uparrow\infty$, these operators converge
strongly, and so do their adjoints, with uniformly bounded norms. Pairing
with the fixed trace-class Weyl operator therefore gives convergence of
the traces. To justify the corresponding limit in the kernel integral,
take $0<h\le1$ and define
\[
 J_h^{\mathrm{abs}}(v)
 =\int_{|w|<2v}h^{-1}|\mathfrak b(v,w/h)|\,|L_g(w/v)|\,dw.
\]
For $0<v<h$, the substitution $w=vz$ and
\eqref{eq:hardy-Lg-bounds} give $J_h^{\mathrm{abs}}(v)\le C_{b,g}v/h$.
For $h\le v\le L$, split the integral into $|w|\le v$ and
$v<|w|<2v$. In the first region $L_g(w/v)$ is bounded, and the
$L^1$ bound in the Fourier variable from
\eqref{eq:hardy-weight-fourier-bounds} gives an $O_{b,g}(1)$ bound.
In the second region, that estimate with $N=3$ and $w=vz$ gives
\[
 \int_{v<|w|<2v}h^{-1}|\mathfrak b(v,w/h)|\,|L_g(w/v)|\,dw
 \le C_b\frac{h^2}{v^2}\int_{1<|z|<2}|L_g(z)|\,dz
 \le C_{b,g}\frac{h^2}{v^2}.
\]
Consequently
\[
 J_h^{\mathrm{abs}}(v)\le C_{b,g}
 \begin{cases}
  v/h,&0<v<h,\\
  1,&h\le v\le L,
 \end{cases}
 \qquad
 \int_0^L J_h^{\mathrm{abs}}(v)\frac{dv}{v}
 \le C_{b,g}\bigl(1+\log(L/h)\bigr)<\infty.
\]
Thus the untruncated kernel product is absolutely integrable for each
fixed $h$. Dominated convergence removes the truncations and proves
\eqref{eq:hardy-exact-weighted-trace}.

For $v>0$, denote the inner integral in
\eqref{eq:hardy-exact-weighted-trace} by $I_h(v)$.
If $0<v<h$, the substitution $w=vz$ and
\eqref{eq:hardy-Lg-bounds} give
\[
 |I_h(v)|\le C_{b,g}\frac vh,
 \qquad
 \left|\int_0^h I_h(v)\frac{dv}{v}\right|\le C_{b,g}.
\]
For $h\le v\le L$, compare $I_h(v)$ with $c_gb(v,0)$. In the region
$|w|\le v$, the Lipschitz bound in \eqref{eq:hardy-Lg-bounds} yields
\[
 \int_{|w|\le v}h^{-1}|\mathfrak b(v,w/h)|
          |L_g(w/v)-c_g|\,dw
 \le C_{b,g}\frac hv.
\]
For $v<|w|<2v$, the decay with $N=3$, followed by $w=vz$, gives
\[
 \int_{v<|w|<2v}h^{-1}|\mathfrak b(v,w/h)|
             (|L_g(w/v)|+|c_g|)\,dw
 \le C_{b,g}\frac{h^2}{v^2}.
\]
The completion of the constant term to the whole real line has the same
bound:
\[
 \int_{|w|\ge2v}h^{-1}|\mathfrak b(v,w/h)|\,dw
 \le C_b\frac{h^2}{v^2}.
\]
Therefore
\[
 I_h(v)=c_gb(v,0)+O_{b,g}\!\left(\frac hv+\frac{h^2}{v^2}\right),
 \qquad h\le v\le L.
\]
Since
\[
 \int_h^L\left(\frac h{v^2}+\frac{h^2}{v^3}\right)dv=O(1),
\]
the trace equals
\[
 c_g\int_h^L\frac{b(v,0)}v\,dv+O_{b,g}(1)
 =c_gb(0,0)\log h^{-1}+O_{b,g}(1),
\]
where the final step uses the smoothness of $b$ at $v=0$.
These estimates require only finitely many seminorms of $b$ on a fixed
support, and establish the asserted uniformity.

Finally, $B_+=I-B$ on $L^2(\mathbb R_+)$.
Apply the proved formula with $g^\vee(\lambda)=g(1-\lambda)$, observing
that $W(g^\vee)=W(g)$.
\end{proof}

\begin{remark}[Variance normalization]
\label{rem:hardy-variance-check}
For $g(\lambda)=\lambda(1-\lambda)$, one has $W(g)=1$ and
\[
 k_g(z)=\frac{z}{8\pi^2\sinh(z/2)},\qquad
 G_g(s,t)=\frac{\log(s/t)}{4\pi^2(s-t)},
\]
with the continuous diagonal value $G_g(s,s)=1/(4\pi^2s)$.
The same kernel follows directly from
\[
 B(1-B)=QP_-(1-Q)P_-Q,\qquad
 G_g(s,t)=\frac1{4\pi^2}\int_0^\infty
                         \frac{du}{(s+u)(t+u)}.
\]
This identity also gives the single-branch variance coefficient
$1/(4\pi^2)$ directly.
\end{remark}

\subsection{External momentum cutoffs and comparison with a finite band}

\begin{lemma}[Two elementary trace-class estimates]
\label{lem:hardy-external-cutoff-tools}
The following estimates hold.
\begin{enumerate}
\item If $a\in C_c^\infty(\mathbb R)$, then
\[
 \|[Q,a(h\mathsf D)]\|_{\mathfrak S_1}\le C_a,
\]
uniformly in $h>0$. The same holds if $a-1\in C_c^\infty$.
\item If $A_1,A_2$ are positive self-adjoint contractions, $0\le A_1,A_2\le I$,
$J$ is bounded, and
$A_1J-JA_2\in\mathfrak S_1$, then
\[
 \|g(A_1)J-Jg(A_2)\|_{\mathfrak S_1}
 \le C_g\|A_1J-JA_2\|_{\mathfrak S_1}.
\]
\end{enumerate}
The first estimate is uniform for symbols in a bounded smooth family on
a fixed compact support, and for their frequency translates.
\end{lemma}

\begin{proof}
For the first assertion, the two off-diagonal blocks have, after reflection
and unitary dilation, kernels proportional to $\widehat a(x+y)$ and
$\widehat a(-(x+y))$, respectively, on $\mathbb R_+^2$, independent of $h$.
Both are trace-class Hankel operators.
Indeed, choose a smooth function $\rho$ which is one on $[0,\infty)$ and
zero on $(-\infty,-1]$. The kernel
$\rho(x)\rho(y)\widehat a(\pm(x+y))$ is Schwartz on $\mathbb R^2$ for either sign.
Factoring its operator on both sides through a sufficiently high negative
power of $1+x^2-\partial_x^2$ proves membership in every Schatten class,
in particular in trace class. Restriction to the positive half-line
preserves this property. Frequency translation gives only unitary
modulation. For $a-1$ compactly supported, subtract the identity.

The second assertion follows from
Lemma~\ref{toolbox:intertwining} with $p=1$, which also applies
to spectra in any fixed compact interval.
\end{proof}

\begin{lemma}[Compressing an external phase-space weight]
\label{lem:hardy-weight-compression}
Let $W_h=\operatorname{Op}_h^w(b)$, where $b$ is smooth and compactly
supported, and let $\psi\in C_c^\infty(\mathbb R)$ be real and equal to
one on a fixed neighborhood of the momentum projection of
$\operatorname{supp}b$.
Set $J=Q\psi(h\mathsf D)Q$ and $W_h^+=QW_hQ$. Then
\begin{equation}
 W_h^+=JW_h^+J+R_h,\qquad \|R_h\|_{\mathfrak S_1}=O(1).
 \label{eq:hardy-weight-compression}
\end{equation}
All constants are uniform under fixed support separation and uniform
bounds on finitely many symbol seminorms.
\end{lemma}

\begin{proof}
We first prove the smooth frequency separation estimate
\[
 W_h=\psi(h\mathsf D)W_h\psi(h\mathsf D)+O_{\mathfrak S_1}(h^M)
\]
for every fixed $M>0$. Define the semiclassical Fourier transform by
\[
 (\mathcal F_hu)(\xi)=(2\pi h)^{-1/2}
       \int_{\mathbb R}e^{-is\xi/h}u(s)\,ds.
\]
The kernel of $\mathcal F_hW_h\mathcal F_h^{-1}$ is
\[
 \widetilde W_h(\xi,\eta)=\frac1{2\pi h}
   \int_{\mathbb R}e^{-is(\xi-\eta)/h}
      b\!\left(s,\frac{\xi+\eta}{2}\right)\,ds.
\]
Let $K_\tau$ be the momentum projection of $\operatorname{supp}b$.
Choose $\delta>0$ such that $\psi=1$ whenever
$\operatorname{dist}(\xi,K_\tau)<\delta$.
On the support of $(1-\psi(\xi))\widetilde W_h(\xi,\eta)$,
the midpoint $(\xi+\eta)/2$ belongs to $K_\tau$ and hence
$|\xi-\eta|\ge2\delta$. Repeated integration by parts in $s$
therefore gives an arbitrary power of $h$ and rapid decay in
$\xi-\eta$. The midpoint remains in a fixed compact set, so this
decay controls both $\xi$ and $\eta$. After any fixed number of
derivatives and polynomial weights in these two variables, the same
argument, with more integrations by parts, gives an $O(h^M)$ bound.
Thus the complementary kernel is $O(h^M)$ in every fixed Schwartz
seminorm, for every $M$.

For completeness, let $\mathcal H_\xi=1+\xi^2-\partial_\xi^2$.
If $T_h$ is an operator with this complementary kernel, then
$\mathcal H_\xi T_h\mathcal H_\xi$ has Hilbert--Schmidt norm
$O(h^M)$ by the kernel bounds, while $\mathcal H_\xi^{-1}$ is
Hilbert--Schmidt. Factoring
\[
 T_h=\mathcal H_\xi^{-1}
     (\mathcal H_\xi T_h\mathcal H_\xi)\mathcal H_\xi^{-1}
\]
proves $\|T_h\|_{\mathfrak S_1}=O(h^M)$.
The same proof applies with the complementary factor $1-\psi(\eta)$
on the right. Returning by the unitary transform $\mathcal F_h$ and using
\[
 W_h-\psi(h\mathsf D)W_h\psi(h\mathsf D)
 =(I-\psi(h\mathsf D))W_h
  +\psi(h\mathsf D)W_h(I-\psi(h\mathsf D))
\]
proves the claimed separation estimate. Only finitely many symbol
seminorms are needed for each fixed $M$, uniformly for fixed $\delta$.

After compression, the difference between
$Q\psi(h\mathsf D)W_h\psi(h\mathsf D)Q$ and $JW_h^+J$ is exactly
\[
 Q\psi(h\mathsf D)(1-Q)W_h\psi(h\mathsf D)Q
 +Q\psi(h\mathsf D)QW_h(1-Q)\psi(h\mathsf D)Q.
\]
Each term contains a trace-class off-diagonal block from
Lemma~\ref{lem:hardy-external-cutoff-tools}; all other factors are
uniformly bounded. Hence the difference is $O_{\mathfrak S_1}(1)$,
which proves \eqref{eq:hardy-weight-compression}.
\end{proof}

\begin{theorem}[Finite-band trace with a general phase-space weight]
\label{thm:finite-band-general-weight}
For $t>0$ put
\[
 P_{t,h}=\mathbf1_{(-t,t)}(h\mathsf D),\qquad
 A_{t,h}=QP_{t,h}Q\big|_{L^2(\mathbb R_+)}.
\]
For $0<\varepsilon<T<\infty$, $b\in C_c^\infty(\mathbb R^2)$, and
$g\in C^\infty([0,1])$ with $g(0)=g(1)=0$,
\begin{equation*}
 \operatorname{Tr}\left[Q\operatorname{Op}_h^w(b)Q\,g(A_{t,h})\right]
 =c_g\bigl[b(0,t)+b(0,-t)\bigr]\log h^{-1}+O(1),
\end{equation*}
uniformly for $\varepsilon\le t\le T$. The uniformity also holds for
bounded smooth symbol families on a fixed compact support.
\end{theorem}

\begin{proof}
First suppose that the momentum support of $b$ lies in a small
neighborhood of $+t$, separated uniformly from $(-\infty,-t]$.
Choose a real smooth momentum cutoff $\psi$, equal to one near that
support and satisfying the same separation. Write
$J=Q\psi(h\mathsf D)Q$, and
let
\[
 R_{t,h}^+=\mathbf1_{(-\infty,t)}(h\mathsf D),\qquad
 B_{t,h}^+=QR_{t,h}^+Q.
\]
For any Fourier projection $P$, the identity
\[
 [QPQ,Q\psi(h\mathsf D)Q]
 =-QP(1-Q)\psi(h\mathsf D)Q
  +Q\psi(h\mathsf D)(1-Q)PQ
\]
and Lemma~\ref{lem:hardy-external-cutoff-tools} imply a uniform trace-norm
bound. In particular,
\[
 \|[A_{t,h},J]\|_1+\|[B_{t,h}^+,J]\|_1=O(1).
\]
Moreover, for
$F_{t,h}=\mathbf1_{(-\infty,-t]}(h\mathsf D)$,
$F_{t,h}\psi(h\mathsf D)=0$ and
\[
 (A_{t,h}-B_{t,h}^+)J
 =-QF_{t,h}[Q,\psi(h\mathsf D)]Q.
\]
The right side is uniformly trace class, and so is its adjoint.
It follows that $A_{t,h}J-JB_{t,h}^+$ is uniformly trace class.
The smooth quasi-commutator estimate, also applied to $[B_{t,h}^+,J]$,
therefore yields
\begin{equation}
 \|J[g(A_{t,h})-g(B_{t,h}^+)]J\|_1=O(1).
 \label{eq:finite-band-hardy-external-comparison}
\end{equation}

With $W_h^+=Q\operatorname{Op}_h^w(b)Q$, use
Lemma~\ref{lem:hardy-weight-compression} and cyclicity of the trace to
obtain
\[
 \operatorname{Tr}W_h^+g(A_{t,h})
 =\operatorname{Tr}W_h^+g(B_{t,h}^+)+O(1).
\]
Indeed, \eqref{eq:finite-band-hardy-external-comparison} gives
\[
 \left|\operatorname{Tr}\!\left(
 W_h^+J[g(A_{t,h})-g(B_{t,h}^+)]J\right)\right|
 \le \|W_h^+\|\,
 \|J[g(A_{t,h})-g(B_{t,h}^+)]J\|_1=O(1).
\]
The unitary modulation $M_tu(s)=e^{its/h}u(s)$ preserves $Q$ and satisfies
\[
 M_t^*B_{t,h}^+M_t=B,\qquad
 M_t^*\operatorname{Op}_h^w(b)M_t
 =\operatorname{Op}_h^w\bigl(b(s,\tau+t)\bigr).
\]
Theorem~\ref{thm:hardy-general-weight} gives the contribution
$c_gb(0,t)\log h^{-1}+O(1)$.

Near the negative endpoint use instead
\[
 B_{t,h}^-=Q\mathbf1_{(-t,\infty)}(h\mathsf D)Q
 =I-Q\mathbf1_{(-\infty,-t]}(h\mathsf D)Q.
\]
The identical external-cutoff comparison applies. Replacing $g$ by
$g^\vee(\lambda)=g(1-\lambda)$ and shifting the threshold to $-t$ gives
$c_gb(0,-t)\log h^{-1}+O(1)$, because
$W(g^\vee)=W(g)$.

It remains to treat weights supported away from both endpoints. Split
their momentum supports into occupied and unoccupied parts. For either
part choose a nested momentum cutoff with
\[
 P_{t,h}\psi(h\mathsf D)=c\,\psi(h\mathsf D),\qquad c\in\{0,1\}.
\]
Then
\[
 A_{t,h}J-cJ=QP_{t,h}[Q,\psi(h\mathsf D)]Q
\]
is uniformly trace class. Since $g(c)=0$, the quasi-commutator estimate
gives $\|g(A_{t,h})J\|_1=O(1)$. Lemma
\ref{lem:hardy-weight-compression} consequently bounds the corresponding
weighted trace by $O(1)$.

A smooth momentum partition into the two endpoint neighborhoods and
their complement completes the proof. Its widths can be fixed in terms
of $\varepsilon$, so all seminorms and all estimates are uniform for
$\varepsilon\le t\le T$.
\end{proof}

\begin{remark}[External localization]
\label{rem:hardy-cutoffs-external}
The comparison acts on the externally localized operators
$Jg(A)J$ and $Jg(B)J$. Its $O(1)$ trace-norm error follows from
the bounds for $[Q,\psi(h\mathsf D)]$ in
Lemma~\ref{lem:hardy-external-cutoff-tools}.
\end{remark}

\subsection{Tangential partial traces and the branch factor}

\begin{corollary}[Tangential variables and a general Hardy weight]
\label{cor:hardy-tangential-weight}
Let $n\ge1$, and let $b(y,s,\eta,\tau)$ be a smooth compactly supported
symbol on $\mathbb R_y^{n-1}\times\mathbb R_s\times
\mathbb R_\eta^{n-1}\times\mathbb R_\tau$. On
$L^2(\mathbb R_y^{n-1}\times\mathbb R_s)$ let
$\mathcal Q=I\otimes Q$. On
$\operatorname{Ran}\mathcal Q\simeq
L^2(\mathbb R_y^{n-1}\times\mathbb R_+)$ let
$\mathcal B=I\otimes B$; the trace below is taken on this range.
Then
\begin{align*}
 \operatorname{Tr}\left[
 \mathcal Q\operatorname{Op}_h^w(b)\mathcal Q\,g(\mathcal B)\right]
 ={}&(2\pi h)^{1-n}c_g\log h^{-1}
       \int_{\mathbb R^{2(n-1)}}b(y,0,\eta,0)\,dy\,d\eta
       +O(h^{1-n}).
\end{align*}
The same conclusion holds for the opposite Hardy orientation. It is
uniform for bounded smooth symbol families on a fixed compact support.
For a classical family $b_h=b_0+O(h)$ in these seminorms, the leading
coefficient may be computed from $b_0$; the resulting
$O(h^{2-n}\log h^{-1})$ error is absorbed in $O(h^{1-n})$.
\end{corollary}

\begin{proof}
Taking the exact tangential Fourier diagonal of the Weyl operator gives
\[
 \operatorname{Tr}\left[
 \mathcal Q\operatorname{Op}_h^w(b)\mathcal Q\,g(\mathcal B)\right]
 =(2\pi h)^{1-n}\int dy\,d\eta\;
 \operatorname{Tr}_{L^2(\mathbb R_+)}
 \left[Q\operatorname{Op}_{h,s}^w
    \bigl(b(y,\cdot,\eta,\cdot)\bigr)Q\,g(B)\right].
\]
For separated symbols the identity follows from the Weyl kernel.
Smooth tensor expansion extends it to general compactly supported
symbols; the uniform normal trace-class bounds justify termwise
integration. Theorem~\ref{thm:hardy-general-weight} then gives the result.
\end{proof}

\begin{corollary}[Frozen finite-band fibers]
\label{cor:frozen-finite-band-branches}
Let $t:\mathbb R^{n-1}\to[0,\infty)$ be measurable, and set
$A_{0,h}=0$. On $L^2(\mathbb R_y^{n-1}\times\mathbb R_+)$ define
the tangential semiclassical Fourier transform by
\[
 (\mathcal F_{h,y}u)(\eta,s)
 =(2\pi h)^{-(n-1)/2}
   \int_{\mathbb R^{n-1}}e^{-iy\cdot\eta/h}u(y,s)\,dy.
\]
The translation-invariant tangential model $\mathcal A_h$ is specified by
\[
 \mathcal F_{h,y}\mathcal A_h\mathcal F_{h,y}^{-1}
 =\int_{\mathbb R^{n-1}}^\oplus A_{t(\eta),h}\,d\eta.
\]
Here $A_{t,h}$ is the half-line compression defined in
Theorem~\ref{thm:finite-band-general-weight}. Its fibers are measurable:
the Fourier multiplier
$\mathbf1_{\{|\tau|<t(\eta)\}}$ is jointly measurable in $(\eta,\tau)$,
and compression by $Q$ preserves measurability.
Each fiber is a positive contraction, so the direct integral defines
a bounded self-adjoint positive contraction. For $n=1$, the tangential
space is $\mathbb R^0$, its integral has unit mass, and
$\mathcal F_{h,y}$ is the identity.
Suppose the smooth compactly supported weight $b$ is supported where
$\varepsilon\le t(\eta)\le T$, with $0<\varepsilon<T<\infty$.
Continuous functional calculus acts fiberwise, as follows first for
polynomials and then by uniform approximation on $[0,1]$.
The exact partial-trace formulas in
Corollary~\ref{cor:hardy-tangential-weight} and
Theorem~\ref{thm:finite-band-general-weight} give
\begin{align*}
 \operatorname{Tr}\left[
 \mathcal Q\operatorname{Op}_h^w(b)\mathcal Q\,g(\mathcal A_h)\right]
 ={}&(2\pi h)^{1-n}c_g\log h^{-1}\int
 \bigl[b(y,0,\eta,t(\eta))+b(y,0,\eta,-t(\eta))\bigr]dy\,d\eta
 \\
 &\hspace{25mm}+O(h^{1-n}).
\end{align*}
For a single-branch weight only its corresponding summand remains.
For the frozen Schr\"odinger model $-h^2\Delta-\kappa_0$, with
$\kappa_0>0$, the tangential semiclassical Fourier transform gives
the direct-integral fibers with
$t(\eta)=\sqrt{\kappa_0-|\eta|^2}$ for $|\eta|^2<\kappa_0$.
For $|\eta|^2\ge\kappa_0$ the momentum band is empty up to a null set,
so the fiber projection is zero and contributes nothing since $g(0)=0$.
The corollary applies on every fixed transverse region
$t(\eta)\ge\varepsilon$ within the occupied region.
\end{corollary}

\begin{remark}[Two branches at each one-dimensional spatial endpoint]
\label{rem:hardy-one-dimensional-factor}
At a transverse spatial endpoint in dimension one there are two Fermi
momenta, $+t$ and $-t$. An external weight equal to one at both points
therefore contributes
\[
 2c_g\log h^{-1}
 =\frac{W(g)}{2\pi^2}\log h^{-1}.
\]
The projector-localization estimates therefore give the coefficient
for a finite union of intervals as
\[
 \frac{\#\partial_*\Omega}{2\pi^2}W(g),
 \qquad\partial_*\Omega=\partial(\operatorname{int}\overline\Omega).
\]
In particular, one interval has variance coefficient $1/\pi^2$.
\end{remark}

\section{The transverse local trace formula}
\label{sec:transverse}
\label{subsec:transverse-quasi-intertwining}

By Theorem~\ref{loc:comparison}, we may work with a smooth
quadratic-confining Hamiltonian.  Throughout this section, $H_h$ is
self-adjoint, its spectral projection
\[
 P_h=\mathbf1_{(-\infty,0]}(H_h)
\]
has finite rank, and the sharp-to-smooth comparison
\eqref{S6:sharp-smooth} is available.  The zero of energy has been
shifted to the Fermi energy.  Theorem~\ref{loc:comparison} and
Lemma~\ref{toolbox:transfer-moments} transfer the conclusions below to
the original Hamiltonian. Writing $T_h^{\mathrm{orig}}$ and $T_h^{\mathrm c}$ for the original
and confining compressions, respectively, we have, for every smooth
$g$ with $g(0)=0$ and every uniformly bounded weight $W_h$,
\[
 \bigl|\Tr\bigl(W_h[g(T_h^{\mathrm{orig}})-g(T_h^{\mathrm c})]\bigr)\bigr|
 \le \|W_h\|\,
       \|g(T_h^{\mathrm{orig}})-g(T_h^{\mathrm c})\|_{\mathfrak S_1}
 =O_g(h^{1-n}).
\]

In a fixed boundary chart use coordinates $(y,s)\in\mathbb R^{n-1}\times\mathbb R$, with
$Q=\mathbf1_{\{s>0\}}$.  The unitary change to these coordinates includes
the square root of the volume density.  Let $p(y,s,\eta,\xi_s)$ be the
principal symbol.  Fix a compact transverse branch, a number
$\varepsilon>0$, and a sufficiently small, fixed phase-space neighborhood
of that branch on which
\[
 |\partial_{\xi_s}p|\geq\varepsilon,
 \qquad \sigma=\operatorname{sgn}(\partial_{\xi_s}p)\in\{-1,1\}.
\]
The zero-energy root on the spatial boundary is denoted by
\[
 p(y,0,\eta,a(y,\eta))=0.
\]
All cutoffs have fixed supports and uniformly bounded symbol
seminorms. The branch and $\varepsilon>0$ are fixed throughout;
constants may depend on both. We assume
$g\in C^\infty([0,1])$ with $g(0)=g(1)=0$, and retain
$W(g)$ from \eqref{eq:W-intro}.

\begin{theorem}[Transverse local trace formula]
\label{thm:transverse-local-trace}
Let $L_h\in\Psi_h^0$ be a fixed-support, properly supported classical
pseudodifferential cutoff whose microsupport is compactly contained in
the transverse neighborhood just described.  Set
\[
 W_h=L_h^*L_h,
 \qquad w_0=|\ell_0|^2,
 \qquad T_h=QP_hQ,
\]
where $\ell_0$ is the principal symbol of $L_h$.  After shrinking the
neighborhood, if necessary, one has
\begin{align}
 \operatorname{Tr}\bigl(W_hg(T_h)\bigr)
 &=\frac{\log h^{-1}}{(2\pi h)^{n-1}}
   \frac{W(g)}{4\pi^2}
   \int w_0\bigl(y,0,\eta,a(y,\eta)\bigr)\,dy\,d\eta
   +O_{g,\varepsilon,L}(h^{1-n}).
 \label{eq:transverse-local-trace}
\end{align}
\end{theorem}

\begin{proof}
\smallskip\noindent
\emph{The boundary-adapted canonical transformation.}
The implicit-function theorem first writes the energy equation as
\[
 \xi_s=\alpha(y,s,\eta,E),
 \qquad p\bigl(y,s,\eta,\alpha(y,s,\eta,E)\bigr)=E.
\]
Solve the noncharacteristic Hamilton--Jacobi initial-value problem
\begin{equation}
 p\bigl(y,s,\partial_yS,\partial_sS\bigr)=\sigma\tau,
 \qquad S(y,0;\eta,\tau)=y\cdot\eta.
 \label{eq:boundary-hj}
\end{equation}
On a sufficiently small fixed chart its smooth solution satisfies
\[
 \det\partial_{(y,s),(\eta,\tau)}^2S\neq0,
 \qquad \det\partial_{y,\eta}^2S\neq0.
\]
Indeed, on $s=0$ the tangential mixed derivative is the identity and
\[
 \partial_s\partial_\tau S
 =\frac{\sigma}{\partial_{\xi_s}p}>0.
\]
Since $\partial_\tau S(y,0;\eta,\tau)=0$, Hadamard's formula gives
\begin{equation}
 \partial_\tau S(y,s;\eta,\tau)
 =s\,c(y,s,\eta,\tau),\qquad c\geq c_\varepsilon>0.
 \label{eq:normal-sign}
\end{equation}
The phase
\[
 S(y,s;\eta,\tau)-z'\cdot\eta-u\tau
\]
therefore generates a local canonical transformation from the model
variables $(z',u;\eta,\tau)$ to the physical variables.  It sends the
model symbol $\sigma\tau$ to $p$, sends $u=0$ to $s=0$, and preserves
the two sides of these hypersurfaces.  At their intersection with the
energy surface, it is explicitly
\begin{equation}
 (z',0;\eta,0)\longmapsto
 (y,0;\eta,a(y,\eta)),\qquad y=z'.
 \label{eq:boundary-restricted-map}
\end{equation}

\smallskip\noindent
\emph{Quantization and microlocal unitarity.}
We fix the localization sets before quantization. Let $\kappa$ be
the canonical map just constructed and choose model neighborhoods
\[
 \mathcal V_0\Subset\mathcal V_1\Subset\mathcal V_2
\]
inside its domain. Here $\operatorname{MS}$ denotes the fixed compact
microsupport containing the full-symbol support modulo residual terms,
and we require
$\kappa^{-1}(\operatorname{MS}(L_h))\Subset\mathcal V_0$.
The neighborhood in the statement is chosen small enough for this nesting.
Choose $\tau_0>0$ such that
\[
 \bigcup_{|t|\le\tau_0}
 \exp(tH_{\sigma\tau})(\overline{\mathcal V_0})
 \Subset\mathcal V_1,\qquad H_{\sigma\tau}=\sigma\partial_u.
\]
Take a smooth compact model cutoff equal to one near this flow tube
and supported in $\mathcal V_1$, and a second cutoff equal to one
near $\overline{\mathcal V_1}$ and supported in $\mathcal V_2$.
Their physical counterparts are transported by $\kappa$.
Construct the normal form on a neighborhood of
$\overline{\mathcal V_1}$ and use the outer cutoffs to extend the FIO
as a bounded, properly supported operator. All these choices are
independent of $h$. In the estimates below, $C_h$ has microsupport
compactly contained in $\mathcal V_0$; the final choice $C_h=c_h^*$
has this property by the support condition on $L_h$.

Quantize the canonical graph generated by \eqref{eq:boundary-hj}, using
\cite[Theorem~11.6]{Zworski}. The transport construction in the proof
of \cite[Theorem~12.3]{Zworski} gives an elliptic order-zero FIO
$U_{0,h}$ associated with this graph. Put
\[
 D_h=\sigma h\mathsf D_u,\qquad
 \mathcal R_h=H_hU_{0,h}-U_{0,h}D_h.
\]
For every $C_h\in\Psi_h^0$ satisfying
$\operatorname{MS}(C_h)\Subset\mathcal V_1$,
\[
 \mathcal R_hC_h=O_{\mathcal S'\to\mathcal S}(h^\infty).
\]
The transport equations are noncharacteristic because
$|\partial_{\xi_s}p|\ge\varepsilon$ on the chosen chart.

Set $G_h=U_{0,h}^*U_{0,h}$. This is a positive elliptic
pseudodifferential operator microlocally on $\mathcal V_1$. The
algebraic identity
\[
 D_hG_h-G_hD_h
 =U_{0,h}^*\mathcal R_h-\mathcal R_h^*U_{0,h}
\]
implies, for self-adjoint $C_{j,h}\in\Psi_h^0$ with
$\operatorname{MS}(C_{j,h})\Subset\mathcal V_1$,
\[
 C_{1,h}[D_h,G_h]C_{2,h}
 =C_{1,h}U_{0,h}^*\mathcal R_hC_{2,h}
  -(\mathcal R_hC_{1,h})^*U_{0,h}C_{2,h}
 =O_{\mathcal S'\to\mathcal S}(h^\infty).
\]
Choose a self-adjoint $\chi_h\in\Psi_h^0$ that is microlocally equal
to $I$ on a neighborhood of $\overline{\mathcal V_1}$ and whose
microsupport lies in the elliptic region of $U_{0,h}$, and define
\[
 \widetilde G_h
 :=\chi_hG_h\chi_h+(I-\chi_h)^2.
\]
Then $\widetilde G_h\in\Psi_h^0$ is self-adjoint and nonnegative,
equals $G_h$ microlocally on $\mathcal V_1$, and equals $I$
microlocally outside a compact set. Ellipticity of $U_{0,h}$ gives
constants $c_0>0$ and, for every $N$, $C_N>0$ such that
\[
 \langle G_h\chi_hu,\chi_hu\rangle
 \ge c_0\|\chi_hu\|^2-C_Nh^N\|u\|^2.
\]
Since
\[
 \|u\|^2\le
 2\bigl(\|\chi_hu\|^2+\|(I-\chi_h)u\|^2\bigr),
\]
we obtain $\widetilde G_h\ge cI$ for all sufficiently small $h$, with
$c>0$ independent of $h$. Semiclassical functional calculus then gives
\[
 \widetilde G_h^{-1/2}\in\Psi_h^0,\qquad
 [D_h,\widetilde G_h^{-1/2}]
 =O_{\mathcal S'\to\mathcal S}(h^\infty)
 \quad\text{microlocally on }\mathcal V_1.
\]
Define $U_h=U_{0,h}\widetilde G_h^{-1/2}$.

Let $C_h\in\Psi_h^0$ satisfy
$\operatorname{MS}(C_h)\Subset\mathcal V_0$, and let
$\Lambda_h\in\Psi_h^0$ satisfy
$\operatorname{MS}(\Lambda_h)\Subset\kappa(\mathcal V_0)$. Then
\begin{align}
 (H_hU_h-U_hD_h)C_h
 &=O_{\mathcal S'\to\mathcal S}(h^\infty), \notag\\
 (U_h^*U_h-I)C_h,\quad C_h(U_h^*U_h-I)
 &=O_{\mathcal S'\to\mathcal S}(h^\infty), \notag\\
 (U_hU_h^*-I)\Lambda_h,\quad
 \Lambda_h(U_hU_h^*-I)
 &=O_{\mathcal S'\to\mathcal S}(h^\infty).
 \label{eq:microunitary-hj}
\end{align}
For the last line, an elliptic FIO associated with a canonical graph
has a microlocal two-sided parametrix. Since $U_h^*U_h=I$
microlocally on the model chart, this parametrix equals $U_h^*$ modulo
a residual operator. We use bounded, properly supported FIO extensions
with smooth compact microlocal cutoffs outside these charts.

\smallskip\noindent
\emph{The spatial projection commutator.}
Let $C_h\in\Psi_h^0$ have fixed compact microsupport in an inner
model chart and put $J_h=U_hC_h$.  Then, for every $p_*>0$,
\begin{equation}
 \|QJ_h-J_hQ_0\|_{\mathfrak S_{p_*}}^{p_*}
 \leq C_{p_*,\varepsilon,C}h^{1-n},
 \qquad Q_0=\mathbf1_{\{u>0\}}.
 \label{eq:direct-boundary-leakage}
\end{equation}
With fixed spatial cutoffs, the kernel of $J_h$ is, modulo a
residual operator,
\[
 (2\pi h)^{-n}\int
 e^{i\Phi(y,s,z',u,\eta,\tau)/h}
 b_h(y,s,z',u,\eta,\tau)\,d\eta\,d\tau,
\]
with a fixed compact amplitude and uniformly bounded derivatives,
where
\[
 \Phi=S(y,s;\eta,\tau)-z'\!\cdot\eta-u\tau.
\]
On the support of the kernel of $QJ_h-J_hQ_0$ the normal variables
have opposite signs.  By \eqref{eq:normal-sign},
\[
 |\partial_\tau\Phi|\geq c(|s|+|u|).
\]
Furthermore,
\[
 S(y,s;\eta,\tau)=y\!\cdot\eta+s\beta(y,s,\eta,\tau)
\]
with $\beta$ smooth, and hence
\[
 \partial_\eta\Phi=y-z'+s\partial_\eta\beta.
\]
Combining the last two estimates gives
\begin{equation}
 |\partial_{(\eta,\tau)}\Phi|
 \geq c'\bigl(|s|+|u|+|y-z'|\bigr).
 \label{eq:boundary-fio-nonstationary}
\end{equation}
Use a smooth dyadic partition in
$r=(s^2+u^2+|y-z'|^2)^{1/2}$, with the innermost term $r\lesssim h$.
On a shell $r\asymp R\geq h$, partition the tangential variables
into cubes of side comparable to $R$.  Only $O(R^{1-n})$ pairs of
such cubes occur, since $|y-z'|=O(R)$ and the tangential support is
fixed. Dilating all spatial variables by $R$ gives the parameter
$\lambda=R/h$, the prefactor $\lambda^n$, and the phase
\[
 \frac{\Phi}{R}
 =\frac{y-z'}{R}\cdot\eta
   +\frac{s}{R}\beta(y,s,\eta,\tau)-\frac{u}{R}\tau.
\]
The centers of each tangential cube pair differ by $O(R)$.
Consequently the rescaled phase has uniformly bounded derivatives.
Its $(\eta,\tau)$ gradient on the cross-boundary shell is bounded
away from zero by
\eqref{eq:boundary-fio-nonstationary}. Insert a smooth mask equal to
one on this part of the shell and supported in a fixed neighborhood
where the same lower bound holds. Compactness and the uniform
lower bound allow the mask to have uniformly bounded rescaled
derivatives. Repeated integration by parts in $(\eta,\tau)$ gives,
for arbitrary $N$ and $m$, a bound $C_{N,m}\lambda^{-N}$ for the
resulting smooth kernel in $C^m$, including its prefactor.
Restriction of this smooth kernel to the two opposite half-spaces
recovers the original cross-boundary kernel.

Fourier expansion of the smooth kernel on a fixed product of
tori gives a rank-one series. Bounds on sufficiently many
derivatives make its coefficients summable in
$\min\{p_*,1\}$th power.
Thus each rescaled block has
$\mathfrak S_{p_*}$ $p_*$th power at most $C_N(R/h)^{-N}$,
after sufficiently many integrations by parts.  For $0<p_*\leq1$,
summing the blocks gives
\[
 C_N\sum_{j\geq0}(2^jh)^{1-n}2^{-jN}
 \leq C h^{1-n}.
\]
The innermost scale $r\lesssim h$ has uniformly smooth rescaled
kernels and $O(h^{1-n})$ tangential blocks, and obeys the same bound.
For $p_*>1$, apply
$\|K\|_{\mathfrak S_{p_*}}^{p_*}
 \le\|K\|^{p_*-1}\|K\|_{\mathfrak S_1}$
with $K=QJ_h-J_hQ_0$ and $\|K\|\le2\|J_h\|=O(1)$.
This proves \eqref{eq:direct-boundary-leakage}.

\smallskip\noindent
\emph{Spectral quasi-intertwining.}
Use the fixed charts, flow tube and $\tau_0$ chosen above.
Thus all translates of the microsupport of $C_h$ for
$|t|\le\tau_0$ remain where \eqref{eq:microunitary-hj} holds.
Choose $\rho\in\mathcal S(\mathbb R)$ with $\int\rho=1$ and
$\operatorname{supp}\widehat\rho\subset(-\tau_0,\tau_0)$, and set
\[
 F=\rho*\mathbf1_{(-\infty,0]},\qquad f_h(E)=F(E/h).
\]
The sharp-to-smooth estimate \eqref{S6:sharp-smooth} gives
\begin{equation}
 \|P_h-f_h(H_h)\|_{\mathfrak S_{p_*}}^{p_*}
 \leq C_{p_*}h^{1-n}\qquad(p_*>0).
 \label{eq:input-sharp-smooth-f5}
\end{equation}

On the range of $C_h$, Duhamel's formula gives
\begin{align*}
 e^{itH_h/h}U_hC_h-U_he^{itD_h/h}C_h
 =\frac{i}{h}\int_0^t e^{i(t-v)H_h/h}
 (H_hU_h-U_hD_h)e^{ivD_h/h}C_h\,dv.
\end{align*}
The nested-chart flow condition makes the integrand residual.
The identity can first be checked on smooth vectors; $C_h$ has
compact microsupport and is smoothing for each fixed $h$, so all
unbounded Hamiltonians in this formula are applied on their
domains.  More explicitly, insert a fixed model cutoff equal to
one on the union of the translates of the microsupport of $C_h$.
The factor $H_hU_h-U_hD_h$ composed with this cutoff is a residual
operator with rapidly decreasing kernels; the complementary
factor acting on $e^{ivD_h/h}C_h$ is residual as well.
Lemma~\ref{app:propagation} supplies the global weighted
Duhamel estimate for the auxiliary quadratic-confining Hamiltonian.
It follows that the propagator discrepancy is $O(h^\infty)$ in
every Schwartz kernel seminorm, uniformly for $|t|\le\tau_0$ and
after every fixed number of $t$-derivatives. The weighted estimates hold to arbitrary order in $h$.

Since $F'=-\rho$, the unitary Fourier convention above gives
$it\widehat F(t)=-\widehat\rho(t)$ in the distributional sense.
More explicitly, the Fourier transform of the negative half-line
indicator and $\int\rho=1$ give
\[
 \widehat F(t)
 =\sqrt{\frac{\pi}{2}}\,\delta_0(t)
 -\widehat\rho(t)\operatorname{pv}\frac1{it}.
\]
To check the sign and the coefficient of $\delta_0$, regularize the
half-line indicator by $e^{\epsilon E}\mathbf1_{\{E\le0\}}$.
Its Fourier transform is
\[
 \frac1{\sqrt{2\pi}(\epsilon-it)}
 =\frac1{\sqrt{2\pi}}
   \left(\frac{\epsilon}{\epsilon^2+t^2}
       +\frac{it}{\epsilon^2+t^2}\right).
\]
As $\epsilon\downarrow0$, the two terms in parentheses converge in
distributions to $\pi\delta_0$ and
$i\operatorname{pv}(1/t)$, respectively.
The convolution identity and
$\widehat\rho(0)=(2\pi)^{-1/2}$ give the displayed formula for $\widehat F$.
Here $\operatorname{pv}$ denotes the Cauchy principal value, and
the second term is multiplication of that distribution by the smooth
function $\widehat\rho$.
Thus $\widehat F$ is a compactly supported distribution of finite order,
with support in $(-\tau_0,\tau_0)$.
Pairing the preceding propagator identity with this distribution proves
\begin{equation}
 f_h(H_h)U_hC_h-U_hf_h(D_h)C_h
 =O_{\mathfrak S_{p_*}}(h^\infty)\qquad(p_*>0).
 \label{eq:smooth-spectral-intertwining-f5}
\end{equation}
For the Schatten estimate, let
$\Lambda_{x,h}=1+|x|^2-h^2\Delta_x$ and define $\Lambda_{z,h}$
in the model coordinates. A global residual satisfies
\[
 \|\Lambda_{x,h}^mR_h\Lambda_{z,h}^m\|=O(h^N)
 \qquad(m,N\ge0).
\]
The oscillator eigenvalue count gives
$\|\Lambda_{x,h}^{-m}\|_{\mathfrak S_{2p_*}}^{2p_*}
+\|\Lambda_{z,h}^{-m}\|_{\mathfrak S_{2p_*}}^{2p_*}
=O(h^{-n})$ for $2mp_*>n$. Factoring on both sides yields
\[
 \|R_h\|_{\mathfrak S_{p_*}}^{p_*}
 \le
 \|\Lambda_{x,h}^{-m}\|_{\mathfrak S_{2p_*}}^{p_*}
 \|\Lambda_{x,h}^mR_h\Lambda_{z,h}^m\|^{p_*}
 \|\Lambda_{z,h}^{-m}\|_{\mathfrak S_{2p_*}}^{p_*}
 =O(h^{Np_*-n}).
\]
Since $N$ is arbitrary, this proves the remainder estimate in
\eqref{eq:smooth-spectral-intertwining-f5}.

Put $P_0=\mathbf1_{(-\infty,0]}(D_h)$.  The two elementary model
estimates are
\begin{equation}
 \|(f_h(D_h)-P_0)C_h\|_{\mathfrak S_{p_*}}^{p_*}
 +\|[P_0,C_h]\|_{\mathfrak S_{p_*}}^{p_*}
 \leq C_{p_*,C}h^{1-n}.
 \label{eq:model-cutoff-estimates-f5}
\end{equation}
Lemma~\ref{toolbox:model-ideals} supplies the normal ideal bounds
and the tangential density used here.
Write $C_h$, modulo a residual, in left quantization with its full
classical symbol of fixed compact support, and expand that symbol
on fixed tangential and normal boxes. The tangential factors obey
the usual fixed-support semiclassical ideal bound $O(h^{1-n})$ for
the $p_*$-power.  In the normal factors,
$f_h(D_h)-P_0
=F(\sigma\mathsf D_u)-\mathbf1_{(-\infty,0]}(\sigma\mathsf D_u)$
is independent of $h$.  Its multiplier is bounded, has rapidly
decaying tails, and has finite moments of every order.  Its product
with a compactly supported smooth multiplier belongs to every
$\mathfrak S_{p_*}$ by the rank-one Fourier expansion in
Lemma~\ref{toolbox:model-ideals}.
For the second estimate, the commutator of the Hardy projection
with a smooth compactly supported multiplier is a pair of Hankel
operators with kernels proportional to $\widehat\phi(v+w)$ and
$\widehat\phi(-(v+w))$, respectively, on $\mathbb R_+\times\mathbb R_+$.  All derivatives of these kernels
decay rapidly in $v+w$.  Partition into unit squares and expand on
each square to obtain membership in every $\mathfrak S_{p_*}$.
The right normal Fourier multiplier in a separated symbol is
uniformly bounded.  Rapid decay of the symbol's Fourier
coefficients absorbs the polynomial seminorms of the modulated
normal factors.  These observations prove
\eqref{eq:model-cutoff-estimates-f5}.

Combining \eqref{eq:input-sharp-smooth-f5}--
\eqref{eq:model-cutoff-estimates-f5} with the identity
\begin{align*}
 P_hJ_h-J_hP_0
 ={}&(P_h-f_h(H_h))U_hC_h\\
 &+\bigl(f_h(H_h)U_hC_h-U_hf_h(D_h)C_h\bigr)\\
 &+U_h(f_h(D_h)-P_0)C_h+U_h[P_0,C_h]
\end{align*}
gives
\begin{equation}
 \|P_hJ_h-J_hP_0\|_{\mathfrak S_{p_*}}^{p_*}
 \leq C_{p_*,\varepsilon,C}h^{1-n}.
 \label{eq:spectral-quasi-intertwining-bound}
\end{equation}

\smallskip\noindent
\emph{Compression and smooth functional calculus.}
Set
\[
 T_{0,h}=Q_0P_0Q_0,\qquad
 E_{P,h}=P_hJ_h-J_hP_0,\qquad
 E_{Q,h}=QJ_h-J_hQ_0.
\]
Equations~\eqref{eq:direct-boundary-leakage} and
\eqref{eq:spectral-quasi-intertwining-bound} give, for each fixed
$p_*>0$,
\begin{equation*}
 \|E_{Q,h}\|_{\mathfrak S_{p_*}}^{p_*}
 +\|E_{P,h}\|_{\mathfrak S_{p_*}}^{p_*}
 \le C_{p_*,\varepsilon,C}h^{1-n}.
\end{equation*}
All compressed operators act on their full ambient Hilbert spaces
and vanish on the complementary half-space. In particular,
$0\le T_h,T_{0,h}\le I$. Direct expansion gives
\begin{equation*}
 \begin{aligned}
 T_hJ_h-J_hT_{0,h}
 &=QP_h(QJ_h-J_hQ_0)+Q(P_hJ_h-J_hP_0)Q_0\\
 &\quad +(QJ_h-J_hQ_0)P_0Q_0\\
 &=QP_hE_{Q,h}+QE_{P,h}Q_0+E_{Q,h}P_0Q_0.
 \end{aligned}
\end{equation*}
Taking $p_*=1$ in the projection-error estimates gives
\[
 \|T_hJ_h-J_hT_{0,h}\|_{\mathfrak S_1}
 \le2\|E_{Q,h}\|_{\mathfrak S_1}+\|E_{P,h}\|_{\mathfrak S_1}
 \le C_{\varepsilon,C}h^{1-n}.
\]
Choose $\widetilde g\in C_c^\infty(\mathbb R)$ with
$\widetilde g=g$ on $[0,1]$, and put
$K_h=T_hJ_h-J_hT_{0,h}$.
Duhamel's formula gives, for $t\in\mathbb R$,
\[
 e^{itT_h}J_h-J_he^{itT_{0,h}}
 =i\int_0^t e^{i(t-s)T_h}K_he^{isT_{0,h}}\,ds,
 \qquad
 \|e^{itT_h}J_h-J_he^{itT_{0,h}}\|_{\mathfrak S_1}
 \le |t|\|K_h\|_{\mathfrak S_1}.
\]
Fourier inversion therefore yields
\begin{equation*}
 \|g(T_h)J_h-J_hg(T_{0,h})\|_{\mathfrak S_1}
 \leq \frac1{\sqrt{2\pi}}\left(\int_{\mathbb R}|t\widehat{\widetilde g}(t)|\,dt\right)
       \|T_hJ_h-J_hT_{0,h}\|_{\mathfrak S_1}
 \leq C_{g,\varepsilon,C}h^{1-n}.
\end{equation*}

\smallskip\noindent
\emph{Transformation of the external weight.}
Define the model pseudodifferential operator
\[
 c_h=U_h^*L_hU_h
\]
with nested microlocal cutoffs equal to one on the displayed
microsupports.  It has fixed compact microsupport, and its principal
symbol is $\ell_0\circ\kappa$, where $\kappa$ is the canonical
map above.  In the preceding argument choose
\[
 C_h=c_h^*,\qquad J_h=U_hc_h^*.
\]
The microsupport of $L_h$ was chosen sufficiently far inside the
fixed charts that this choice satisfies all the required nesting
conditions. The localized microlocal-unitarity identities give
\[
 J_h-L_h^*U_h
 =(U_hU_h^*-I)L_h^*U_h
 =O_{\mathfrak S_1}(h^\infty).
\]
Moreover,
\[
 J_h^*J_h-c_hc_h^*
 =c_h(U_h^*U_h-I)c_h^*
 =O_{\mathfrak S_1}(h^\infty).
\]
Consequently,
\begin{align*}
 J_hJ_h^*-L_h^*L_h
 ={}&(J_h-L_h^*U_h)J_h^*
   +L_h^*U_h(J_h^*-U_h^*L_h)\\
 &+L_h^*(U_hU_h^*-I)L_h
 =O_{\mathfrak S_1}(h^\infty).
\end{align*}
Thus
\begin{equation}
 J_hJ_h^*=L_h^*L_h+O_{\mathfrak S_1}(h^\infty),
 \qquad
 J_h^*J_h=c_hc_h^*+O_{\mathfrak S_1}(h^\infty).
 \label{eq:weight-orientation-f5}
\end{equation}
All compositions are localized on nested microsupports; the
residual identities in \eqref{eq:weight-orientation-f5} are global
trace-ideal statements after these localizations.

The operators $c_h$ and $J_h$ are Hilbert--Schmidt for each fixed
$h$, so the following products are trace class. Cyclicity gives
\[
 \begin{aligned}
 &\operatorname{Tr}(J_hJ_h^*g(T_h))
  -\operatorname{Tr}(J_h^*J_hg(T_{0,h}))\\
 &\qquad=\operatorname{Tr}\bigl(
 J_h^*[g(T_h)J_h-J_hg(T_{0,h})]\bigr).
 \end{aligned}
\]
The absolute value is bounded by
$\|J_h\|\,\|g(T_h)J_h-J_hg(T_{0,h})\|_{\mathfrak S_1}=O(h^{1-n})$.
Using \eqref{eq:weight-orientation-f5}, we obtain
\begin{equation}
 \operatorname{Tr}\bigl(L_h^*L_hg(T_h)\bigr)
 =\operatorname{Tr}\bigl(c_hc_h^*g(T_{0,h})\bigr)
  +O_{g,\varepsilon,L}(h^{1-n}).
 \label{eq:nonlinear-transverse-trace-comparison}
\end{equation}
To summarize the roles of the localized estimates,
\eqref{eq:direct-boundary-leakage} controls the spatial projection,
\eqref{eq:spectral-quasi-intertwining-bound} controls the spectral
projection, and their compression identity gives the smooth
functional-calculus comparison. Equation~\eqref{eq:weight-orientation-f5}
then converts that comparison to
\eqref{eq:nonlinear-transverse-trace-comparison} with the prescribed
external weight. Each step uses the same fixed inner support and
the flow tube specified above.

The fixed nested cutoffs and the classical composition calculus give
\[
 c_hc_h^*=\Op_h^w(b_h)+R_h,\qquad
 \|R_h\|_{\mathfrak S_1}=O(h^\infty),
\]
where $b_h\in C_c^\infty(T^*\mathbb R^n)$ is a classical symbol
with support in a fixed compact set and uniformly bounded derivatives.
The residual satisfies
\[
 \bigl|\operatorname{Tr}(R_hg(T_{0,h}))\bigr|
 \leq\|R_h\|_{\mathfrak S_1}\|g\|_{L^\infty([0,1])}
 =O_g(h^\infty).
\]
The principal symbol $b_0$ satisfies, by
\eqref{eq:boundary-restricted-map},
\[
 b_0(y,0,\eta,0)
 =w_0\bigl(y,0,\eta,a(y,\eta)\bigr).
\]
Corollary~\ref{cor:hardy-tangential-weight} applies to
$\Op_h^w(b_h)$. For the normal compression, write
\[
 \mathsf B_\sigma
 =Q_0\mathbf1_{(-\infty,0]}(\sigma h\mathsf D_u)Q_0
 \big|_{L^2(\mathbb R_+)},\qquad
 \mathsf B_{-\sigma}=I-\mathsf B_\sigma.
\]
With $g^\vee(t)=g(1-t)$,
\[
 g(\mathsf B_{-\sigma})=g^\vee(\mathsf B_\sigma),\qquad
 W(g^\vee)=\int_0^1\frac{g(1-t)}{t(1-t)}\,dt=W(g).
\]
Since $g(0)=g(1)=0$, the functional-calculus expressions extend by zero
to the complementary half-space. Both signs have the same coefficient.
The resulting formula, together with
\eqref{eq:nonlinear-transverse-trace-comparison}, is
\eqref{eq:transverse-local-trace}.
\end{proof}

\begin{remark}[Boundary compatibility of the normal form]
The canonical transformation preserves the two sides of the
spatial boundary, and the Hamiltonian is reduced to
$\sigma h\mathsf D_u$ modulo a residual operator.  At the operator level,
boundary compatibility is quantified by
\eqref{eq:direct-boundary-leakage} for each localized FIO.  The
$O(h^{1-n})$ error in that estimate is sufficient for
\eqref{eq:nonlinear-transverse-trace-comparison} and is negligible
relative to $h^{1-n}\log h^{-1}$.

The Hardy model in this argument is $I_y\otimes\mathsf B_\sigma$
in the boundary-compatible canonical coordinates. Its weighted trace
is evaluated by Corollary~\ref{cor:hardy-tangential-weight}, which in turn
uses Theorem~\ref{thm:hardy-general-weight}. For a variable potential,
the passage from the original compression to this model is justified by
the spatial projection estimate \eqref{eq:direct-boundary-leakage},
the spectral projection estimate
\eqref{eq:spectral-quasi-intertwining-bound}, and the resulting
nonlinear trace comparison
\eqref{eq:nonlinear-transverse-trace-comparison}.
The direct-integral fiberization in
Corollary~\ref{cor:frozen-finite-band-branches} applies to the
translation-invariant frozen model and uses
Theorem~\ref{thm:finite-band-general-weight}.
\end{remark}

\section{Global trace asymptotics}
\label{sec:gluing}
\label{sec:positive-global-completion}

The transverse trace formula and a positive partition of phase space
give the global asymptotic for smooth functions. The variance estimate
controls the complementary weight, and the fractional moment estimate
extends the formula to the endpoint class in Theorem~\ref{thm:main-target}.

\subsection{Boundary coefficients and preliminary estimates}

Assume the hypotheses of Theorem~\ref{thm:main-target}, with the notation
of Section~\ref{sec:introduction}. In dimension one, set
\[
 \Omega_{\mathrm{reg}}=\operatorname{int}\overline\Omega.
\]
The symmetric difference of these two sets is finite, and hence
\begin{equation}
 \mathbf1_\Omega=\mathbf1_{\Omega_{\mathrm{reg}}}
 \quad\hbox{as multiplication operators on }L^2(\mathbb R).
 \label{eq:glue-interval-regularization}
\end{equation}
There is a representation
\[
 \Omega_{\mathrm{reg}}=\bigcup_{j=1}^{m}(a_j,b_j),\qquad
 a_j<b_j<a_{j+1}\quad(1\le j<m),
\]
whose interval closures are pairwise disjoint. Thus
$\partial_*\Omega=\{a_1,b_1,\ldots,a_m,b_m\}$ in dimension one.

Relative to $L^2=\operatorname{Ran}Q\oplus\ker Q$,
\[
 T_h=(T_h|_{\operatorname{Ran}Q})\oplus0,\qquad
 g(T_h)=g(T_h|_{\operatorname{Ran}Q})\oplus0\quad\text{if }g(0)=0.
\]
Define
\[
 R(x)=\sqrt{\mu-V(x)}\quad(x\in\partial_*\Omega).
\]
Compactness and $\Omega\Subset D$ give constants
\begin{equation}
 0<R_0\le R(x)\le R_1<\infty.
 \label{eq:glue-positive-fermi-radius}
\end{equation}
Let $\nu(x)$ denote a choice of unit normal to $\partial_*\Omega$.
Using the Euclidean metric to identify tangential vectors and covectors,
write
\[
 p(x,\eta)=\sqrt{R(x)^2-|\eta|^2},\qquad
 \xi_\sigma(x,\eta)=\eta+\sigma p(x,\eta)\nu(x),
 \quad \sigma\in\{+1,-1\},\quad |\eta|<R(x).
\]
For a bounded measurable weight $w$ on the boundary Fermi set, put
\begin{equation}
 \mathcal A(w)
 =\frac1{4\pi^2}\int_{\partial_*\Omega}
   \int_{|\eta|<R(x)}
   \sum_{\sigma=\pm1}w\bigl(x,\xi_\sigma(x,\eta)\bigr)
   \,d\eta\,d\sigma(x).
 \label{eq:glue-boundary-functional}
\end{equation}
Here $d\eta$ is Euclidean volume on $T_x^*\partial_*\Omega$.
This measure agrees with that in the local trace theorem.
Indeed, if $y$ are boundary coordinates, $G(y)$ is the induced metric,
and $\eta_{\mathrm{can}}$ are the dual coordinate covectors, then
\[
 d\sigma(x)=\sqrt{\det G(y)}\,dy,\qquad
 d\eta_{\mathrm{can}}
   =\sqrt{\det G(y)}\,d\eta_{\mathrm{orth}},\qquad
 dy\,d\eta_{\mathrm{can}}
   =d\sigma(x)\,d\eta_{\mathrm{orth}}.
\]
Thus the canonical density $dy\,d\eta$ in
Theorem~\ref{thm:transverse-local-trace} becomes the density in
\eqref{eq:glue-boundary-functional}, with the half-density factors
included in the unitary coordinate transformation.
For $n=1$, the zero-dimensional tangential fiber has unit mass.  The two momentum branches contribute
\[
 \int_{|\eta|<R(x)}\sum_{\sigma=\pm1}1\,d\eta
 =2c_{n-1} R(x)^{n-1}.
\]
Consequently,
\begin{equation}
 \mathcal A(1)=C_{\Omega,1}.
 \label{eq:glue-geometric-constant}
\end{equation}

By Lemma~\ref{toolbox:transfer-moments},
\[
\|E_h-E_h^{\mathrm c}\|_{\mathfrak S_1}
=O(h^{1-n}).
\]
Since $V_{\mathrm c}=V$ near $\overline\Omega$, the boundary
coefficients for the two potentials coincide. Applying
\cite[Theorem~1]{DeleporteLambert} to $V_{\mathrm c}$ and using
\eqref{eq:glue-geometric-constant}, we obtain
\begin{equation}
	E_h\in\mathfrak S_1,\qquad
	\operatorname{Tr}E_h
	=M_h\mathcal A(1)+O(h^{1-n}).
	\label{eq:glue-variance-input}
\end{equation}
For $n=1$, we apply that theorem to $\Omega_{\mathrm{reg}}$,
using \eqref{eq:glue-interval-regularization}.

Fix $\varepsilon>0$ and a transverse chart on which
$|\xi\cdot\nu(x)|\ge\varepsilon$ along the boundary Fermi set.
Let $a\in C_c^\infty(T^*\mathbb R^n)$ be supported in a
sufficiently small neighborhood within this chart, and set
$Z_h=\operatorname{Op}_h^w(a)$, with a fixed proper-support
cutoff. For every $g\in C^\infty([0,1])$ satisfying
$g(0)=g(1)=0$, Lemma~\ref{toolbox:transfer-moments} gives
\[
\begin{aligned}
	&\left|
		\operatorname{Tr}\bigl(Z_hg(T_h)Z_h^*\bigr)
		-\operatorname{Tr}\bigl(Z_hg(T_h^{\mathrm c})Z_h^*\bigr)
	\right|\\
	&\qquad\le
		\|Z_h\|^2
	\|g(T_h)-g(T_h^{\mathrm c})\|_{\mathfrak S_1}
	=O_{a,g}(h^{1-n})
	=o(M_h).
\end{aligned}
\]
Hence Theorem~\ref{thm:transverse-local-trace} yields
\begin{equation}
		\operatorname{Tr}\bigl(Z_hg(T_h)Z_h^*\bigr)
	=M_hW(g)\mathcal A(|a|^2)+o_{a,g}(M_h).
	\label{eq:glue-local-input}
\end{equation}
The chart and the constants may depend on the fixed
parameter $\varepsilon$.

Finally, Theorem~\ref{S6:main} and
Lemma~\ref{toolbox:transfer-moments} imply that, for every
fixed $q\in(0,1)$,
\begin{equation}
	E_h^q\in\mathfrak S_1,\qquad
	\operatorname{Tr}E_h^q\le C_qM_h
	\qquad(0<h<h_q).
	\label{eq:glue-fractional-input}
\end{equation}
Indeed, with $X_h=(I-Q)P_hQ$,
\begin{equation}
	E_h=X_h^*X_h,\qquad
	[P_h,Q]=
	\begin{pmatrix}
		0&-X_h^*\\
		X_h&0
	\end{pmatrix},
	\qquad
	\|[P_h,Q]\|_{\mathfrak S_{2q}}^{2q}
	=2\operatorname{Tr}E_h^q,
	\label{eq:glue-commutator-identity}
\end{equation}
where the matrix representation is relative to
$L^2(\mathbb R^n)=\operatorname{Ran}Q\oplus\ker Q$.

\subsection{Positive localization weights}

\begin{lemma}[Positive transverse partition]
\label{lem:glue-positive-partition}
Suppose $n\ge2$ and fix $0<\varepsilon<R_0/4$.
There exist finitely many real smooth compactly supported symbols
$a_{\nu,\varepsilon}$, each supported in a chart admissible for
\eqref{eq:glue-local-input}, such that
\begin{equation}
 w_\varepsilon=\sum_\nu|a_{\nu,\varepsilon}|^2\le1
 \label{eq:glue-symbol-partition}
\end{equation}
everywhere, and $w_\varepsilon=1$ on the boundary Fermi set wherever
$|\xi\cdot\nu(x)|\ge2\varepsilon$.  There is a constant
$K_\varepsilon\ge0$ such that, for all sufficiently small $h$,
\begin{equation}
 \begin{split}
 A_{\nu,h}&=\operatorname{Op}_h^w(a_{\nu,\varepsilon}),
 \qquad \kappa_{\varepsilon,h}=(1+K_\varepsilon h)^{-1},\\
 W_{\varepsilon,h}
 &=\kappa_{\varepsilon,h}\sum_\nu A_{\nu,h}^*A_{\nu,h},
 \qquad R_{\varepsilon,h}=I-W_{\varepsilon,h},
 \end{split}
 \label{eq:glue-exact-weights}
\end{equation}
satisfy $0\le W_{\varepsilon,h}\le I$ and
$0\le R_{\varepsilon,h}\le I$ exactly.
Moreover,
\begin{equation}
 0\le\mathcal A(1)-\mathcal A(w_\varepsilon)
 \le C\varepsilon^2,
 \label{eq:glue-missing-volume}
\end{equation}
with $C$ independent of $\varepsilon$.
\end{lemma}

\begin{proof}
The part of the compact boundary Fermi set satisfying
$|\xi\cdot\nu(x)|\ge2\varepsilon$ is covered by finitely many relatively
compact transverse charts with
$|\xi\cdot\nu(x)|>\varepsilon$.
Choose $\chi_\varepsilon\in C_c^\infty$, with
$0\le\chi_\varepsilon\le1$, supported in their union and equal to one
on that compact set.  A smooth square partition subordinate to those
charts gives symbols $a_{\nu,\varepsilon}$ with
$\sum_\nu|a_{\nu,\varepsilon}|^2=\chi_\varepsilon^2$.
For example, multiply $\chi_\varepsilon$ by
$\theta_\nu/(\sum_\rho\theta_\rho^2)^{1/2}$ on a neighborhood of its
support, for a finite subordinate family with positive denominator there.
This proves \eqref{eq:glue-symbol-partition}.

For this fixed $\varepsilon$, semiclassical composition gives
\[
 \sum_\nu A_{\nu,h}^*A_{\nu,h}
 =\operatorname{Op}_h^w(w_\varepsilon)+O_{L^2\to L^2,\varepsilon}(h).
\]
Since $1-w_\varepsilon\ge0$, sharp G\aa rding
\cite[Theorem~4.32]{Zworski} gives
\[
 \sum_\nu A_{\nu,h}^*A_{\nu,h}
 \le(1+K_\varepsilon h)I
\]
after increasing $K_\varepsilon$ if necessary. Hence the operators in
\eqref{eq:glue-exact-weights} satisfy $0\le W_{\varepsilon,h},
R_{\varepsilon,h}\le I$.

On either boundary Fermi branch, $1-w_\varepsilon$ can be nonzero only
when $p(x,\eta)<2\varepsilon$.  For $n\ge2$, the volume of this set in
the tangential fiber is
\[
 c_{n-1}\left[R(x)^{n-1}-\bigl(R(x)^2-4\varepsilon^2\bigr)^{(n-1)/2}\right]
 \le C_{n,R_0,R_1}\varepsilon^2.
\]
The mean value theorem and \eqref{eq:glue-positive-fermi-radius}
give this estimate for every $n\ge2$.
Summing the two branches and integrating over the compact spatial
boundary proves \eqref{eq:glue-missing-volume}.
\end{proof}

\subsection{The positive remainder and the smooth trace formula}

\begin{lemma}[Positive weighted domination]
\label{lem:glue-positive-domination}
Let $0\le T\le I$ be self-adjoint, put $E=T(1-T)$, and assume
$E\in\mathfrak S_1$.  Suppose that $g:[0,1]\to\mathbb C$ is bounded and Borel measurable,
$g(0)=g(1)=0$, and
\[
 |g(t)|\le C_g t(1-t)\qquad(0\le t\le1).
\]
Then $g(T)\in\mathfrak S_1$.  For every bounded $R\ge0$,
\begin{equation}
 \bigl|\operatorname{Tr}(Rg(T))\bigr|
 \le\|R^{1/2}g(T)R^{1/2}\|_{\mathfrak S_1}
 \le C_g\operatorname{Tr}(RE).
 \label{eq:glue-positive-domination}
\end{equation}
\end{lemma}

\begin{proof}
Define $b(t)=g(t)/(t(1-t))$ for $0<t<1$, and set $b(0)=b(1)=0$.
Then $\|b\|_\infty\le C_g$ and
$g(T)=E^{1/2}b(T)E^{1/2}$, so $g(T)$ is trace class.
The operator $F=R^{1/2}E^{1/2}$ is Hilbert--Schmidt and
\[
 R^{1/2}g(T)R^{1/2}=Fb(T)F^*,\qquad
 \|F\|_{\mathfrak S_2}^2=\operatorname{Tr}(RE).
\]
The trace norm ideal inequality and cyclicity for trace-class products
give \eqref{eq:glue-positive-domination}.
\end{proof}

\begin{proposition}[Smooth trace asymptotics]
\label{prop:glue-smooth-completion}
Assume $n\ge2$ and the geometric and operator hypotheses above.
For every $g\in C^\infty([0,1])$ with $g(0)=g(1)=0$,
\begin{equation}
 \lim_{h\downarrow0}\frac{\operatorname{Tr}g(T_h)}{M_h}
 =\mathcal A(1)W(g).
 \label{eq:glue-smooth-limit}
\end{equation}
\end{proposition}

\begin{proof}
Set
\[
 C_g=\sup_{0<t<1}\frac{|g(t)|}{t(1-t)}<\infty.
\]
Lemma~\ref{lem:glue-positive-domination} makes all traces below well defined.
Fix $\varepsilon\in(0,R_0/4)$ and use
Lemma~\ref{lem:glue-positive-partition}.  Trace cyclicity gives the exact
identity
\begin{equation*}
 \operatorname{Tr}\bigl(W_{\varepsilon,h}g(T_h)\bigr)
 =\kappa_{\varepsilon,h}
  \sum_\nu\operatorname{Tr}
  \bigl(A_{\nu,h}g(T_h)A_{\nu,h}^*\bigr).
\end{equation*}
The sum is finite for fixed $\varepsilon$ and
$\kappa_{\varepsilon,h}\to1$.  Hence \eqref{eq:glue-local-input} yields
\begin{equation}
 \frac{\operatorname{Tr}(W_{\varepsilon,h}g(T_h))}{M_h}
 \longrightarrow W(g)\mathcal A(w_\varepsilon).
 \label{eq:glue-weighted-limit}
\end{equation}

Apply the same formula to $g_2(t)=t(1-t)$, for which $W(g_2)=1$.
Subtracting it from \eqref{eq:glue-variance-input} gives
\begin{equation}
 \frac{\operatorname{Tr}(R_{\varepsilon,h}E_h)}{M_h}
 \longrightarrow
 D_\varepsilon:=\mathcal A(1)-\mathcal A(w_\varepsilon),
 \qquad 0\le D_\varepsilon\le C\varepsilon^2.
 \label{eq:glue-variance-deficit}
\end{equation}
Because $I=W_{\varepsilon,h}+R_{\varepsilon,h}$ exactly,
\begin{equation*}
 \operatorname{Tr}g(T_h)
 =\operatorname{Tr}(W_{\varepsilon,h}g(T_h))
  +\operatorname{Tr}(R_{\varepsilon,h}g(T_h)).
\end{equation*}
Lemma~\ref{lem:glue-positive-domination} and
\eqref{eq:glue-variance-deficit} give
\[
 \limsup_{h\downarrow0}
 \frac{|\operatorname{Tr}(R_{\varepsilon,h}g(T_h))|}{M_h}
 \le C_gD_\varepsilon.
\]
Together with \eqref{eq:glue-weighted-limit}, this yields
\[
 \limsup_{h\downarrow0}
 \left|\frac{\operatorname{Tr}g(T_h)}{M_h}
       -\mathcal A(1)W(g)\right|
 \le \bigl(C_g+|W(g)|\bigr)D_\varepsilon
 \le C\bigl(C_g+|W(g)|\bigr)\varepsilon^2.
\]
The limit $h\downarrow0$ is taken with $g$ and $\varepsilon$ fixed.
Letting $\varepsilon\downarrow0$ then proves
\eqref{eq:glue-smooth-limit}.
\end{proof}

\subsection{Dimension one}

When $n=1$, the boundary Fermi set is the finite set
\[
 \mathcal B=\{(x,+R(x)),(x,-R(x)):x\in\partial_*\Omega\}.
\]
Every point is transverse by
\eqref{eq:glue-positive-fermi-radius}.  Choose a fixed finite square
partition in transverse charts whose sum $w$ is one at every point of
$\mathcal B$ and is at most one everywhere. The sharp G\aa rding
normalization in the proof of Lemma~\ref{lem:glue-positive-partition},
which is valid in every dimension, produces $0\le W_h\le I$ and
$R_h=I-W_h\ge0$.
In this case
\[
 \mathcal A(w)=\mathcal A(1),\qquad
 \operatorname{Tr}(R_hE_h)=o(\log h^{-1})
\]
by \eqref{eq:glue-variance-input} and the local formula for $g_2$.
Lemma~\ref{lem:glue-positive-domination} gives
\[
 |\operatorname{Tr}(R_hg(T_h))|\le C_g\operatorname{Tr}(R_hE_h)
 =o(\log h^{-1}),
\]
which proves \eqref{eq:glue-smooth-limit}.

There are two normal momentum branches at each spatial endpoint, with
coefficient $W(g)/(4\pi^2)$ per branch.
Since $c_0=1$, the counting measure on the essential boundary gives
\begin{equation}
 C_{\Omega,1}=\frac{\#\partial_*\Omega}{2\pi^2}
             =\frac{m}{\pi^2}.
 \label{eq:glue-one-dimensional-coefficient}
\end{equation}
For one interval, $C_{\Omega,1}=1/\pi^2$.

\subsection{Removal of the spectral endpoint cutoff}

\begin{lemma}[Weighted smooth approximation]
\label{lem:glue-weighted-approximation}
Suppose $g\in C([0,1];\mathbb C)$, $g(0)=g(1)=0$, and for some
$\alpha_0,\alpha_1>0$, $C>0$, and $\delta_0\in(0,1/2)$,
\begin{equation}
 |g(t)|\le Ct^{\alpha_0}\quad(0<t<\delta_0),\qquad
 |g(t)|\le C(1-t)^{\alpha_1}\quad(1-\delta_0<t<1).
 \label{eq:glue-endpoint-assumption}
\end{equation}
Choose
\begin{equation}
 0<q<\min\{1,\alpha_0,\alpha_1\}.
 \label{eq:glue-strict-q}
\end{equation}
There are functions $g_\delta\in C_c^\infty((0,1);\mathbb C)$ and
numbers $\epsilon_\delta\downarrow0$ such that
\begin{equation}
 |g(t)-g_\delta(t)|
 \le\epsilon_\delta[t(1-t)]^q\qquad(0\le t\le1).
 \label{eq:glue-weighted-approximation}
\end{equation}
Consequently,
\begin{equation}
 |W(g)-W(g_\delta)|\le\epsilon_\delta B(q,q),
 \qquad B(q,q)=\int_0^1[t(1-t)]^{q-1}\,dt<\infty.
 \label{eq:glue-coefficient-approximation}
\end{equation}
\end{lemma}

\begin{proof}
The function
\[
 f(t)=\frac{g(t)}{[t(1-t)]^q}\quad(0<t<1),\qquad f(0)=f(1)=0,
\]
is continuous on $[0,1]$.  Its vanishing at both endpoints follows from
the strict inequalities in \eqref{eq:glue-strict-q}.
Approximate $f$ uniformly by $f_\delta\in C_c^\infty((0,1);\mathbb C)$, using a
cutoff followed by ordinary smooth approximation on the interior, and
set
\[
 g_\delta(t)=[t(1-t)]^q f_\delta(t).
\]
Since $f_\delta$ is supported in a compact subset of $(0,1)$,
$g_\delta\in C_c^\infty((0,1);\mathbb C)$.
After passing to a subsequence, put
$\epsilon_\delta=\|f-f_\delta\|_\infty\downarrow0$.
This gives \eqref{eq:glue-weighted-approximation}; division by
$t(1-t)$ and integration give \eqref{eq:glue-coefficient-approximation}.
\end{proof}

\begin{theorem}[Global Widom formula]
\label{thm:glue-final-theorem}
Under the geometric and operator hypotheses above, let
$g\in C([0,1];\mathbb C)$ satisfy $g(0)=g(1)=0$ and
\eqref{eq:glue-endpoint-assumption} for positive endpoint exponents.
Then $g(T_h)$ is trace class for all sufficiently small $h$, $W(g)$ is
absolutely convergent, and
\begin{equation}
 \begin{aligned}
 \lim_{h\downarrow0}
 \frac{\operatorname{Tr}g(T_h)}{(2\pi h)^{1-n}\log h^{-1}}
 &=C_{\Omega,1}W(g)\\
 &=\frac{c_{n-1}}{2\pi^2}
   \left(\int_{\partial_*\Omega}
       (\mu-V(x))^{(n-1)/2}\,d\sigma(x)\right)
   \left(\int_0^1\frac{g(t)}{t(1-t)}\,dt\right).
 \end{aligned}
 \label{eq:glue-final-widom-formula}
\end{equation}
For $n=1$, surface measure counts the essential endpoints
$\partial_*\Omega=\partial\Omega_{\mathrm{reg}}$, and the coefficient is
given by \eqref{eq:glue-one-dimensional-coefficient}.
\end{theorem}

\begin{proof}
Choose $q$ as in \eqref{eq:glue-strict-q}.  The continuous extension
of $g(t)/[t(1-t)]^q$ is bounded.  Functional calculus and
\eqref{eq:glue-fractional-input} therefore give $g(T_h)\in\mathfrak S_1$.
The same endpoint inequalities imply absolute convergence of $W(g)$.

Lemma~\ref{lem:glue-weighted-approximation} and the spectral theorem give
\begin{equation}
 \|g(T_h)-g_\delta(T_h)\|_{\mathfrak S_1}
 \le\epsilon_\delta\operatorname{Tr}E_h^q
 \le C_q\epsilon_\delta M_h.
 \label{eq:glue-endpoint-trace-error}
\end{equation}
For every fixed $\delta$, Proposition~\ref{prop:glue-smooth-completion}
and the one-dimensional argument above give
\[
 \lim_{h\downarrow0}
 \frac{\operatorname{Tr}g_\delta(T_h)}{M_h}
 =\mathcal A(1)W(g_\delta).
\]
Combining this with \eqref{eq:glue-endpoint-trace-error} and
\eqref{eq:glue-coefficient-approximation} yields
\[
 \limsup_{h\downarrow0}
 \left|\frac{\operatorname{Tr}g(T_h)}{M_h}
       -\mathcal A(1)W(g)\right|
 \le\epsilon_\delta
      \bigl(C_q+\mathcal A(1)B(q,q)\bigr).
\]
Letting $\delta\downarrow0$ proves
\eqref{eq:glue-final-widom-formula}.

For $n\ge2$ the limits are taken in the order
$h\downarrow0$, $\varepsilon\downarrow0$, $\delta\downarrow0$;
for $n=1$ they are $h\downarrow0$, $\delta\downarrow0$.
The exponent $q$ in \eqref{eq:glue-strict-q} remains fixed.
This proves Theorem~\ref{thm:main-target}.
\end{proof}

\section{Entanglement entropy and particle-number statistics}
\label{sec:applications}

We prove the entropy and entanglement-spectrum corollaries stated in
the introduction, derive the commutator and counting-statistics
asymptotics, and compute the coefficients for intervals and a disk. All parameters remain fixed
as $h\downarrow0$.

\subsection{Spectral integrals and proofs of the corollaries}

\begin{proof}[Proof of Corollary~\ref{cor:entropy-intro}]
Write $s=s_1$.
The function $s$ satisfies the endpoint power bounds with every exponent
in $(0,1)$.  Expanding the logarithm and integrating the resulting
absolutely convergent series gives
\[
 W(s)=-\int_0^1\frac{\log t}{1-t}\,dt
       -\int_0^1\frac{\log(1-t)}t\,dt
      =2\sum_{k=1}^\infty\frac1{k^2}
      =\frac{\pi^2}{3}.
\]
For $\alpha>0$, $\alpha\ne1$,
\[
 |s_\alpha(t)|\le C_\alpha[t(1-t)]^{\min\{1,\alpha\}},
 \qquad 0\le t\le1.
\]
Put $t=(1+e^x)^{-1}$ and
\[
 L(x)=\log(1+e^x)-x_+,\qquad x_+=\max\{x,0\}.
\]
Then
\[
 \int_{\mathbb R}L(x)\,dx
 =2\int_0^\infty\log(1+e^{-x})\,dx
 =2\sum_{k=1}^\infty\frac{(-1)^{k+1}}{k^2}
 =\frac{\pi^2}{6}.
\]
The linear terms cancel under this substitution, and hence
\[
 \begin{aligned}
 W(s_\alpha)
 &=\frac1{1-\alpha}
   \int_{\mathbb R}\bigl(L(\alpha x)-\alpha L(x)\bigr)\,dx\\
 &=\frac{\pi^2}{6}\frac{\alpha^{-1}-\alpha}{1-\alpha}
  =\frac{\pi^2}{6}(1+\alpha^{-1}).
 \end{aligned}
\]
Theorem~\ref{thm:main-target} gives \eqref{eq:entropy-asymptotic-intro}.

Integration of the first two determinantal correlation functions gives
\[
 \begin{aligned}
 \operatorname{Var}N_\Omega
 &=\int_\Omega P_h(x,x)\,dx
   -\int_{\Omega\times\Omega}|P_h(x,y)|^2\,dx\,dy\\
 &=\operatorname{Tr}(T_h-T_h^2)=\operatorname{Tr}E_h.
 \end{aligned}
\]
By \eqref{eq:glue-variance-input},
\[
 \operatorname{Var}N_\Omega=C_{\Omega,1}M_h+o(M_h).
\]
Since $\Omega$ is nonempty and $\overline\Omega\subset D$,
the geometric coefficient $C_{\Omega,1}$ is strictly positive.
Division in \eqref{eq:entropy-asymptotic-intro} proves
\eqref{eq:entropy-variance-intro}, including $\alpha=1$.
\end{proof}

\begin{proof}[Proof of Corollary~\ref{cor:entanglement-spectrum}]
Define
\[
 g_\varphi(t)=\varphi\!\left(\log\frac{1-t}{t}\right)
 \quad(0<t<1),\qquad g_\varphi(0)=g_\varphi(1)=0.
\]
Since $\varphi$ has compact support, $g_\varphi$ is continuous and
vanishes in neighborhoods of both endpoints. It therefore satisfies
the assumptions of Theorem~\ref{thm:main-target}. The substitution
$u=\log((1-t)/t)$, with $du=-dt/[t(1-t)]$, gives
\[
 W(g_\varphi)=\int_{\mathbb R}\varphi(u)\,du.
\]
Since $g_\varphi(0)=g_\varphi(1)=0$,
\[
 \operatorname{Tr}g_\varphi(T_h)=\sum_j\varphi(\varepsilon_{h,j}).
\]
For $0<\delta<(b-a)/2$, choose $\varphi_\delta^\pm\in C_c(\mathbb R)$
such that
\[
 0\le\varphi_\delta^-\le\mathbf1_{[a,b]}\le\varphi_\delta^+,
 \qquad
 b-a-2\delta\le\int\varphi_\delta^-
 \le\int\varphi_\delta^+\le b-a+2\delta.
\]
Writing $\mathcal N_h[a,b]=\#\{j:a\le\varepsilon_{h,j}\le b\}$,
\eqref{eq:entanglement-spectrum-vague} gives
\[
 \begin{aligned}
 C_{\Omega,1}(b-a-2\delta)
 &\le\liminf_{h\downarrow0}\frac{\mathcal N_h[a,b]}{M_h}\\
 &\le\limsup_{h\downarrow0}\frac{\mathcal N_h[a,b]}{M_h}
 \le C_{\Omega,1}(b-a+2\delta).
 \end{aligned}
\]
Letting $\delta\downarrow0$ proves \eqref{eq:entanglement-window-count}.
\end{proof}

\subsection{Spatial coupling and counting statistics}

The commutator measures the coupling induced by $P_h$ between
$L^2(\Omega)$ and $L^2(\Omega^c)$. The following formula determines its
singular-value moments for every fixed positive exponent.

\begin{corollary}[Schatten asymptotics for the spatial commutator]
\label{cor:commutator-asymptotics}
Under the hypotheses of Theorem~\ref{thm:main-target}, for every fixed $p>0$,
\begin{equation*}
 \|[P_h,Q]\|_{\mathfrak S_p}^{p}
 =2C_{\Omega,1}B(p/2,p/2)M_h+o(M_h),
\end{equation*}
where $B(a,b)=\int_0^1t^{a-1}(1-t)^{b-1}\,dt$ is the beta function.
If $\Omega$ is nonempty, then
\[
 \frac{\|[P_h,Q]\|_{\mathfrak S_1}}{\operatorname{Var}N_\Omega}
 \longrightarrow 2\pi.
\]
For $p=2$ one has the exact identity
$\|[P_h,Q]\|_{\mathfrak S_2}^{2}=2\operatorname{Var}N_\Omega$.
\end{corollary}

\begin{proof}
Relative to $QL^2\oplus(I-Q)L^2$, the operator $X_h$ in
\eqref{eq:glue-commutator-identity} gives
\[
 [P_h,Q]=\begin{pmatrix}0&-X_h^*\\ X_h&0\end{pmatrix},
 \qquad X_h^*X_h=E_h\big|_{QL^2}.
\]
Since
\[
 [P_h,Q]^*[P_h,Q]
 =\begin{pmatrix}X_h^*X_h&0\\0&X_hX_h^*\end{pmatrix},
\]
and $X_h^*X_h$, $X_hX_h^*$ have the same positive eigenvalues,
for every $p>0$,
\[
 \|[P_h,Q]\|_{\mathfrak S_p}^{p}=2\operatorname{Tr}E_h^{p/2}.
\]
Apply Theorem~\ref{thm:main-target} to
$g_p(t)=[t(1-t)]^{p/2}$, for which $W(g_p)=B(p/2,p/2)$.
The special cases follow from $B(1/2,1/2)=\pi$ and
$\operatorname{Tr}E_h=\operatorname{Var}N_\Omega$.
\end{proof}

This determines the leading coefficient in Theorem~\ref{S6:main}.

\begin{corollary}[Centered logarithmic moment-generating function]
\label{cor:counting-statistics}
Under the hypotheses of Theorem~\ref{thm:main-target}, for every fixed $t\in\mathbb R$,
\begin{equation}
 \log\mathbb E e^{t(N_\Omega-\mathbb EN_\Omega)}
 =\frac{t^2}{2}C_{\Omega,1} M_h+o(M_h).
 \label{eq:counting-statistics-asymptotic}
\end{equation}
\end{corollary}

\begin{proof}
Since $P_h$ has finite rank, $T_h$ is a finite-rank positive
contraction. Let $\lambda_1,\ldots,\lambda_{r_h}$ be its nonzero
eigenvalues, counted with multiplicity, where
$r_h=\operatorname{rank}T_h$.
By \cite[Theorem~7]{HoughEtAl}, applied to the determinantal process
restricted to $\Omega$, $N_\Omega$ has the distribution of a sum of
independent Bernoulli variables with parameters $\lambda_j$.
Taking their moment-generating functions gives
\[
 \begin{aligned}
 \mathbb E e^{tN_\Omega}
 &=\prod_{j=1}^{r_h}\bigl(1+(e^t-1)\lambda_j\bigr)
   =\det\bigl(I+(e^t-1)T_h\bigr),\\
 \mathbb EN_\Omega&=\sum_{j=1}^{r_h}\lambda_j=\operatorname{Tr}T_h.
 \end{aligned}
\]
Set
\[
 g_t(\lambda)=\log\bigl(1+(e^t-1)\lambda\bigr)-t\lambda.
\]
For each real $t$, the logarithm has positive argument on $[0,1]$,
and $g_t$ is smooth with $g_t(0)=g_t(1)=0$.  The centered logarithm
is therefore $\operatorname{Tr}g_t(T_h)$.
Differentiation of the convergent scalar integral gives
\[
 \frac{d}{dt}W(g_t)
 =(e^t-1)\int_0^1\frac{d\lambda}{1+(e^t-1)\lambda}
 =t,
\]
where the identity at $t=0$ follows by continuity.  Since $W(g_0)=0$,
we obtain $W(g_t)=t^2/2$.  Applying
Theorem~\ref{thm:main-target} proves
\eqref{eq:counting-statistics-asymptotic}.
\end{proof}

The leading logarithmic generating function is quadratic for each fixed
real parameter, describing exponentially weighted particle-number
fluctuations at scale $M_h$.

\begin{remark}[Fixed higher cumulants]
\label{rem:higher-cumulants}
For each fixed integer $m\ge3$, the $m$-th cumulant of the particle
count satisfies
\begin{equation}\label{eq:higher-cumulants}
 \kappa_m(N_\Omega)=o(M_h).
\end{equation}
Indeed, the Bernoulli cumulant polynomials obey
\[
 p_1(t)=t,\qquad p_{m+1}(t)=t(1-t)p_m'(t).
\]
Differentiating the exact finite-rank determinant identity, or using
the representation of the count as a sum of independent Bernoulli
variables, gives $\kappa_m(N_\Omega)=\operatorname{Tr}p_m(T_h)$.
For $m\ge2$, $p_m(0)=p_m(1)=0$, and for $m\ge3$,
\[
 W(p_m)=\int_0^1p_{m-1}'(t)\,dt
       =p_{m-1}(1)-p_{m-1}(0)=0.
\]
Theorem~\ref{thm:main-target}, applied to this fixed polynomial, proves
\eqref{eq:higher-cumulants}.
\end{remark}

\subsection{Explicit geometries}

\begin{example}[One-dimensional domains]
\label{cor:one-dimensional-entropies}
Suppose $n=1$, and let
$m=\tfrac12\#\partial_*\Omega$ be the number of component intervals
of $\operatorname{int}\overline\Omega$.  Then
\begin{align*}
 \operatorname{Tr}\bigl(T_h(1-T_h)\bigr)
 &=\frac{m}{\pi^2}\log h^{-1}+o(\log h^{-1}),
 \\
 \operatorname{Tr}s(T_h)
 &=\frac{m}{3}\log h^{-1}+o(\log h^{-1}),
 \\
 \operatorname{Tr}s_\alpha(T_h)
 &=\frac{m}{6}(1+\alpha^{-1})\log h^{-1}
   +o(\log h^{-1})
\end{align*}
for every fixed $\alpha>0$, $\alpha\ne1$.
\end{example}

\begin{proof}
Substitute $C_{\Omega,1}=m/\pi^2$ from
\eqref{eq:glue-one-dimensional-coefficient} and the spectral integrals
$W(t(1-t))=1$, $W(s)=\pi^2/3$, and
$W(s_\alpha)=\pi^2(1+\alpha^{-1})/6$ into the trace formula.
\end{proof}

\begin{example}[A disk in a harmonic potential]
\label{cor:harmonic-disk}
Let $n=2$, $V(x)=|x|^2$, and $\Omega=\{x:|x|<R\}$, where
$R>0$ and $\mu>R^2$ are fixed.  Then
\begin{align*}
 \operatorname{Tr}\bigl(T_h(1-T_h)\bigr)
 &=\frac{R\sqrt{\mu-R^2}}{\pi^2h}\log h^{-1}
   +o(h^{-1}\log h^{-1}),
 \\
 \operatorname{Tr}s(T_h)
 &=\frac{R\sqrt{\mu-R^2}}{3h}\log h^{-1}
   +o(h^{-1}\log h^{-1}).
\end{align*}
\end{example}

\begin{proof}
Here $c_1=2$, the boundary length is $2\pi R$, and the Fermi radius
on the boundary is $\sqrt{\mu-R^2}$.  Hence
\[
 C_{\Omega,1}=\frac{2R}{\pi}\sqrt{\mu-R^2}.
\]
The conclusions follow from the main trace formula and
\eqref{eq:entropy-asymptotic-intro}, with
$M_h=(2\pi h)^{-1}\log h^{-1}$.
\end{proof}

\section{A counterexample under weighted integrability}
\label{sec:endpoint-counterexample}

The proof of Theorem~\ref{thm:endpoint-counterexample} uses smooth
interpolation on finite spectral sets with arbitrarily small integrals
against $dt/[t(1-t)]$.

\begin{lemma}[Smooth interpolation on a finite spectral set]
\label{lem:finite-spectral-bumps}
Let $F$ be a finite subset of $(0,1)$ and let $\eta>0$.
There is a function $\phi\in C_c^\infty(0,1)$ such that
\[
 0\le\phi\le1,\qquad \phi(t)=\phi(1-t),\qquad
 \phi|_F=1,\qquad
 \int_0^1\frac{\phi(t)}{t(1-t)}\,dt\le\eta.
\]
\end{lemma}

\begin{proof}
The finite set $F\cup(1-F)$ lies in a compact subinterval of $(0,1)$.
Since $dt/[t(1-t)]$ is locally finite and has no atoms, there is a
symmetric open neighborhood $U\Subset(0,1)$ of this set with
\[
 \int_U\frac{dt}{t(1-t)}<\eta.
\]
Choose $\psi\in C_c^\infty(U)$ with $0\le\psi\le1$ and
$\psi=1$ near $F\cup(1-F)$. Then
$\phi(t)=(\psi(t)+\psi(1-t))/2$ has the required symmetry, is
supported in $U$, and equals one on $F$.
Its weighted integral is at most the weighted measure of $U$.
\end{proof}

\begin{proof}[Proof of Theorem~\ref{thm:endpoint-counterexample}]
The eigenvalues of the harmonic oscillator in
\eqref{eq:counterexample-model} are $(2m+1)h$, $m=0,1,\ldots$;
its eigenfunctions are Hermite polynomials times a Gaussian, with the
semiclassical scaling described in
\cite[Section~6.1, Theorem~6.2]{Zworski}.
Consequently, for
\[
 h_N=\frac1{2N},
\]
the projection $P_{h_N}$ has rank exactly $N$.
Every nonzero vector in its range is a Gaussian multiplied by a
nonzero polynomial. Such a function cannot vanish almost everywhere
on any nonempty open interval. Since both $\Omega$ and its complement
contain nonempty open intervals,
\begin{equation*}
 0<\|Qu\|_2^2<\|u\|_2^2
 \qquad(0\ne u\in\operatorname{Ran}P_{h_N}).
\end{equation*}
Thus the finite-dimensional operator
$P_{h_N}QP_{h_N}|_{\operatorname{Ran}P_{h_N}}$ has all its
eigenvalues strictly between zero and one.

To relate this finite matrix to $T_{h_N}$, set
$A=Q|_{\operatorname{Ran}P_{h_N}}$, considered as a map into
$L^2(\mathbb R)$. Then
\[
 A^*A=P_{h_N}QP_{h_N}|_{\operatorname{Ran}P_{h_N}},
 \qquad AA^*=T_{h_N}.
\]
The nonzero eigenvalues of $A^*A$ and $AA^*$ agree, including
multiplicity. Hence $T_{h_N}$ has exactly $N$ nonzero eigenvalues,
all in $(0,1)$. It also has an infinite-dimensional zero eigenspace
on $L^2(\mathbb R)$.

Now set
\[
 N_k=4^k,\qquad h_k=\frac1{2\cdot4^k},\qquad k\ge1,
\]
and denote the nonzero eigenvalues of $T_{h_k}$, with multiplicity,
by $\lambda_{k,1},\ldots,\lambda_{k,N_k}$.
Apply Lemma~\ref{lem:finite-spectral-bumps} to their finite set,
with $\eta=2^{-k}$, to obtain
$\phi_k\in C_c^\infty(0,1)$ satisfying
\begin{equation*}
 \begin{gathered}
 0\le\phi_k\le1,\qquad \phi_k(t)=\phi_k(1-t),\qquad
 \phi_k(\lambda_{k,j})=1\quad(1\le j\le N_k),\\
 \int_0^1\frac{\phi_k(t)}{t(1-t)}\,dt\le2^{-k}.
 \end{gathered}
\end{equation*}
Extend each $\phi_k$ by zero to $[0,1]$ and define
\begin{equation*}
 g_0(t)=\sum_{k=1}^\infty 2^{-k}\phi_k(t).
\end{equation*}
The series converges uniformly by the Weierstrass test. Therefore
$g_0\in C([0,1])$ is nonnegative, symmetric, and zero at both
endpoints. Tonelli's theorem gives
\begin{equation*}
 0<W(g_0)
 =\sum_{k=1}^\infty2^{-k}
       \int_0^1\frac{\phi_k(t)}{t(1-t)}\,dt
 \le\sum_{k=1}^\infty4^{-k}=\frac13.
\end{equation*}
Strict positivity follows because at least one smooth function
$\phi_k$ equals one at an interior point.

The function $g_0$ is independent of $h$. Since $g_0(0)=0$,
$g_0(T_h)$ has finite rank for every $h>0$. At $h=h_k$,
\begin{equation}
 \operatorname{Tr}g_0(T_{h_k})
 =\sum_{j=1}^{N_k}g_0(\lambda_{k,j})
 \ge 2^{-k}N_k=2^k.
 \label{eq:counterexample-trace-lower}
\end{equation}
Since $\log(h_k^{-1})=(2k+1)\log2$,
\begin{equation}
 \frac{\operatorname{Tr}g_0(T_{h_k})}{\log(h_k^{-1})}
 \ge\frac{2^k}{(2k+1)\log2}
 \longrightarrow+\infty.
 \label{eq:counterexample-subsequence}
\end{equation}

For the endpoint condition \eqref{eq:counterexample-slow-endpoints}, set
\begin{equation*}
 b(t)=\frac1{\log^2(e/[t(1-t)])}\quad(0<t<1),
 \qquad b(0)=b(1)=0,
 \qquad g=g_0+b.
\end{equation*}
The function $b$ is continuous, nonnegative, and symmetric.
For $0<t\le1/2$, one has $t/2\le t(1-t)\le t$, and therefore
\[
 \frac{b(t)}{t(1-t)}
 \le\frac{2}{t\log^2(e/t)}.
\]
The right-hand side is integrable at zero, so symmetry proves
$W(b)<\infty$. Since $\log(e/[t(1-t)])\sim\log(1/t)$ as
$t\downarrow0$, one has $b(t)/t^\alpha\to+\infty$ for every
$\alpha>0$.
Thus $g$ satisfies \eqref{eq:counterexample-slow-endpoints} and has
$0<W(g)<\infty$. Since $g\ge g_0$, its trace retains
\eqref{eq:counterexample-trace-lower} and
\eqref{eq:counterexample-subsequence}. This proves all assertions.
\end{proof}

\begin{remark}[Scope of the counterexample]
Here $C_{\Omega,1}W(g)<\infty$, whereas
\eqref{eq:counterexample-subsequence} diverges. Thus continuity and
weighted integrability do not imply the formula in
\cite[Conjecture~1]{DeleporteLambert}, for any finite normalization
of its coefficient. The lower bound in \eqref{eq:counterexample-subsequence} is supplied by the spectral-peak term $g_0$.

The function classes in \cite{KulikovLarsen2026} satisfy integrability
conditions on monotone endpoint envelopes and a trace-class admissibility
condition. For the present
finite-rank compressions, every continuous $g$ with $g(0)=0$ already
satisfies $g(T_h)\in\mathfrak S_1$ for each $h>0$.
\end{remark}
\begin{remark}[Normalization]\label{rem:DL-normalization}
The coefficient in \cite[Conjecture~1]{DeleporteLambert},
in arXiv:2405.07796v1, is written as $2C_\Omega$.
Since $W(t(1-t))=1$, substitution of $g(t)=t(1-t)$ gives an
extra factor of two relative to \cite[Theorem~1]{DeleporteLambert}.
We use the normalization $C_{\Omega,1}$ fixed by that variance theorem,
which also agrees with the entropy prediction
\cite[(1.7)]{DeleporteLambert}.
\end{remark}

\appendix
\section{Global Schwartz remainders for the auxiliary Hamiltonian}
\label{sec:global-schwartz}\label{app:global-remainders}

Recall that $V_{\mathrm c}(x)=\mu+1+|x|^2$ outside a compact set.
Choose $C\ge0$ so large that
$\inf_{x\in\mathbb R^n}(V_{\mathrm c}(x)-\mu+C)\ge1$, and set
\[
 A_h=H_h^{\mathrm c}-\mu,\qquad
 \mathcal L_h=A_h+C,\qquad
 \mathcal B_h=-h^2\Delta+|x|^2+b,\qquad b=1+C,
\]
with $\mathcal B_1=-\Delta+|x|^2+b$. Then
\[
 \mathcal L_h=\mathcal B_h+w,\qquad
 w(x)=V_{\mathrm c}(x)-\mu-1-|x|^2
      \in C_c^\infty(\mathbb R^n;\mathbb R),
 \qquad \mathcal L_h\ge I.
\]
Constants in this appendix are uniform for $0<h\le1$.

\begin{lemma}[Graph norms and propagation]\label{app:graphnorms}\label{app:propagation}
For every integer $k\ge0$,
\begin{equation}\label{app:graph-norms}
 \mathcal D(\mathcal L_h^k)=\mathcal D(\mathcal B_h^k),\qquad
 \|\mathcal L_h^ku\|_2\asymp_k\|\mathcal B_h^ku\|_2.
\end{equation}
Writing $\mathcal U_h(t)=e^{-itA_h/h}$, one also has, for integers
$k,r\ge0$,
\begin{equation}\label{app:propagation-bound}
 \sup_{t\in\R}
 \|\mathcal B_1^k\partial_t^r\mathcal U_h(t)u\|_2
 \le C_{k,r}h^{-2k-r}\|\mathcal B_1^{k+r}u\|_2,
 \qquad u\in\mathcal S(\R^n).
\end{equation}
\end{lemma}
\begin{proof}
The creation and annihilation operators of the harmonic oscillator give
\begin{equation}\label{app:oscillator-words}
 \|\mathcal B_h^ku\|_2\asymp_k
 \sum_{|\alpha|+|\beta|\le2k}
       \|x^\alpha(h\mathsf D)^\beta u\|_2.
\end{equation}
The constants are uniform: in the scaled Hermite basis every word of
degree at most $2k$ is a finite sum of shifts with coefficients bounded
by the $k$th power of the initial oscillator energy, and
$\mathcal B_h\ge I$. The reverse estimate follows by expanding
$\mathcal B_h^k$; the commutators introduce only powers of $h\le1$.
The Leibniz formula consequently gives
$\|\mathcal B_h^kwu\|_2\le C_k\|\mathcal B_h^ku\|_2$.
Initially these statements hold on $\mathcal S$, and they extend by
closure to $\mathcal D(\mathcal B_h^k)$.

Since $w$ is bounded and preserves each $\mathcal D(\mathcal B_h^k)$,
induction in $\mathcal L_hu=\mathcal B_hu+wu$ gives
$\mathcal D(\mathcal L_h^k)=\mathcal D(\mathcal B_h^k)$.
The norm induction follows from
\[
\begin{aligned}
 \|\mathcal B_h^ku\|_2
 &\le C_k\bigl(\|\mathcal L_h^ku\|_2+
                  \|\mathcal L_h^{k-1}u\|_2\bigr),\\
 \|\mathcal L_h^ku\|_2
 &\le C_k\bigl(\|\mathcal B_h^ku\|_2+
                  \|\mathcal B_h^{k-1}u\|_2\bigr),
 \qquad k\ge1,
\end{aligned}
\]
and $\mathcal L_h,\mathcal B_h\ge I$.

Equation~\eqref{app:oscillator-words} also gives
\[
 \|\mathcal B_1^kv\|_2\le C_kh^{-2k}\|\mathcal B_h^kv\|_2,
 \qquad
 \|\mathcal B_h^kv\|_2\le C_k\|\mathcal B_1^kv\|_2.
\]
By the domain equality, $\bigcap_k\mathcal D(\mathcal L_h^k)=\mathcal S$.
The spectral theorem therefore implies that $\mathcal U_h(t)$ preserves
$\mathcal S$, every $\mathcal L_h$ graph norm, and that
$\partial_t^r\mathcal U_h(t)=(-i/h)^rA_h^r\mathcal U_h(t)$ on
$\mathcal S$. Since $A_h=\mathcal L_h-C$, these facts prove
\eqref{app:propagation-bound}. The $\mathcal B_1$ graph norms generate the Schwartz topology.
\end{proof}

The estimate applies to kernels in the $x$ variable, with $L^2_y$ as a
Hilbert-space parameter. Thus, if $K_h(t)$ has an
$O(h^\infty)$ Schwartz kernel uniformly on $|t|\le t_0$, with all fixed
time derivatives, so does
\[
 h^{-1}\int_0^t \mathcal U_h(t-s)K_h(s)\,ds.
\]
Indeed, spatial weights and derivatives in $y$ commute with $\mathcal U_h$,
while time derivatives introduce only the factors controlled by
\eqref{app:propagation-bound} and endpoint values of $K_h$.
Weighted Sobolev embedding gives the assertion in every Schwartz
seminorm. Pairing such a residual with a compactly supported distribution
in $t$ preserves the same conclusion, since that distribution has finite
order.

\begin{lemma}[Fixed smooth functional calculus]\label{app:fixed-calculus}
For every fixed $f\in C_c^\infty(\R)$ there is a classical symbol
$b_f(x,\xi;h)$ supported in a fixed compact phase-space set such that
\[
 f(A_h)=\Op_h^w(b_f)+R_{f,h},\qquad
 b_f\sim\sum_{j\ge0}h^jb_{f,j},\qquad
 b_{f,0}(x,\xi)=f(|\xi|^2+V_{\mathrm c}(x)-\mu).
\]
The kernel of $R_{f,h}$ is $O(h^\infty)$ in every Schwartz seminorm.
In particular, $R_{f,h}=O_{\mathfrak S_p}(h^\infty)$ for every fixed $p>0$.
\end{lemma}
\begin{proof}
Let $m=1+|x|^2+|\xi|^2$ and $a=|\xi|^2+V_{\mathrm c}-\mu$.
Here $S(m)$ denotes the global semiclassical symbol class associated
with the tempered order function $m$. Then $a\in S(m)$ is real and
$|a\pm i|\ge c m$.
The fixed-function calculus applies; see
\cite[Theorem~8.7]{DimassiSjostrand} and
\cite[Section~12.6, Theorem~83]{GuilleminSternberg}.
Its resolvent construction gives coefficients consisting of derivatives
of $a$ multiplied by derivatives of $f$ evaluated at $a$.
Their supports lie in the compact set $a^{-1}(\operatorname{supp}f)$.
Taking a Borel sum in one fixed compact phase-space set yields
\begin{equation}\label{app:l2-residual}
 \|R_{f,h}\|_{L^2\to L^2}=O(h^N)\quad\text{for every }N.
\end{equation}

By \eqref{app:graph-norms},
\[
 \mathcal B_h^j\mathcal L_h^{-j}\in\mathcal L(L^2),\qquad
 (\mathcal B_h^j\mathcal L_h^{-j})^*
 =\mathcal L_h^{-j}\mathcal B_h^j,
\]
with norms bounded uniformly in $h$. Hence
\[
\begin{aligned}
 \mathcal B_h^j f(A_h)\mathcal B_h^j
 &=(\mathcal B_h^j\mathcal L_h^{-j})
   [\mathcal L_h^{2j}f(A_h)]
   (\mathcal L_h^{-j}\mathcal B_h^j),\\
 \|\mathcal B_h^j f(A_h)\mathcal B_h^j\|
 &\le C_j\sup_{E\in\R}(E+C)^{2j}|f(E)|\le C_{j,f}.
\end{aligned}
\]
Weighted products are understood as bounded extensions from $\mathcal S$. The same bound holds for $\Op_h^w(b_f)$:
Weyl composition with powers of the quadratic oscillator gives
compactly supported symbols with uniformly bounded derivatives.
Consequently $\|\mathcal B_h^jR_{f,h}\mathcal B_h^j\|\le C_{j,f}$.
For $N>1$, let
\[
 \Pi_{h,N}=\mathbf1_{[1,N]}(\mathcal B_h).
\]
The projection $\Pi_{h,N}$ has finite rank and commutes with every
complex power of $\mathcal B_h$. Hence
\[
 \Phi_{h,N}(z)
 =\mathcal B_h^{2jz}\Pi_{h,N}R_{f,h}\Pi_{h,N}
  \mathcal B_h^{2jz},
 \qquad 0\le\operatorname{Re}z\le1,
\]
is operator-valued and entire on the finite-dimensional range of
$\Pi_{h,N}$, and
\[
 \sup_{t\in\mathbb R}\|\Phi_{h,N}(it)\|
 \le\|R_{f,h}\|,\qquad
 \sup_{t\in\mathbb R}\|\Phi_{h,N}(1+it)\|
 \le\|\mathcal B_h^{2j}R_{f,h}\mathcal B_h^{2j}\|.
\]
The three-lines theorem at $z=1/2$ gives
\[
 \|\Pi_{h,N}\mathcal B_h^jR_{f,h}\mathcal B_h^j\Pi_{h,N}\|
 \le \|R_{f,h}\|^{1/2}
       \|\mathcal B_h^{2j}R_{f,h}\mathcal B_h^{2j}\|^{1/2}.
\]
Since $\Pi_{h,N}\to I$ strongly and
$\mathcal B_h^jR_{f,h}\mathcal B_h^j$ is bounded, passage to the limit
$N\to\infty$ yields the same inequality without $\Pi_{h,N}$.
Together with \eqref{app:l2-residual}, this proves
$\|\mathcal B_h^jR_{f,h}\mathcal B_h^j\|=O(h^\infty)$ for each
integer $j\ge0$. To obtain unequal weights, fix integers
$k,\ell\ge0$ and take $j\ge\max\{k,\ell\}$. Then
\[
 \mathcal B_h^kR_{f,h}\mathcal B_h^\ell
 =\mathcal B_h^{k-j}
   (\mathcal B_h^jR_{f,h}\mathcal B_h^j)
   \mathcal B_h^{\ell-j}.
\]
Since $\mathcal B_h\ge I$, the two exterior negative powers have
operator norm at most one. Thus the unequal-weight product is also
$O(h^\infty)$ in operator norm. Lemma~\ref{app:graphnorms} now gives
\[
 \|\mathcal B_1^kR_{f,h}\mathcal B_1^\ell\|
 \le C_{k,\ell}h^{-2k-2\ell}
       \|\mathcal B_h^kR_{f,h}\mathcal B_h^\ell\|
 =O(h^\infty).
\]
For an integer $s>n/2$, the operator $\mathcal B_1^{-s}$ is
Hilbert--Schmidt, and therefore
\[
 \|\mathcal B_1^kR_{f,h}\mathcal B_1^\ell\|_{\mathfrak S_2}
 \le \|\mathcal B_1^kR_{f,h}\mathcal B_1^{\ell+s}\|
       \|\mathcal B_1^{-s}\|_{\mathfrak S_2}
 =O(h^\infty).
\]
Writing $K_{R_{f,h}}$ for the Schwartz kernel, the reality of the
oscillator coefficients gives
\[
 \|\mathcal B_{1,x}^k\mathcal B_{1,y}^{\ell}K_{R_{f,h}}\|_{L^2(\mathbb R^{2n})}
 =\|\mathcal B_1^kR_{f,h}\mathcal B_1^{\ell}\|_{\mathfrak S_2}
 =O(h^\infty).
\]
Weighted Sobolev embedding in $(x,y)$ proves the Schwartz assertion.
Let $\{\varphi_\alpha\}_{\alpha\in\mathbb N_0^n}$ be the standard
Hermite basis, and put
\[
 r_{\alpha\beta}(h)=\langle R_{f,h}\varphi_\beta,\varphi_\alpha\rangle.
\]
The Schwartz kernel estimates imply, for all integers $N,M\ge0$,
\[
 |r_{\alpha\beta}(h)|
 \le C_{N,M}h^N(1+|\alpha|)^{-M}(1+|\beta|)^{-M}.
\]
For $0<p\le1$, choose $Mp>n$. Rank-one summation gives
\[
 \begin{aligned}
 \|R_{f,h}\|_{\mathfrak S_p}^p
 &\le\sum_{\alpha,\beta}|r_{\alpha\beta}(h)|^p\\
 &\le C_{N,M,p}h^{Np}
 \left(\sum_{\alpha\in\mathbb N_0^n}(1+|\alpha|)^{-Mp}\right)^2
 \le C_{N,M,p}h^{Np}.
 \end{aligned}
\]
For $p>1$, use
$\|R_{f,h}\|_{\mathfrak S_p}\le\|R_{f,h}\|_{\mathfrak S_1}$.
Since $N$ is arbitrary, this proves the assertion for every fixed $p>0$.
\end{proof}

\section*{Acknowledgments}

\noindent\textbf{Funding.} The authors were partially supported by the
National Natural Science Foundation of China (Grant numbers
12171484 and 12571140), the Natural Science Foundation of Hunan Province
(Grant number 2023JJ20056), the Science and Technology Innovation Program
of Hunan Province (Grant number 2023RC3028), the Central South University
Innovation-Driven Research Programme (Grant number 2023CXQD032), and
the Hunan Basic Science Research Center for Mathematical Analysis
(2024JC2002).

\smallskip
\noindent\textbf{AI Statement.}
The author used OpenAI’s ChatGPT (GPT-5.6 Sol) for language editing,
LaTeXformatting,
and limited assistance with local mathematical reasoning. The proof strategy
andmathematical
development are the author’ own; all mathematical content was
independentlywritten and checked by the author, who take full responsibility for
it.

\end{document}